\documentclass[usegraphicx,usenatbib]{mn2e}

\usepackage{psfig,dynamics}

\title[Dynamics of Core Accretion]{Dynamics of Core Accretion}

\author[Andrew F. Nelson and Maximilian Ruffert]
       {Andrew F. Nelson$^1$\thanks{Former fellow of the United Kingdom 
        Astrophysical Fluid Facility (UKAFF)}
        and Maximilian Ruffert$^{2}$\\
$^{1}$ Los Alamos National Laboratory, XCP-2 MS T087,
               Los Alamos NM, 87545, USA \\
$^{2}$ School of Mathematics and Maxwell Institute, University of Edinburgh,
                          Edinburgh Scotland EH9~3JZ\\
     E-mail: andy.nelson@lanl.gov (AFN); M.Ruffert@ed.ac.uk (MR)} 

\date{Accepted 2345, February 31. Received 2346, February 32; in original form
2347, February 33}

\begin{document}

\maketitle

\label{firstpage}

\begin{abstract}

We perform 3-dimensional hydrodynamic simulations of gas flowing
around a planetary core of mass \mplan=10\me\, embedded in a near
Keplerian background flow, using a modified shearing box
approximation. We assume an ideal gas behavior following an equation
of state with a fixed ratio of the specific heats, $\gamma=1.42$,
consistent with the conditions of a moderate temperature background
disk with solar composition. No radiative heating or cooling is
included in the models. We employ a nested grid hydrodynamic code
implementing the `Piecewise Parabolic Method' with as many as six
fixed nested grids, providing spatial resolution on the finest grid
comparable to the present day diameters of Neptune and Uranus. 

We find that a strongly dynamically active flow develops such that no
static envelope can form. The activity is not sensitive to plausible
variations in the rotation curve of the underlying disk. It is
sensitive to the thermodynamic treatment of the gas, as modeled by
prescribed equations of state (either `locally isothermal' or `locally
isentropic') and the temperature of the background disk material. The
activity is also sensitive to the shape and depth of the core's
gravitational potential, through its mass and gravitational softening
coefficient. Each of these factors influence the magnitude and
character of hydrodynamic feedback of the small scale flow on the
background, and we conclude that accurate modeling of such feedback is
critical to a complete understanding of the core accretion process.

The varying flow pattern gives rise to large, irregular eruptions of
matter from the region around the core which return matter to the
background flow: mass in the envelope at one time may not be found in
the envelope at any later time. No net mass accretion into the
envelope is observed over the course of the simulation and none is
expected, due to our neglect of cooling. Except in cases of very rapid
cooling however, as defined by locally isothermal or isentropic
treatments, any cooling that does affect the envelope material will
have limited consequences for the dynamics, since the flow quickly
carries cooled material out of the core's environment entirely. The
angular momentum of material in the envelope, relative to the core,
varies both in magnitude and in sign on time scales of days to months
near the core and on time scales a few years at distances comparable
to the Hill radius. The dynamical activity contrasts with the largely
static behavior typically assumed within the framework of the core
accretion model for Jovian planet formation.

We show that material entering the dynamically active environment may
suffer intense heating and cooling events the durations of which are
as short as a few hours to a few days. Shorter durations are not
observable in our work due to the limits of our resolution. Peak
temperatures in these events range from $T \sim 1000$~K to as high as
$T \sim 3-4000$~K, with densities $\rho\sim
10^{-9}-10^{-8}$~g/cm$^3$. These time scales, densities and
temperatures span a range consistent with those required for chondrule
formation in the nebular shock model. We therefore propose that
dynamical activity in the Jovian planet formation environment could be
responsible for the production of chondrules and other annealed
silicates in the solar nebula.

\end{abstract}

\begin{keywords}
hydrodynamics, planets: formation, planetary systems: formation
\end{keywords}

\section{Introduction}

The classical `core accretion' model for Jovian planet formation
\citep{PP4_WGL} is characterized by the idea that a 5--15 earth mass
(\me) rock/ice core forms at a distance of $\sim$5~AU or more from
the central star \citep{Boss95}, with a gaseous envelope surrounding
it. At some critical point during the growth, the gaseous envelope
becomes unstable and collapses, ultimately accreting as much as
several times the mass of Jupiter.

In outline, the core accretion model consists of three distinct phases
\citep{Pol96}. First, a sold core accretes from smaller solid bodies
like comets and asteroids in the solar nebula. The mass of this core
may be as much as 20-30 \me, depending on the model, though values of
5-15\me\  are much more typically expected. The relatively rapid
growth of this core is cut off when it depletes its `feeding zone', at
which point begins a much slower period of growth of both the core,
from material migrating into the feeding zone from other parts of the
disk, and of the surrounding hydrostatic gaseous envelope. In this
second phase, the gas accretion rate is regulated by the balance of
radiative cooling and the continued heating of the envelope by a low
rate of solid body accretion. When the combined mass of the core and
envelope reaches a (model dependent) critical value of $\sim30$\me,
the hydrostatic envelope becomes unstable and period of rapid gas
accretion follows, defining the third, `core instability', stage of
evolution. The planet's final mass is determined during this phase.

The most critical problem afflicting the core accretion model is that
the second stage may require as much as 6 million years or more
\citep{Pol96}, but the circumstellar disk out of which the planet
purportedly forms may have a lifetime of only 2-4 million years
\citep{HLL01}. More critically, when the dynamical consequences of the
interaction between the forming planet and disk are taken into
account, the planet's lifetime may shrink to only a few $\times10^4$
years before it migrates inwards toward the star and is accreted by it
\citep[see e.g.][and references therein]{PP4_WH}. While some migration
is certainly necessary, since the vast majority of extra solar planets
so far discovered have been observed in orbits less than (often much
less than) the $\sim$5~AU where they are expected to form, the rates
expected from theory would imply that no planets could outlive the
disks in which they form.

\subsection{Newer wrinkles in core accretion}\label{sec:new-coreacc} 

\citet{Pol96} show that among the most sensitive input parameters for
the formation time scale is how the planetesimal accretion rate varies
with time, especially in the second phase of growth after the core has
grown substantially and depleted its feeding zone. The cause of the
sensitivity is the continued heating of the envelope gas by the solids
as they pass through and are dissolved in the envelope when they
accrete. The heating then stalls both further envelope contraction and
gas accretion onto the core. 

These models were performed in the context of a single, isolated core
in a background disk where no other cores competed for the same
resources and no migration through the disk occurred. The time scale can
be shortened considerably when either assumption is relaxed.
\citet{AMB04} have shown that including migration can shorten the
required lifetime to $\la1$Myr even in a much less massive disk than
originally studied because, due to the migration the planet does not
deplete its feeding zone for solid bodies. Instead, its heavy element
core continues to grow rapidly, triggering runaway gas accretion much
more quickly than otherwise. A possibly undesirable consequence of the
model is that critical core mass is higher, perhaps in conflict with
the data for the observed core mass of Jupiter. 

Although not including migration, \citet{HBL05} have simulated the
effects of competing embryos with a `cutoff' of the solid body
accretion when the core reaches a predetermined mass, modeling the
fact that the solids at more than some critical distance from one core
have been accreted by another. In addition, they study the sensitivity
of their models to the treatment of opacity in the envelope, which
governs the rate at which cooling of the envelope (followed by
contraction and further accretion) can proceed. Previous studies have
used opacities that are appropriate for interstellar grains and
therefore do not properly model the processing expected during the
evolution of the circumstellar disk and later the gaseous planetary
envelope itself. Such processing, especially in the envelope, means
that opacities will be reduced as grains coagulate and `rain out' of
the forming planet's upper envelope. \citet{HBL05} model such
processing with a temperature dependent reduction of the interstellar
opacity value and, coupled with the cutoff mass, find that the
required formation time scale is reduced to as little as 1 Myr, with
critical core masses of 5-10\me. 

\subsection{Shortcomings in the models and goals for this
work}\label{sec:goals} 

An important underlying assumption made in core accretion models is
that a hydrostatic envelope structure exists, so that a one
dimensional approximation may be used to reproduce the evolution. On
the other hand, previous numerical studies of planets embedded in
disks have shown that this hydrostatic assumption may not be
realistic, even for very low mass cores. For example, \citet{DHK02}
show that in 2D nested grid simulations, spiral structure exists in
the circumplanetary disks of cores as small as 3\me, and that for
cores of $\ga10$\me\  strong spiral shocks are present. \citet{BLOM}
perform similar simulations in 3D and also observe spiral structures
and rotation, though they do not find shock structures as strong as in
the 2D work. Their simulations implemented an isothermal equation of
state however, which is certainly inappropriate for such dense regions
as the envelope of a forming planet and may result in flows that do
not reproduce the true picture of the evolution. Later work of
\citet[][hereafer AB09]{AB09} relaxes the isothermal assumption in
some of their simulations, instead using a flux limited radiative
transfer model of a small cutout region of the disk whose evolution
was simulated with a Smoothed Particle Hydrodynamics (SPH) code with a
resolution of $3\times10^4$ SPH particles. Their conclusions are
similar to the isothermal results in the sense that in both cases,
spiral structures and/or envelope asymmetries develop, but are weaker
in the radiative transfer models, and accretion rates onto the core
are lower.

Although these studies are good first steps, to date there has been
little exploration of the region of parameter space between the two
extremes of, at one end, detailed modeling of the disk with very
coarse modeling of the core and envelope, and at the other, of
detailed modeling of the core and envelope with coarse or non-existent
modeling of the disk. The simulations of AB09 for example, which are
among those with the highest claimed resolution of the region around
the core, employ a total of only $3\times10^4$ or $1\times10^5$ (in
different simulations) particles over the entire volume. In order to
fully understand the core accretion scenario, detailed models of the
flow around and onto the core at much higher resolution are necessary.

Models of accretion onto compact objects have been published
\citep[see e.g.][and references therein]{R97,R99}, and in many cases
large scale instabilities in the flow develop \citet{FR99} that are
often characterized by `flip-flop' behavior in the sense that angular
momentum accretion can temporarily switch sign. While 3D models
produce uniformly less active flow than 2D models, both produce such
behavior and in most cases, the accretion rate of various flow
variables is comparable to the classical Bondi-Hoyle rate even in
flows with transverse gradients. 

The applicability of these models and their results to planet
formation would appear to be quite limited however because they
typically assume a central body which accretes matter with perfect
efficiency (e.g. a `black hole'). In the context of planet formation,
the accretor is of course a rocky planetary core, onto which gas
accretion is forbidden and an envelope structure develops. Moreover,
while the background gas flow in previous studies may include some
transverse velocity or density gradients, it is uni-directional. In
contrast, the flow around a forming planet is quite complex, with a
flow that circulates in opposite directions relative to the planet
inside and outside its orbit, but which librates in a region near its
orbit. The existence of the planet may drive complex interactions
between each of the three regions and with planet's envelope as well. 

With such large differences in the accretion assumptions and in the
background flow pattern, will similar sorts of dynamical activity as
in the simpler cases occur? How does the existence of dynamical
activity (if indeed it is present) affect the time scale required for
the core accretion model to succeed? In other words, can dynamical
activity somehow short circuit the slow, near hydrostatic evolutionary
process assumed in the second phase of planet growth, replacing it
with a more rapid growth phase that avoids the problems of rapid
\tone\ migration? Or does it instead lengthen the process? 

The goal of this paper is to begin an investigation of the accretion
process in order to determine the character of any dynamical activity
that may occur. We will extend the previous comparatively low
resolution, two and three dimensional calculations exploring planet
migration to very high resolution three dimensional calculations
specifically studying the mass accreting onto a 10\me\  core. The
models in our study will consist of a small cubic `cutout' region of a
circumstellar disk and centered on the core. They will therefore not
include migration of the planet itself, but will include some effects
of the background on the accretion flow, such as the effects of
different disk density and temperature distributions. In section
\ref{sec:phys-numerics} we define the physical conditions assumed in
our models and numerical methods used to realize the evolution, as
well as the initial conditions we implement for our numerical models.
Section \ref{sec:results} contains the basic results of our
simulations and introduces some metrics by which those results may be
compared more quantitatively. Section \ref{sec:sensitivity} expands on
these analyses with further studies of the sensitivity of the results
to various parameters. Finally, section \ref{sec:chondrules} proposes
a potential link between the dynamical activity around a forming
Jovian planet and the formation of chondrules. Our concluding remarks
are found in section \ref{sec:discuss}, where we summarize our
conclusions and the implications they have for Jovian planet
formation and compare our results with previous work. Finally, in 
section \ref{sec:questions}, we discuss a number of questions that 
may be profitably addressed by future work similar to our own.

\section{The physical model}\label{sec:phys-numerics} 

In this section we describe the physical models we use to study the
flow pattern of gas around the forming planetary core, the numerical
code used to evolve the models forward in time and the initial
conditions of those models. Our study will include or exclude a
variety of physical properties of the system in different simulations.
Therefore we will outline the most physically inclusive models, and in
later discussion note specifically which properties have been included
in which simulation. In general, the properties we study are either
substitutive, eg. either one equation of state or another, or
cumulative, e.g. gravity from different components of the system, so
that separate, excluded components can simply be neglected in the
specification of the initial state.

\subsection{The mathematical framework, numerical realization and
physical assumptions}\label{sec:physics}

In previous works studying disks, it has been customary to implement
either a fully curvilinear coordinate system (usually cylindrical) or
a `shearing sheet' approximation. The latter form is an approximation
in which locally Cartesian coordinate system is used rather than the
full equations for curvilinear coordinates which simplifies the
analysis, especially in the context of linearized instability
analyses. It neglects most of the non-uniform geometric features of
the coordinate system and instead adds a number of terms to model
approximately some of its effects, and the rotation of the coordinate
frame itself.

We have implemented neither procedure here, but rather an intermediate
step between them which we call a `modified shearing sheet' coordinate
system. As does the standard shearing sheet coordinate system, this
modified shearing sheet system neglects the geometric terms in the
full curvilinear system of equations. Its main difference from the
standard form is that it retains a more precise treatment of the
gravity and fictitious forces on the gas, as noted below. Due to our
somewhat unusual treatment of the coordinate system, we shall describe
here in some detail both the derivation of the background analytic and
numerical frameworks, as well as the initial and boundary conditions
used in our models. 

\subsubsection{Mapping cylindrical to modified shearing sheet
coordinates}\label{sec:mapping}

Like the shearing sheet, our mathematical formalism requires that the
true cylindrical coordinate system be mapped onto a Cartesian
coordinate system centered on the planet. We define the mapping so
that the three pairs ($x$, $r$), ($y$, $r\phi$) and ($z$, $z$)
each correspond as follows. Mapping the cylindrical coordinates of the
disk to the Cartesian coordinates of our grid requires the three
identifications
\begin{equation}\label{eq:rtox}
x = r - a_{\rm pl};\ \ y = r \phi ; \ \ z=z
\end{equation}
with $a_{\rm pl}$ defined as the semi-major axis of the planet. This
identification means that each $x$ grid coordinate is defined to be at
the same distance from the star even though this condition would not
be true for a true Cartesian grid centered on the planet. The errors
that are made by such an identification will be small as long as the
extent of the coordinate grid itself is small, compared to the
distance to the origin.

In our models, we typically simulate the gas flow in a region
$\sim0.45$~AU (4 `Hill radii'  from a 10\me\ object) in each coordinate
direction from a point centered 5.2~AU away from the origin. In a
cylindrical coordinate system, two `cubic' volume elements at the
inner and outer edges of a grid of this size (i.e. $r dr d\phi dz$,
with each side of the volume identical in length) will differ in
actual volume by a factor $\sim1.4$, whereas in our Cartesian
coordinate system they will be equal. Because the flow is largely
azimuthally directed, very little gas will undergo (or fail to undergo
in our coordinate system) expansion or contraction by such a factor
and therefore only small errors will develop from this source. 

\subsubsection{The equations of hydrodynamics}\label{sec:hydro-eq}

Here we state explicitly the fully expanded form of the equations we
solve. The hydrodynamic variables for mass, momentum and energy
conservation are defined with the usual symbols with $\rho$ referring
to the mass density, $v_x$, $v_y$ and $v_z$ to the velocities in each
of the three coordinate directions and $E=\rho(u + v^2/2)$ to the
total energy, where $v$ is the magnitude of the total velocity and $u$
is the specific internal energy.

The continuity equation for mass conservation is given by
\begin{equation}\label{eq:continuity}
         {{\partial \rho      }\over{\partial t}}  + 
         {{\partial (\rho v_x)}\over{\partial x}}  +
         {{\partial (\rho v_y)}\over{\partial y}}  +
         {{\partial (\rho v_z)}\over{\partial z}}  = 0,
\end{equation}
and the equations of motion in the three Cartesian directions are
given by
\begin{eqnarray}\label{eq:force-x}
      {{\partial (\rho v_x    )}\over{\partial t}} + 
      {{\partial (\rho v_x v_x)}\over{\partial x}} +  
      {{\partial (\rho v_x v_y)}\over{\partial y}} +  
      {{\partial (\rho v_x v_z)}\over{\partial z}} -
      {{ \rho V_y^2 }\over{ a_{\rm pl} + x}}
      & & \nonumber \\
    = -       {{\partial p   }\over{\partial x}}
      - \rho {{\partial \Phi }\over{\partial x}},
      & &
\end{eqnarray}
\begin{eqnarray}\label{eq:force-y}
      {{\partial (\rho v_y    )}\over{\partial t}} + 
      {{\partial (\rho v_y v_x)}\over{\partial x}} +
      {{\partial (\rho v_y v_y)}\over{\partial y}} +
      {{\partial (\rho v_y v_z)}\over{\partial z}} + 
             {{\rho v_x v_y }\over{a_{\rm pl} + x}}
      & & \nonumber \\
            +2{{\rho v_x V_y }\over{a_{\rm pl} + x}}
    = -      {{\partial p    }\over{\partial y}}
      - \rho {{\partial \Phi }\over{\partial y}}
      & &
\end{eqnarray}
and
\begin{eqnarray}\label{eq:force-z}
      {{\partial (\rho v_z    )}\over{\partial t}} + 
      {{\partial (\rho v_z v_x)}\over{\partial x}} +
      {{\partial (\rho v_z v_y)}\over{\partial y}} +
      {{\partial (\rho v_z v_z)}\over{\partial z}}  
      \ \ \ \ \ \ \ \ \ \ \ \ \ \
      & & \nonumber \\
   =  -      {{\partial p    }\over{\partial z}} 
      - \rho {{\partial \Phi }\over{\partial z}}.
\end{eqnarray}
For models in which an equation for the conservation of energy  
is required (i.e. models without locally isothermal or locally
isentropic equations of state), the energy conservation equation is
\begin{eqnarray}\label{eq:energ}
{{\partial E }\over{\partial t}}   
+{{\partial ( v_x (E + p))}\over {\partial x}}
+{{\partial ( v_y (E + p))}\over {\partial y}}
+{{\partial ( v_z (E + p))}\over {\partial z}}
   & &  \nonumber \\
       =  - \rho \left( v_x {{\partial \Phi}\over{\partial x}} +
                        v_y {{\partial \Phi}\over{\partial y}} +   
                        v_z {{\partial \Phi}\over{\partial z}}\right).
\end{eqnarray}
In equations \ref{eq:force-x} and \ref{eq:force-y} above, $V_y$ is
defined as
\begin{equation}\label{eq:vy-full}
V_y=v_y + (a_{\rm pl} + x)\Omega_{\rm pl}
\end{equation}
where $\Omega_{\rm pl}$ is the orbital angular velocity of the planet.
$V_y$ corresponds to the total azimuth velocity in a non-rotating
cylindrical coordinate frame. An equation of state for the gas closes
the system of equations and is described in section \ref{sec:eos}. The
form of the total gravitational potential, $\Phi$, is discussed in
section \ref{sec:gravity}, below.

Note that the above set of equations include no explicit contribution
due to any viscous properties of the fluid. Such effects are well
known to be present in the background circumstellar disk, with their
origin commonly ascribed to turbulence generated by the so-called
Magneto-Rotational instability \citep{BHS-MRI96}. Both the spatial and
time scales relevant for this source of dissipation are inapplicable
to the immediate environment of a core embedded in a circumstellar
disk however, and we therefore neglect turbulence as a source of
dissipation in our simulations. We similarly neglect other, less
specific, sources of dissipation in favor of modeling only those
characteristics of the flow to which we can define explicitly. Coupled
with our numerical code (section \ref{sec:code}), which itself
contributes little dissipation to its solution of the flow equations,
our simulations will therefore be highly inviscid. The only
significant sources of dissipation will originate in shocks that
develop in the flow, which the numerical scheme is known to handle
well.

For reference, we point out the differences between our treatment and
both the fully curvi-linear coordinate system and the shearing sheet,
in Appendix \ref{sec:equation-diffs}.

\subsubsection{The thermodynamic treatment of the gas}\label{sec:eos}

The system of hydrodynamic equations are closed assuming an ideal gas
equation of state given by:
\begin{equation}\label{eq:ideal-eos3d}
p=(\gamma - 1)\rho u
\end{equation}
where $\gamma$ is the ratio of specific heats and $p$, $\rho$ and $u$
are respectively the pressure, density and specific internal energy of
the gas. In order to derive the global initial condition for our
simulations, we employ a vertically integrated (2D) equation of state
of the form:
\begin{equation}\label{eq:ideal-eos2d}
P=(\gamma - 1)\Sigma u
\end{equation}
where $P$ and $\Sigma$ are the vertically integrated pressure and the
mass surface density respectively. This form is exactly analogous to
that used in 3D for the simulations themselves, so that no adjustments
to the rotation curve or any other quantities are required to establish
a steady state initial condition. Finally, the sound speed and
temperature are related by the equation
\begin{equation}\label{eq:cspd}
c^2_s = { { \gamma R T }\over{\mu}}
\end{equation}
where $R$ is the gas constant and $\mu=2.31$ is the mean molecular
weight of a gas of solar composition. 

The choice of $\gamma$ is not obvious in the range of temperatures and
densities we study because in various regimes, some or all of a number
of internal molecular states may be important. In order to avoid such
complications, we assume that the gas is of solar composition with an
effective single component accounting for both hydrogen and helium, so
that the value for $\gamma=1.42$. This value implies that the
rotational degrees of freedom in hydrogen gas are active
(corresponding to gas temperatures $\ga 100$~K) but its vibrational
degrees of freedom are inactive (corresponding to temperatures $\la
800-1000$~K), and will be most representative of moderate temperature
regions of the disk expected throughout most regions of our
simulations. 

Apart from heating from hydrodynamic processes like $PdV$ work and
shocks, we impose no other assumptions regarding the heating and
cooling processes that are active in the disk. Of particular
importance in this regard is our neglect of radiative heating and
cooling, which will undoubtedly modify the behavior of the models to
some extent. As a first, crude attempt to model such effects, we also
perform a number of simulations using either a locally isothermal or
locally isentropic equation of state, defined respectively as
\begin{equation}\label{eq:iso-eos}
p= \rho c_s^2
\end{equation}
and 
\begin{equation}\label{eq:adi-eos}
p= K \rho^\gamma 
\end{equation}
with the value of $\gamma$ set to 1.01 or 1.42 for the locally
isothermal or locally isentropic cases respectively. The physical
interpretation these equations of state in terms of their implications
for heating and cooling are discussed in \citet{DynII} and
\citet{Pick03}. The values of $c_s^2$ and $K$ are set by the initial
conditions of the simulation and are thereafter fixed for its
duration. As with the ideal gas case, vertically integrated forms are
employed to determine the global initial condition, prior to the
beginning of the actual simulation.

\subsubsection{The treatment of gravity}\label{sec:gravity}

Gravity from the central star, from the planet core and from the
circumstellar disk are included in our simulations. The contribution
from the disk is further split into two parts, representing the
portion of the disk inside and outside the simulation box.

The total gravitational potential is defined as a sum of several
components from the star $\Phi_*$, the planet $\Phi_{\rm pl}$, the
background disk $\Phi_D$, and the mass inside our simulation cube
$\Phi_{\rm loc}$ as: 
\begin{equation}\label{eq:gpot-sum}
\Phi = \Phi_* + \Phi_{\rm pl} + \Phi_D + 
                         \Phi_{\rm loc} - \Phi^0_{\rm loc}.
\end{equation}
The last term, $\Phi^0_{\rm loc}$, describing the initial value of the
local gravitational potential due to material in the simulation cube,
is required to avoid double counting disk mass that is otherwise
computed in both the background disk and then again locally in the
simulation cube, as we discuss in more detail below.

We assume that the central star is fixed at all times during our
simulations, which means that its contribution to the gravitational
force on the gas is constant. The gravity from the planetary core is
also fixed in time, relative to the grid. Both the star and the planet
are modeled as softened point masses with a softening that is similar
in form to a Plummer law \citep{R94} with a potential given by  
\begin{equation}\label{eq:pm-grav}
\Phi =  { { -G M}\over{( r^2 +
\epsilon^2\delta^2\exp[-r^2/\epsilon^2\delta^2])^{1/2} }}
\end{equation}
where $G$ is the gravitational constant and the quantities $\Phi$ and
$M$ are replaced by with variants subscripted for the star or the
planet planet as appropriate. The softening parameter $\epsilon$ is a
tunable constant multiplying the grid spacing $\delta$, and in our
simulations is set to unity.  The exponential term in the denominator
causes the softening to decrease to zero at distances of more than a
few grid spacings from the point mass, where it is not needed.

The distances, $r$, between the star or planet and each individual
grid cell are determined from the grid spacing itself, with one
important modification. In order to simplify our initial condition
(see section \ref{sec:local-init}), we neglect the $y$ and $z$
components of stellar gravity, so that the distance from the star to
any point in the grid includes only the $x$ component, effectively
treating the $x$ coordinate (plus the offset, $a_{\rm pl}$, defining
the planet's semi-major axis in the underlying circumstellar disk) as
a radial coordinate. The planet's contribution includes all three
Cartesian components of the force. 

The gravitational potential of the disk is split into two components,
one from the disk as a whole (the `global' component), and a second
from the portion of the disk included in our cubic cutout region (the
`local' component). The global disk self gravity is computed once for
the initial state, and thereafter remains constant for the duration of
the simulation. The local potential is dependent on the time dependent
flow through the disk and must therefore be recomputed at each time
step. 

We calculate the global component from the specified global mass
distribution of the disk (see \ref{sec:global-init} below) using a
Fourier transform based gravitational potential solver in cylindrical
coordinates \citep[see e.g.][]{GalDyn}. Although the disk is
axisymmetric, so that we would expect that a one dimensional solution
for the potential with such a solver would be adequate, we have found
that in practice this solver requires that a number of azimuth
cells be used in order to converge to a well determined potential
value. As discussed in section \ref{sec:global-init} below, the radial
grid for our global initial condition requires as many as
approximately 30000 cells for our highest resolution models. We have
found that with this radial resolution we require at least 1024
azimuth grid cells be included in the potential solver to produce a
potential solution that is accurate to numerical double precision
(i.e. one part in $\sim10^{15}$) over the entire radial range of 0.5
to 20 AU included in the global initial condition. Although we have
made no specific investigation of the reasons for the required number
of azimuth cells, it seems likely to be due to the existence of a
small self-contribution to the potential as obtained from the Fourier
transform technique in cylindrical coordinates \citep{masset_thesis},
for which we attempted no correction. As for the stellar gravity, the
background disk gravity is mapped to the $x$ coordinate only.

The gravitational potential of material on the grid, $\Phi_{\rm loc}$,
is determined from the Poisson equation 
\begin{equation}\label{eq:poisson}
\left({{\partial^2 }\over{\partial x^2}}  +
      {{\partial^2 }\over{\partial y^2}}  +
      {{\partial^2 }\over{\partial z^2}} \right)\Phi_{\rm loc} 
                                                 = 4\pi G\rho.
\end{equation}
We compute its value at each time using a 3D Cartesian coordinate FFT
based potential solver. Due to the nested grid structure and the
inclusion of the global disk potential there are two additional
complications regarding the calculation of the potential that do not
occur in the standard method of solution on a single grid. Both
complications can be circumvented by straightforward application of
the fact that the Poisson equation is linear, so that separate terms
may be added and subtracted independently.

The first complication is that the potential calculation on each of
the nested grids require contributions from mass outside of the
sub-cube on which the potential is to be calculated. \citet{R92}, notes
that the full potential can be obtained from a sum of the contribution
from a given fine grid, and from a second potential calculation that
is performed on the overlying coarse grid, in which the mass on the
finer grid is temporarily deleted. The portion of this second
potential (on a given coarse grid) that overlaps the fine grid is
mapped onto that fine grid as a background term and summed with the
fine grid computation to obtain a correct final value. 

The second complication arises because the global gravitational
potential of the disk accounts for the contribution due to mass
located both inside and outside of our local simulation box. This
means that simply adding the contribution from the mass distribution
within the simulation box would result in that matter being accounted
for twice in the net gravitational potential. To determine the correct
local potential, we first compute the gravitational potential of
matter in the box due to the mass distribution of the initial state
and save this value for the life of the simulation. The initial
contribution is subtracted from the potential calculated at all later
times, so that effectively only the difference in the mass
distribution of the later and initial states is accounted for in the
local potential and the double counting is avoided. 

\subsubsection{The numerical code}\label{sec:code}

To model the flow at the extremely high spatial resolution required we
have modified the code of \citet{R92}. This code is based on the
`Piecewise Parabolic Method' (PPM) of \citet{ColWood84}, in which a
high order polynomial interpolation is used to determine cell edge
values used in calculating a second order solution to a Riemann
problem at each cell boundary.  The interpolation is modified in
regions of sharp discontinuities to track shocks and contact
discontinuities more closely and retain their sharpness, while a
monotonizing condition smoothes out unphysical oscillations. The
solution to the one-dimensional Riemann problem is then used to
calculate fluxes and advance the solution in time. Since the Riemann
problem solution explicitly models the physical dissipation present in
shock structures, no artificial viscosity is required for stability of
the code, and none is included. 

This code solves the hydrodynamic equations on a three dimensional
Cartesian grid, and includes the possibility of including a series of
statically generated nested grids. Each grid contains an identical
number of zones in each of the three coordinate directions, but the
grid spacing decreases by a factor of two when proceeding from a
coarse to fine grid. 

We have modified the original, strictly Cartesian code to account for
centrifugal and Coriolis forces required in a coordinate system
rotating with the planet core, but have neglected terms required to
reproduce a fully curvi-linear coordinate system, as noted in section
\ref{sec:hydro-eq}, above. This approximation will remain valid so
long as the region modeled is relatively small, so that the curvature
is unimportant.

Depending on the simulation (see table \ref{tab:sims} below), between
three and six nested grids each with $64^3$ zones are used to model a
cubic region around the planet. Successively finer grids have the same
number of zones, but half of the linear dimensions of the immediately
coarser grid and are also centered on the planet. When the simulation
volume is defined by a region $\pm4$ Hill radii (section \ref{sec:radii}) 
around a 10\me\  planet, the linear resolution on the finest grid of our
highest resolution models (6 nested grids) is 
$\delta x \approx 6.5\times10^{9}$~cm in size, about 30\% larger than
the diameter of Neptune or Uranus and the spatial volume is resolved
with a total of approximately 1.4 million zones.

\subsubsection{The Hill and accretion radii}\label{sec:radii}

In the discussion throughout this paper, two quantities will appear
repeatedly, and so we introduce them here. They are the accretion radius
of the planet core, defined as
\begin{equation}\label{eq:acc-r}
R_{\rm A} = {{ G M_{\rm pl}} \over{c_s^2}}
\end{equation}
and its Hill radius
\begin{equation}\label{eq:Hill-r}
R_{\rm H} = a_{\rm pl}\left({{M_{\rm pl}}\over{3M_*}}\right)^{1/3}
\end{equation}
where $G$ is the gravitational constant, \mplan\  and $M_*$ are the
mass of the planet core and the central star, respectively, $a_{\rm
pl}$ is the semi-major axis at which the core orbits the star and
$c_s^2$ is the square of the sound speed of the gas in the unperturbed
(initial) disk near the core. 

Taken in its original context of the Bondi accretion model, the
accretion radius defines the distance from the core at which the
magnitude of the internal energy of an isothermal gas and its
gravitational potential energy become comparable. It therefore
represents a measure of the spatial scale over which exchanges between
the fluid's thermal energy and energy bound up in the bulk kinetic
motion derived from its infall onto the core might be expected to lead
to dynamical activity. Together with the mass of the core, the
background temperature, through its presence in the relation for the
sound speed, completely specifies the accretion radius.

The Hill radius defines another spatial scale of interest. Its value
is derived from the Jacobi integral, a quantity which plays a role
similar to the total energy in other (inertial) systems, and which is
the only conserved quantity in the restricted three body problem. It
relates the gravitational potential energies imposed by the core and
star on a test particle with that particle's kinetic energy as it
travels through the combined system and a fourth `centrifugal
potential' term quantifying the effects of the non-inertial reference
frame assumed for the system. It also defines the inner and outer
extents of the Roche lobe of the planet, which is the full 3D volume
around the core for which the combined gravitational influences
of the core and star are comparable to the influence of the fictitious
forces arising from the rotating reference frame.

\subsection{Initial conditions}\label{sec:initconds}

We develop the initial conditions for our simulations in a two stage
process. First, we define a set of global conditions that describe an
entire circumstellar disk in two spatial dimensions ($r,\phi$). From
these conditions, we extract a three dimensional cubic `cutout' region
centered on a point mass, assumed to be embedded within the disk. This
dual specification requires a small note of clarification for readers,
in that in various places throughout the text, `mass density' might
refer to either a mass per unit area, or to a mass per unit volume.
While in most contexts the distinction will be clear, we will
typically refer to each as either `surface' or `volume' density,
respectively.

\subsubsection{The Global Model}\label{sec:global-init}

\begin{figure*}
\rotatebox{-90}{
 \includegraphics[width=14cm]{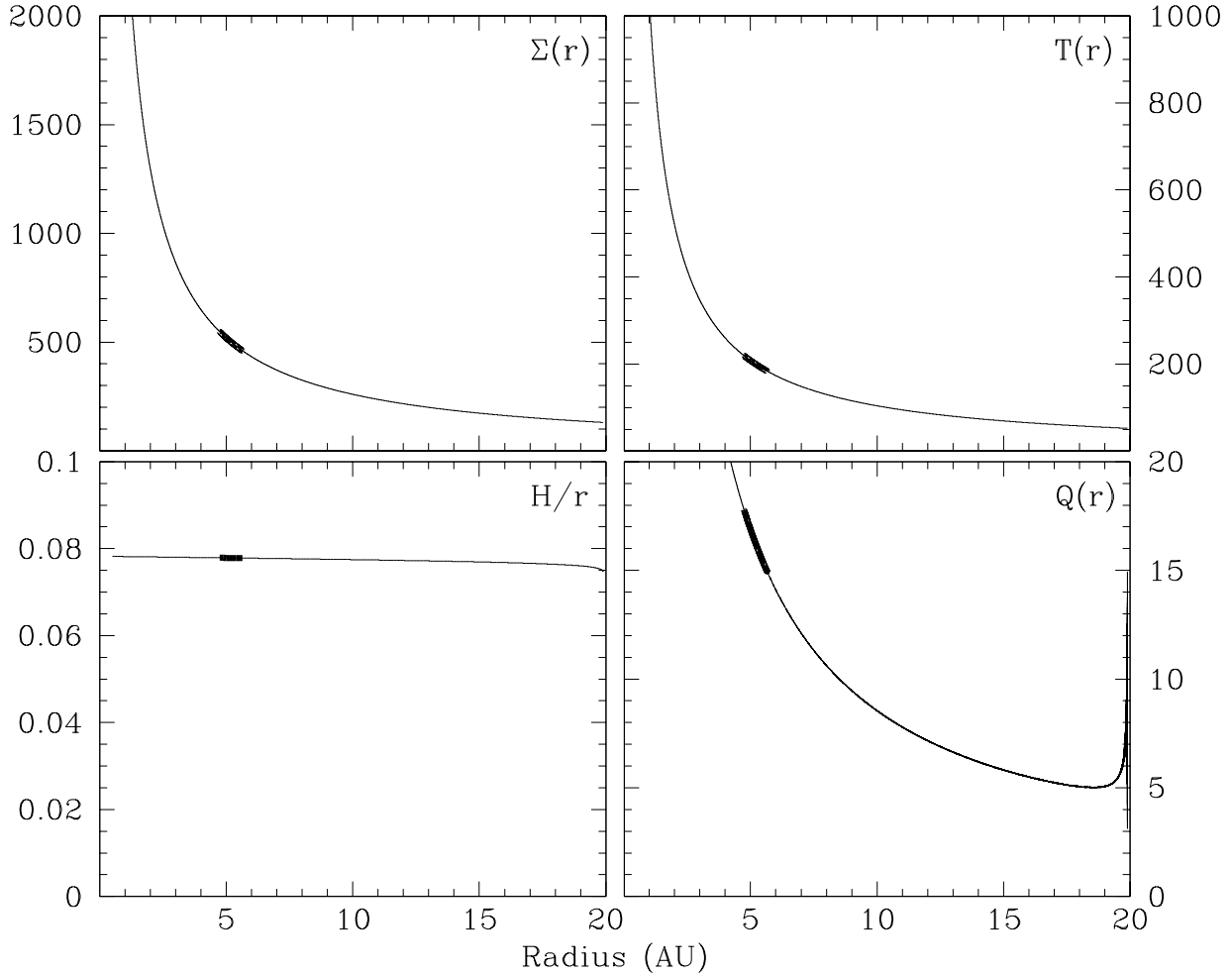}}
\caption
{\label{fig:disk-init} 
Global initial conditions for the disks underlying our simulations.
Clockwise from the top left, the four panels of the image are the
disk surface density, the temperature, the Toomre $Q$ value and the
dimensionless scale height. The heavy portion of each line corresponds
to the portion of the initial condition considered in deriving the
local initial condition.} 
\end{figure*}

The initial conditions are developed from a global model of an
accretion disk quite similar in morphology to the disks modeled in
\citet{JovI}. Specifically, we assume that a point mass representing
the central star is surrounded by a circumstellar disk, whose
inner and outer radial dimensions are $R_i=0.5$~AU and $R_o=20$~AU.

The mass in the disk is distributed according to a surface density
power law of the form
\begin{equation}\label{eq:dlaw-mig}
\Sigma(r) = \Sigma_{\rm pl} \left({{a_{\rm pl}}\over{r}}\right)^p,
\end{equation}
with a similar power law for temperature taking the form
\begin{equation}\label{eq:tlaw-mig}
T(r) = T_{\rm pl} \left({{a_{\rm pl}}\over{r}}\right)^q.
\end{equation}
$\Sigma_{\rm pl}$ and $T_{\rm pl}$ are the surface density and
temperature at the assumed orbit radius, $a_{\rm pl}$ of the planet,
respectively. The power law exponents are ordinarily set to $p=q=1.0$.
In order to study the sensitivity of our results to the mass and
temperature distributions however, we have also performed simulations
with other exponents as well (see table \ref{tab:sims}). 

The surface density coefficient is set to $\Sigma_{\rm
pl}=500$~g/cm$^2$. This value is somewhat smaller than that implied
by the \citet{Pol96} model, for which 6~Myr were required to grow a
Jovian mass planet. It is chosen so as to be consistent with the
minimum mass solar nebula (MMSN), given the dimensions of the disk we
assume. With the given spatial dimensions of the disk this surface
density implies a disk mass of \mdisk=0.0358\msun, comparable to the
traditional MMSN value, but well below the \mdisk$\sim0.1-0.14$\msun
required for the classical core accretion model \citep{Pol96} to
produce a fully formed Jovian planet within $\sim6$Myr.

The temperature exponent value of $q=1$ is chosen to ensure that the
scale height of the disk is approximately constant. It is steeper than
the value usually expected from observations of circumstellar disks
\citep[e.g. ][]{BSCG}, but over the limited radial range we simulate,
we do not expect the temperature gradient to play a large role. The
principle benefit of choosing this value will be to simplify the
correspondence between the global and local disk models, as discussed
below. The temperature scale is set to $T_{\rm pl}=200$~K, comparable
to (but slightly warmer than) the ice condensation temperature
expected to be important for the rapid coagulation of grains that
begins the planet formation process. With this background temperature,
the derived isothermal scale height of the disk, $H=c_s/\Omega$,
normalized by the distance from the star is $H/r=0.08$, as shown in
figure \ref{fig:disk-init}.

We map each of the thermodynamic quantities onto a radial grid in
cylindrical coordinates, equally spaced in $\ln(r)$. In order to
ensure the most precise initial condition for each sub-grid, the local
initial conditions are derived separately for each sub-grid. The grid
spacing of the 1D global grid is adjusted at each derivation to be
equal to the spacing of the current local grid. This means that over
the full extent of the disk (0.5 to 20 AU) we require as many as 30000
radial grid points to determine the global initial condition. Since
the global initial condition is determined only at the beginning of
the simulation, used to determine a corresponding local condition and
thereafter discarded, such high resolution poses a negligible cost in
terms of the total time required for the simulation.

From the predefined temperature, surface density and equation of
state, we determine a pressure gradient for the disk. The pressure
gradient is then used in conjunction with the gravitational forces
from the star and disk to determine a rotational velocity, under the
assumption that radial accelerations due to pressure gradients,
stellar and global disk self gravity and centrifugal forces exactly
cancel each other and no radial motion exists. No initial
perturbations are assumed in any of the models. 

Finally, we combine the various quantities from our specification of
the global disk conditions together to calculate a value of the Toomre
$Q$ parameter, defined as:
\begin{equation}\label{eq:toomre-q}
Q = {{\kappa c_s}\over{ \pi G \Sigma}},
\end{equation}
where $\kappa$ is the local epicyclic frequency and $c_s$ is the sound
speed. Theoretically \citep[see e.g.][]{GalDyn}, values of $Q<1$
define the conditions for which self gravitating, axisymmetric
instabilities in disks are unstable to linear perturbations. Numerical
simulations \citep[e.g.][]{DynI,Pick98} have shown that self
gravitating spiral instabilities can still grow for values $Q\la2-3$,
but that disks with higher values are stable.

The conditions assumed for most of our simulations ensure the global
stability of the disk to large scale self gravitating disturbances, as
measured by the value of its $Q$ parameter. However, in several
simulations we will also study disks with background temperatures as
low as $T_{\rm pl}=$50~K. Since, in the context of a global initial
condition, the value of $Q$ in such a disk would certainly fall to
values low enough (in the outer disk) to allow instabilities to grow
so that other planet formation mechanisms to proceed, such values may
seem rather unphysical or at least irrelevant. The simulations remain
interesting however because of their importance in outlining the
behavior of the flow as a function of the ratio between the Hill and
accretion radii, which we will find is an important measure of the
dynamical activity in the flow.

A graphical summary of the global initial conditions for our prototype
model (discussed in section \ref{sec:prototype} below) is shown in
figure \ref{fig:disk-init}. As indicated, the lower right panel shows
the Toomre stability parameter, $Q$, value for the simulations at each
radius. Because $Q>5$ for all radii, we expect that although global
self gravity is included, self gravitating disk instabilities would
not develop in the background flow, validating our assumption that it
remains smooth. Disk self gravity remains important to consider
however, first because, from a global perspective, it shifts the
position of the corotation resonance relative to the planet, and
second, from a local perspective, because we do not wish to discount
the possibility that self gravitating instabilities could develop in
the forming planet's envelope, even if they would not in the disk as a
whole.

\subsubsection{The Local Model}\label{sec:local-init}

We use the global initial condition to derive the local initial
condition inside our simulation cube in the following manner. First,
we assume a planet core (modeled as a point mass) is embedded in the
disk and orbits the star at a distance of $a_{\rm pl}=5.2$~AU. Its
orbit velocity is determined from the assumption that its orbit is
circular and that it is affected by the gravitational forces from the
star and from the disk. The planet's mass is set to 10\me\ and no
accretion onto the core is allowed. The core is placed at the corner
of eight adjacent grid zones which are not treated specially in any
way, so that the core itself is effectively unresolved by the grid. We
assume that no circumplanetary envelope exists around the core in the
initial state, so that initial state is defined by the background
flow, unperturbed by the planet. In order to allow the later flow to
adapt more gracefully to this condition, we begin with a core mass of
1\me\ and allow it to grow linearly with time to its final mass over
the course of the first 10 years of evolution. We have not observed
any side effects of this adjustment over the rest of the simulation's
lifetime.

The simulation cube is centered on the core and is co-moving with it.
The dimensions of the cube are set to $\pm4$\rh\ in each direction,
corresponding to a cube of size $\sim0.897$~AU on each side. In order
to explore sensitivity to the core mass, we have performed one
simulation with the core mass set to 20\me, and we have performed
several others with the cube volume as large as $\pm6$\rh, depending
on the goals of the simulation. Table \ref{tab:sims} below defines the
initial parameters for each simulation.

The value of the $x$ and $z$ velocities ($v_x$ and $v_z$--each set
initially to zero), corresponding to the radial and vertical
velocities of the global initial condition,  and the internal energy
of the gas are mapped directly from the global initial condition onto
the local condition. Since the grid spacing of the 1D global initial
state is preset to be identical to that on the corresponding 3D local
grid, no interpolations are necessary. The values for the $y$
velocity and volume density $\rho$ require additional specification to
complete the mapping from global to local.

The local $y$ velocity is set from the global azimuth velocities of
the gas in the disk and the planet core. Because we assume a
simulation cube that is fixed relative to the core, in order to obtain
the $y$ velocity, we must first subtract off the angular velocity
of the frame 
\begin{equation}\label{eq:vy}
v_y = V_\phi - (a_{\rm pl} + x)\Omega_{\phi {\rm pl}}
\end{equation}
where $V_\phi$ is the orbit velocity of disk matter at each orbit
radius and $\Omega_{\phi {\rm pl}}$ is the orbit velocity of the
planet (and the rotation rate of the reference frame), and $x$ is the
offset from the planet's radial position, $a_{\rm pl}$.

The volume density remains to be determined. In a complete model of a
circumstellar disk, we expect to find gradients of most of the
hydrodynamic quantities in both the radial direction and in the $z$
direction, due to the near Keplerian character of the disk and to the
radiative heating and cooling processes acting on it. In order to
simplify our study however, we have chosen to neglect the $z$
component of both stellar gravity and global disk self gravity, so
that effectively no vertical structure in the disk exists. The
symmetry is broken only by the planet's gravitational force so that
the flow around the core remains fully three dimensional.

Since the global initial condition is defined by a surface density
rather than a volume density, we require a conversion to completely
specify the local condition. For this conversion we introduce the
small inconsistency in our models that we use an estimate of the disk
scale height (which of course depends on the existence of the $z$
component of stellar and disk gravity), and the relation
$\rho=\Sigma/H$ to specify the correspondence. The scale height is in
turn defined by $H=c_s/\Omega$ and $\Omega$ is the local rotation
velocity in the disk. Its value of $\sim0.4$~AU at the assumed orbit
radius of 5.2~AU is approximately identical to the distance from the
core to the edge of the simulation cube.

\subsection{Boundary Conditions}\label{sec:bound}

In the shearing box formalism, the principle underlying the choice of
boundary conditions is that no location in a disk is different than
any other, excepting only that a shear term due to the background
Keplerian motion must be accounted for explicitly. The box itself
therefore represents one of many, essentially identical, replicas of a
small subvolume of that disk. In this context, the common choice of
implementing periodic boundary conditions for the shearing box
\citep[see e.g.][]{HGB95} means that a simulation of one small
subvolume will be sufficient to represent the evolution of the entire
disk. Clearly this principle does not hold for the present case, where
our simlation volume contains a core and is therefore unique.

Instead, we implement a slighly more restrictive principle. Namely, we
assume that the underlying disk background state is unchanging and is
not affected by changes inside our simulation volume. We therefore set
values for each state quantity in the boundary zones of the $x$ and
$y$ coordinate directions of the local model (corresponding to the
radial and azimuthal directions in the disk) to the initial values
determined from the global model in the same manner that they would be
had they instead been included in the simulation volume itself. 

For the $y$ boundaries, the conditions are fixed to their initial
values for the duration of the simulation. Because the simulation
volume is assumed to be a cutout taken from a circumstellar disk, the
$y$ boundary quantities are not identical at different $x$ positions
in the simulation, since they include the radial gradients of density,
temperature and the rotation curve. This is an important consideration
because at different $x$ locations on the same boundary surface of the
simulation box, corresponding to different orbit radii in the
underlying circumstellar disk, $y$ velocities may be directed either
into or out of the simulation volume, with differing magnitudes. This
behavior is a consequence of the initial near Keplerian flow of the
underlying disk model, for which the $y$ velocity varies as a function
of the $x$ position relative to the planet.

Boundary conditions in the $x$ coordinate are set assuming the same
fixed initial condition. Flow into the grid is permitted but the value
of each hydrodynamic quantity (including $v_x$, the velocity
perpendicular to the boundary) is reset to its initial value at each
time step, so that the assumption of a fixed background state is
maintained. Flow out of the grid is also allowed, should conditions
require, by allowing $v_x$ to take on either zero or the value just
inside the boundary if that velocity is directed outwards. 

This condition compensates for the development of large scale `wakes',
in the flow, corresponding to spiral structures in the underlying
circumstellar disk, which include some radial motion directed away
from the planet in each direction. If a completely fixed condition is
implemented, such structures can be reflected strongly and
unphysically back into the simulation volume. Some reflection still
occurs with the limited outflow condition, although at the level it is
present, it proves essentially harmless because the amount of
reflected material is both small and is quickly carried out of the
grid through the $y$ boundary.

Initial simulations of mass propagating through simulation cube not
containing a planet core demonstrated that inflow/outflow boundaries
in the $z$ coordinate resulted in large scale eddies forming in the
flow. In order to avoid this problem, we use a reflecting condition on
the $z$ boundaries instead. Employing this condition together with the
ones for the $x$ and $y$ directions resulted in a quiescent flow
throughout the simulation volume for as long (several hundred years)
as we cared to follow the evolution, and we did not pursue the origin
of the eddies further.

\section{Results of Simulations}\label{sec:results}

Using the initial conditions and physical models described above we
have performed a number of numerical experiments of the flow around a
10\me\ core. One series of simulations vary the character of
the background flow, by varying the components of gravity due to the
disk that are included in describing the flow and by varying the
background power laws of density and temperature. A second series of
simulations vary the thermodynamic treatment of the gas, by changing
the equation of state assumed for the gas and by changing the absolute
scale of the background temperature. In table \ref{tab:sims} we
summarize several important parameters of the simulations.

Columns 1-3 in the table specify a distinct label for each simulation
and its resolution both in terms of the physical extent of the
simulation cube, the number of grid zones and depth of the nesting.
Columns 4-7 define the power law exponents, $p$ and $q$, for the
surface density and temperature used to define the initial conditions
of the global disk system, the assumed background temperature and a
reference to the assumed equation of state for each model. Columns
8-10 specify whether or not gravitational forces arising from the
disk components described in sections \ref{sec:global-init} and
\ref{sec:local-init} have been included in the interactions with
each other and the planet core in each simulation. The last column 
is the duration of each simulation in years. 

\begin{table*}
\caption{Simulation Parameters}
\label{tab:sims}
\begin{tabular}{cccccrccccr}
\hline
Label & Resolution   & Simulation & $p$ & $q$ & $T_{\rm pl}$ & EOS                      & Global Disk & Global Disk   & Local Disk    &  T$_{\it end}$  \\
      &              & Box Size   &     &     &              &                          &  on Planet  & on Local Disk & on Local Disk &                 \\
\hline
lo10 & $3\times64^3$ & $\pm4$\rh  & 1.0 & 1.0 & 200~K        & eq. \ref{eq:ideal-eos3d} & no         &  no           & no             &    1600~yr      \\ 
ft10 & $5\times64^3$ & $\pm4$\rh  & 0.5 & 0.5 & 200~K        & eq. \ref{eq:ideal-eos3d} & no         &  no           & no             &     100~yr      \\
fl10 & $5\times64^3$ & $\pm4$\rh  & 0.5 & 1.0 & 200~K        & eq. \ref{eq:ideal-eos3d} & no         &  no           & no             &     100~yr      \\
br10 & $5\times64^3$ & $\pm4$\rh  & 1.0 & 1.0 & 200~K        & eq. \ref{eq:ideal-eos3d} & no         &  no           & no             &     100~yr      \\
bp10 & $5\times64^3$ & $\pm4$\rh  & 1.0 & 1.0 & 200~K        & eq. \ref{eq:ideal-eos3d} & yes        &  no           & no             &     100~yr      \\
bs10 & $5\times64^3$ & $\pm4$\rh  & 1.0 & 1.0 & 200~K        & eq. \ref{eq:ideal-eos3d} & yes        &  yes          & no             &     100~yr      \\
sg10 & $5\times64^3$ & $\pm4$\rh  & 1.0 & 1.0 & 200~K        & eq. \ref{eq:ideal-eos3d} & yes        &  yes          & yes            &     100~yr      \\
so10 & $5\times64^3$ & $\pm4$\rh  & 1.0 & 1.0 & 200~K        & eq. \ref{eq:ideal-eos3d} & yes        &  yes          & no             &     100~yr      \\
sg20 & $5\times64^3$ & $\pm4$\rh  & 1.0 & 1.0 & 200~K        & eq. \ref{eq:ideal-eos3d} & yes        &  yes          & yes            &     100~yr      \\
b05h & $6\times64^3$ & $\pm6$\rh  & 1.0 & 1.0 &  50~K        & eq. \ref{eq:ideal-eos3d} & yes        &  yes          & no             &     100~yr      \\
tm05 & $6\times64^3$ & $\pm6$\rh  & 1.0 & 1.0 &  50~K        & eq. \ref{eq:ideal-eos3d} & yes        &  yes          & yes            &     100~yr      \\
tm10 & $6\times64^3$ & $\pm5$\rh  & 1.0 & 1.0 & 100~K        & eq. \ref{eq:ideal-eos3d} & yes        &  yes          & yes            &     100~yr      \\
tm20 & $6\times64^3$ & $\pm4$\rh  & 1.0 & 1.0 & 200~K        & eq. \ref{eq:ideal-eos3d} & yes        &  yes          & yes            &     100~yr      \\
tm40 & $6\times64^3$ & $\pm4$\rh  & 1.0 & 1.0 & 400~K        & eq. \ref{eq:ideal-eos3d} & yes        &  yes          & yes            &     100~yr      \\
iso1 & $6\times64^3$ & $\pm4$\rh  & 1.0 & 1.0 & 200~K        & eq. \ref{eq:iso-eos}     & yes        &  yes          & yes            &     100~yr      \\
iso2 & $6\times64^3$ & $\pm4$\rh  & 1.0 & 1.0 & 100~K        & eq. \ref{eq:iso-eos}     & yes        &  yes          & yes            &      15~yr      \\
iso3 & $6\times64^3$ & $\pm4$\rh  & 1.0 & 1.0 &  50~K        & eq. \ref{eq:iso-eos}     & yes        &  yes          & yes            &      15~yr      \\
adi1 & $6\times64^3$ & $\pm4$\rh  & 1.0 & 1.0 & 200~K        & eq. \ref{eq:adi-eos}     & yes        &  yes          & yes            &     150~yr      \\
\hline
\end{tabular}
\end{table*}

In the discussion below, we will first describe the evolution of a
`prototype' model {\it tm20}, that is typical of the results obtained
in many of our simulations. Then we will describe the results of our
experiments that varied the rotation curve of the underlying disk,
followed by experiments that varied the equation of state and
background temperature of the disk. 

\subsection{Morphological evolution of our prototype model}\label{sec:prototype}

In the discussion that follows and throughout this paper, we designate
the model labeled {\it tm20} as our `prototype' model. Its features
are typical of most of our simulations and the physical model
underlying it is the most inclusive. This model was run with an ideal
gas equation of state, with $\gamma=1.4$ and included gravity from
both the star and the core, as well as from both the global and local
disk material.

\begin{figure*} 
\rotatebox{0}{
\includegraphics[width=120mm]{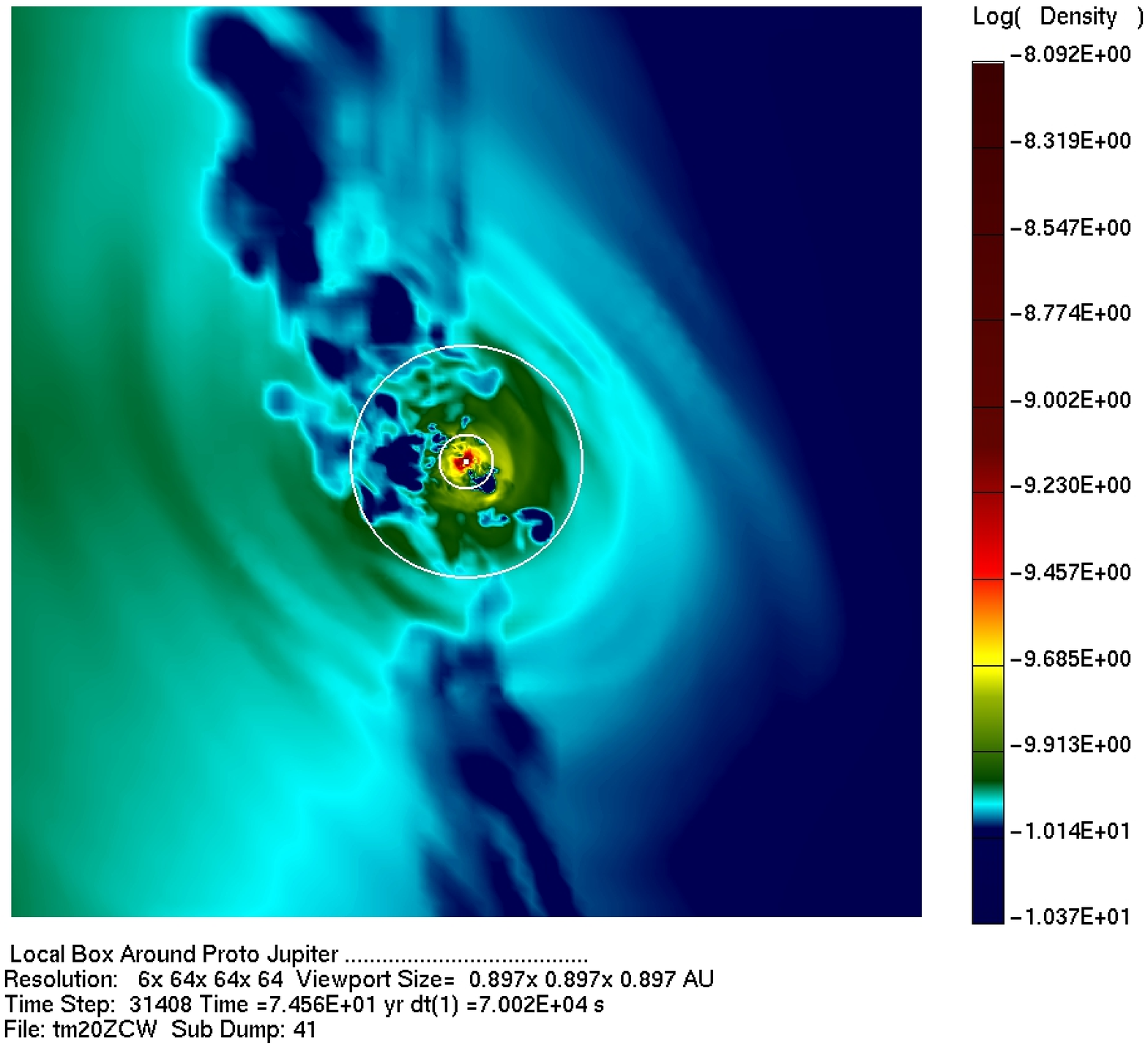}}
\rotatebox{0}{
\includegraphics[width=120mm]{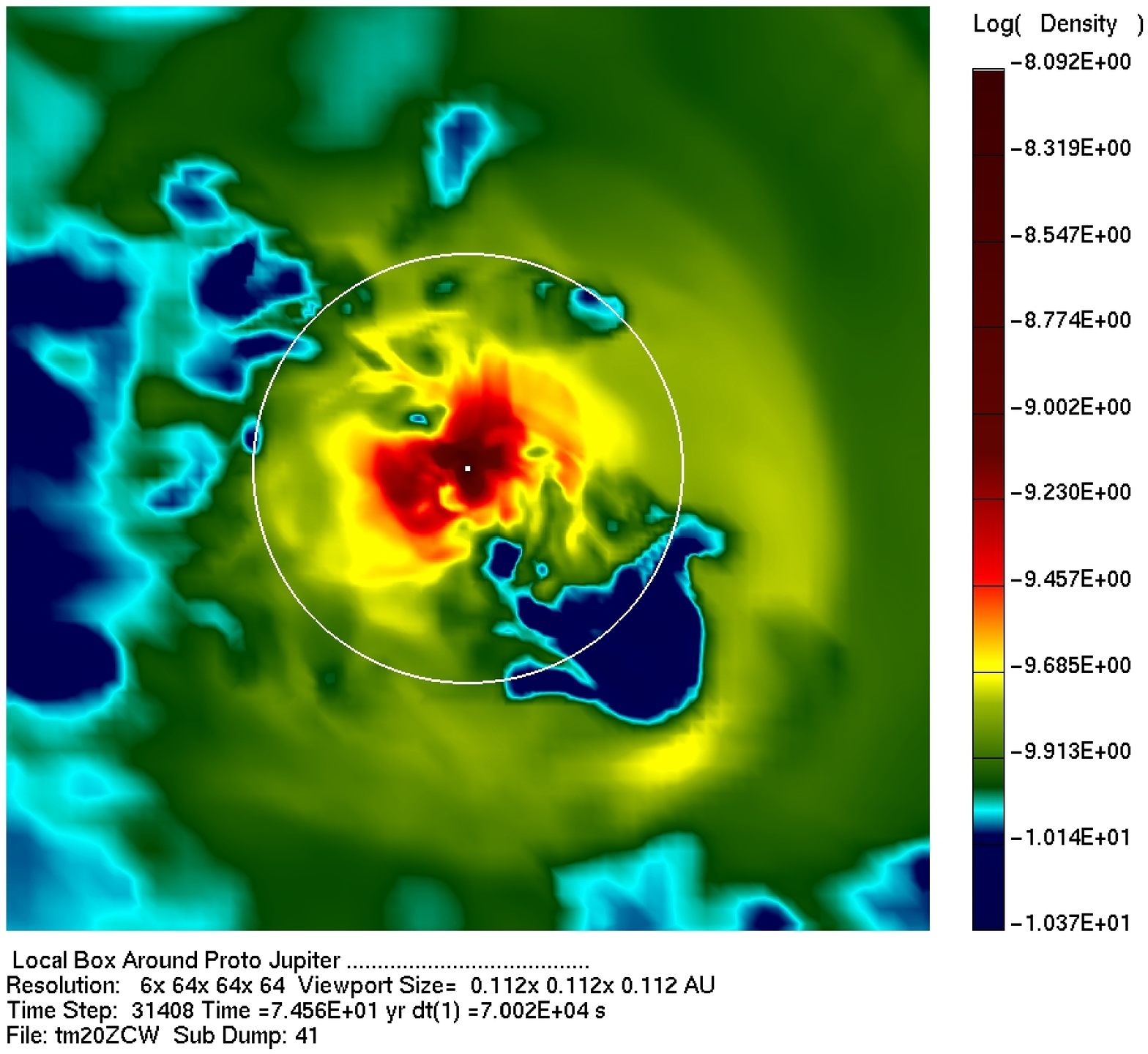}}
\caption{\label{fig:cutout-mid-dens}
The volume density in a 2D slice taken through the disk midplane for
the full simulation volume of $\pm4$\rh\ (top), and a blowup of the
region within $\pm1/2$\rh\ of the core (bottom) The spatial extent of
each frame, in AU, is defined in its caption and labeled ``Grid
Size''. Each cutout intersects the planet core at the origin. The
white circles define the radius of the accretion sphere $R_A=GM_{\rm
pl}/c^2_s$ (small circle) and the Hill radius (large circle). The
color scale is logarithmic and extends from $\sim10^{-10}$ to
$\sim10^{-8}$ g~cm$^{-3}$.}
\end{figure*}

\begin{figure*} 
\rotatebox{0}{
\includegraphics[width=120mm]{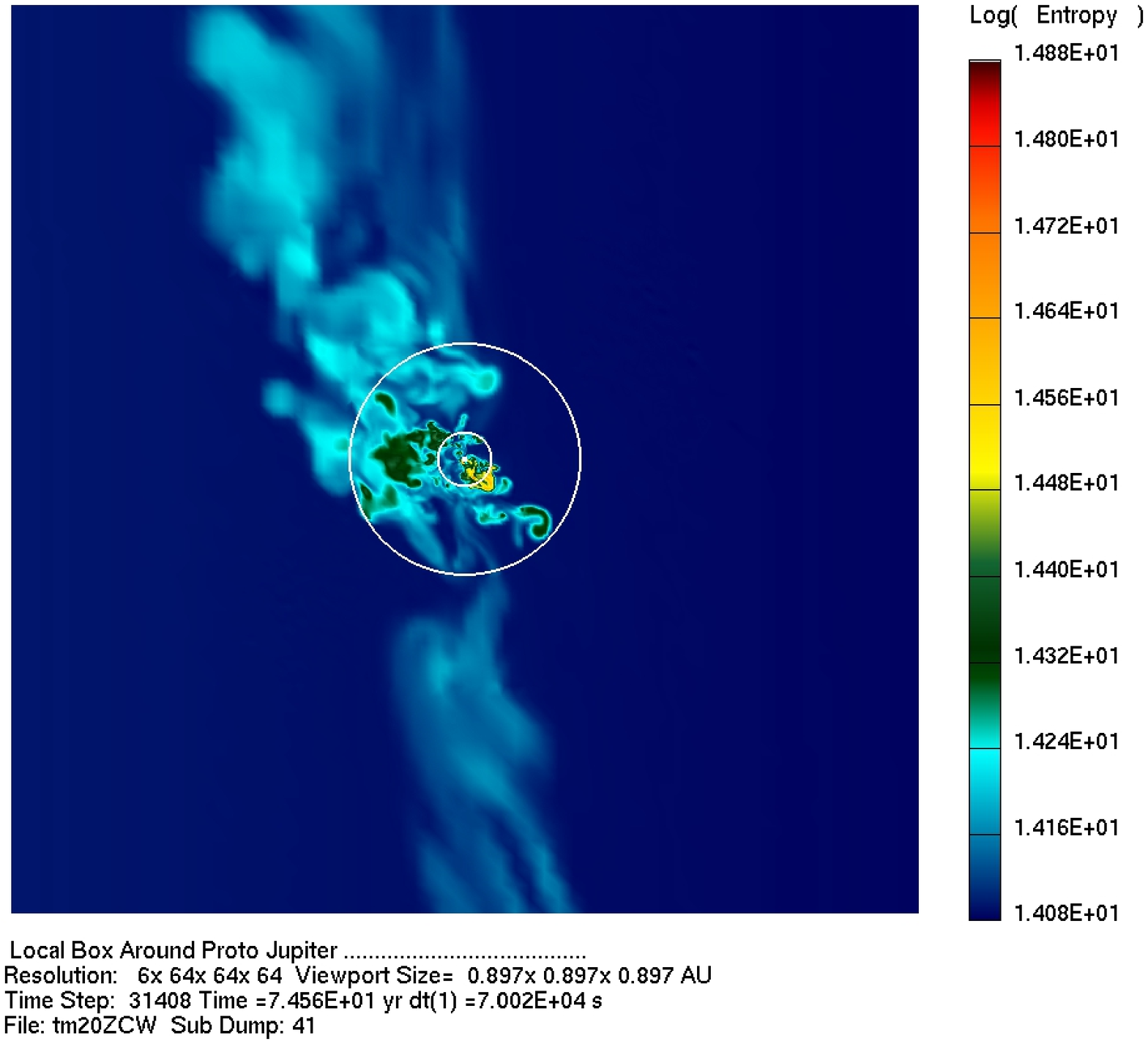}}
\rotatebox{0}{
\includegraphics[width=120mm]{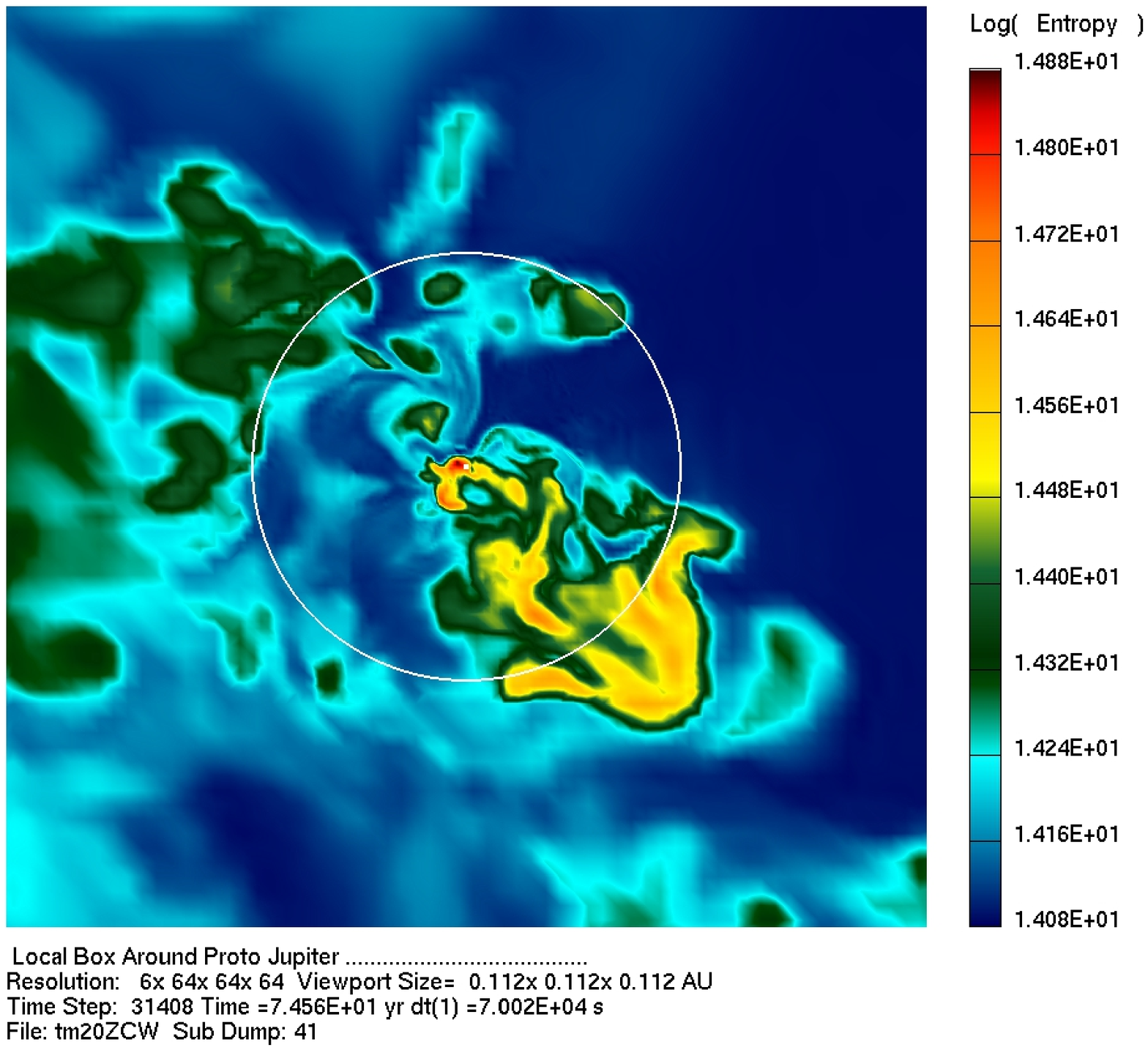}}
\caption{\label{fig:cutout-mid-entr}
As in Figure \ref{fig:cutout-mid-dens}, but showing entropy. The
background flow is characterized by a small radial entropy gradient.
Entropy increases by a factor $\sim 6$ over this value in the highly
shocked neighborhood of the core, then decreases, as it mixes in with
the low entropy background material in post-encounter evolution.} 
\end{figure*}

\begin{figure*} 
\rotatebox{0}{
\includegraphics[width=120mm]{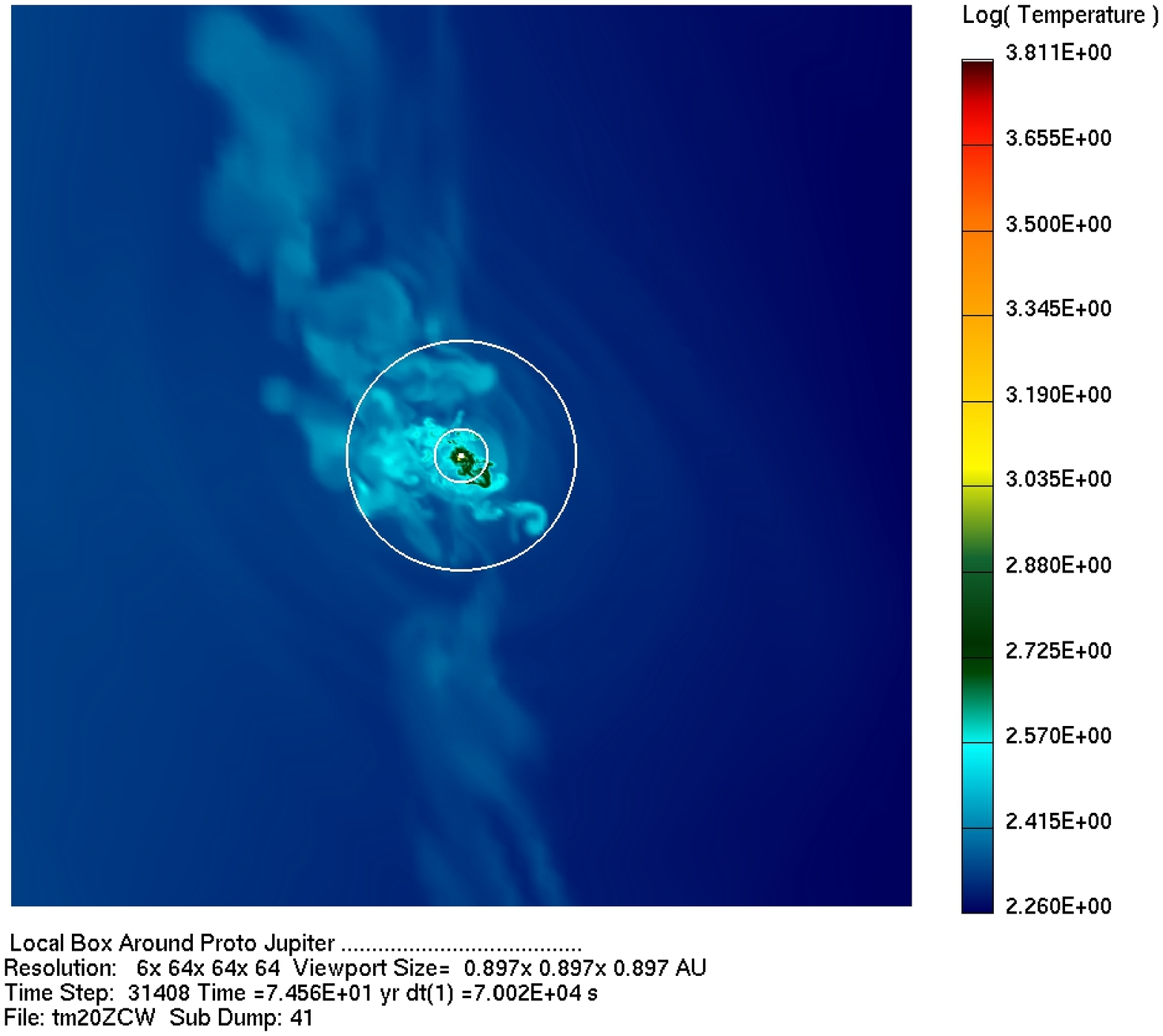}}
\rotatebox{0}{
\includegraphics[width=120mm]{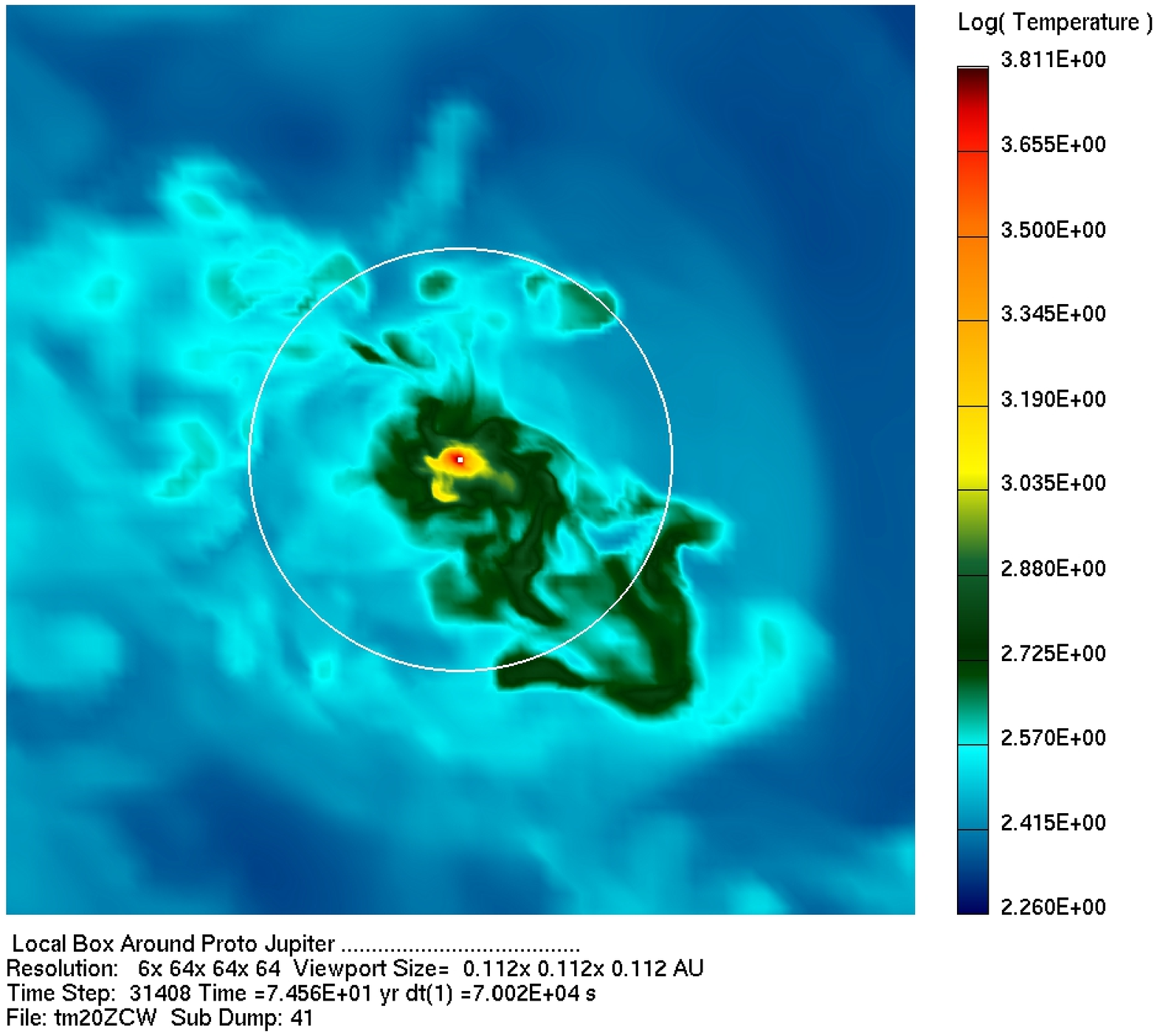}}
\caption{\label{fig:cutout-mid-temp}
As in Figure \ref{fig:cutout-mid-dens}, but showing temperature. The
color scale is logarithmic and extends from 180 to 6500 K.
Temperatures as high as 3-5000~K are common very close to the core,
with temperatures decreasing rapidly to the background $\sim$ 200~K
value at increasing distance. }
\end{figure*}

\begin{figure*}
\rotatebox{0}{
  \includegraphics[width=10.5cm]{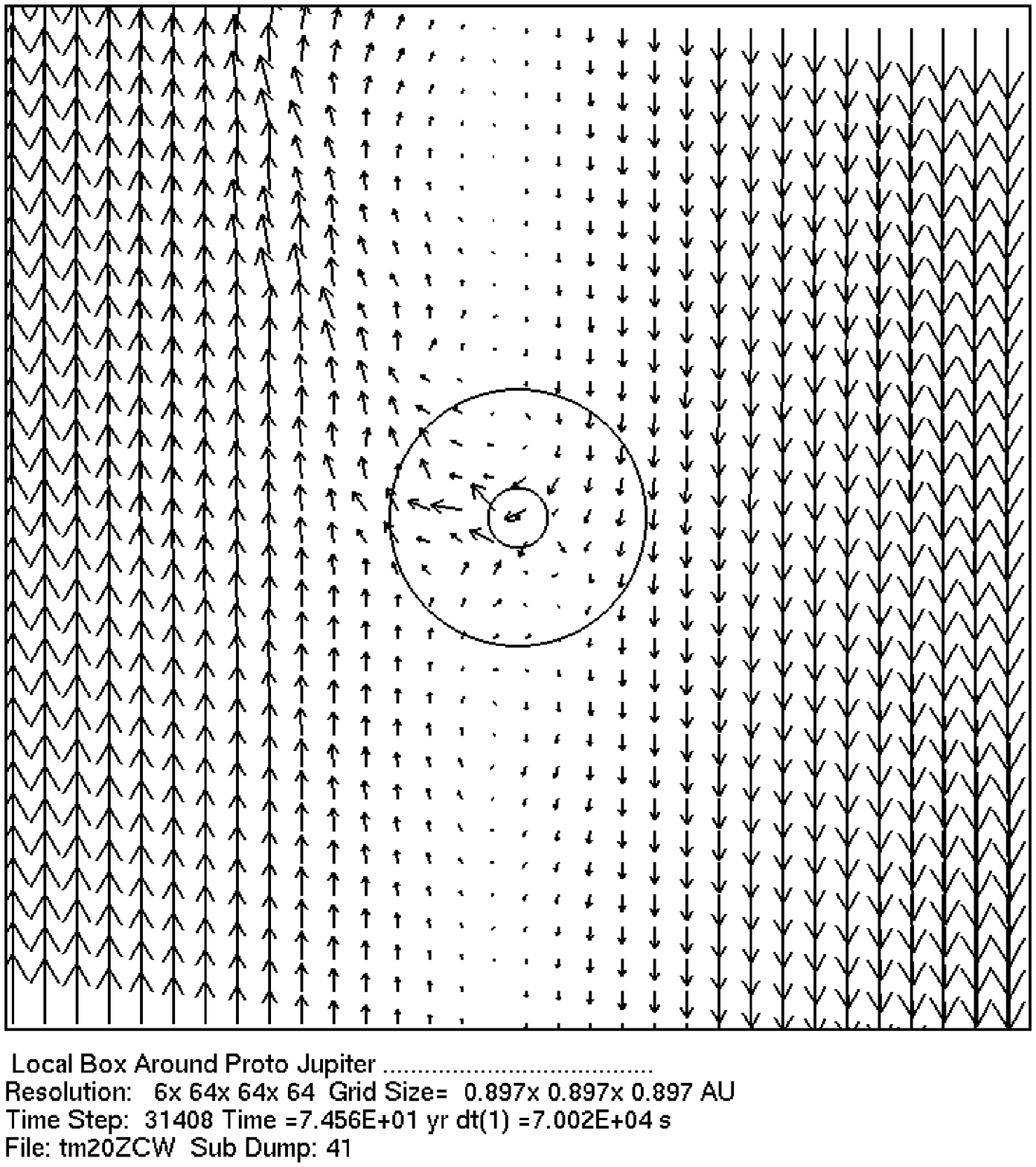}}
\rotatebox{0}{
  \includegraphics[width=10.1cm]{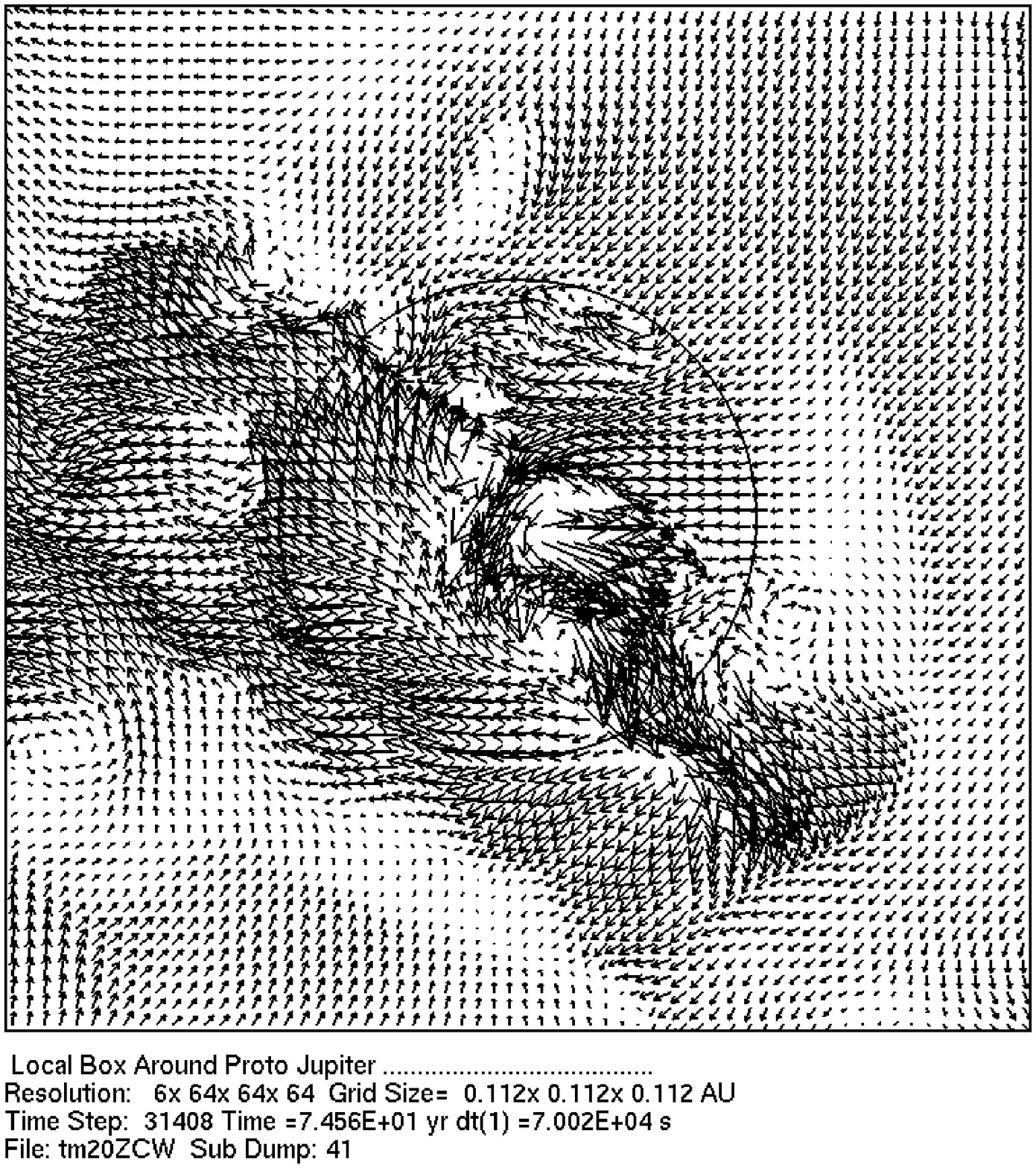}}
\caption{\label{fig:cutout-mid-vvect}
Velocity vectors shown projected onto the disk midplane on the coarsest 
grid in the top panel, and on the fourth nested grid in the bottom panel.}
\end{figure*}

In figures \ref{fig:cutout-mid-dens}--\ref{fig:cutout-mid-vvect}, we
show 2D slices of the simulation volume, taken through the disk
midplane at a time 74~yr after the beginning of the simulation. In
successive figures, we show the gas density, its entropy as defined by
the expression
\begin{equation}
\log S = \log\left( {{ p}\over{\rho^\gamma}} \right),
\end{equation}
its temperature and its velocity vectors projected onto the disk
midplane.

The density structures are highly inhomogeneous and become
progressively more so closer to the core. Densities both above and
below that of the background flow exist due to the development of hot
`bubbles' in the flow near the core, which then expand outwards. One
such bubble is particularly visible in the plots of the temperature
distribution, emerging to the lower right. Such structures are common
over the entire the duration of the simulation and emerge in all
directions, depending on details of the flow at each time. Those
details vary with time because of the very strong feedback cycle that
is generated as outgoing bubbles perturb the incoming gas flow that is
responsible for generating later activity. Activity persists for as
long as we have simulated the evolution without significant decay or
growth. Also, although we have not included figures illustrating so
explicitly, we note that plots of slices through our simulations in
directions perpendicular to the midplane show similar features:
density structures which are highly inhomogenous, which become 
progressively more so closer to the core.

Several ring-like structures surrounding the core are visible and are
remnants of intermittent large scale mass outflows (`eruptions') from
the environment of the core. These rings are sheared into an ovoid
shape in the disk midplane, but are near spherical in the radial
plane. The eruptions themselves appear at least in part to be the
result of hot bubbles that develop, expand and escape into the
background flow and are due to shocks and compressional heating very
close to the core.

The structures within 1--2\ra\  of the core show little evidence of
spiral arms seen in some previous work \citep[e.g.][]{lsa99,dhk03},
but evidence for the development of other irregular dynamical
structures is ubiquitous. Material orbiting the core or falling
towards it encounters other material on different trajectories,
generates shocks and heats the gas, causing bubbles and other complex
morphological structures to develop. In actuality, such `orbits' are
more often little more than single passes past the core as mass from
the surrounding disk first falls into the core's gravitational
influence, then returns to the disk and is swept away.

We attribute the origin of the bubbles to the combination of four
conditions of our simulations. First, we include only a very small
gravitational softening coefficient for the core, which means that the
depth of its gravitational potential well is very deep and that the
energy available for conversion into heat is very large. Second, we
include no explicit viscosity in our formulation of the fluid
equations, so that features of the flow are not blurred or otherwise
damped out by dissipative effects beyond our ability to model with
precision. Also, our numerical method exhibits both very low numerical
dissipation and very high fidelity for capturing the physical
dissipation due to shocks that develop in the flow. Third, we do not
include any mechanisms for energy to be lost from the gas once it is
converted to thermal energy from gravitational potential energy. The
most important such mechanism is of course, the losses that occur as
the gas cools through radiation of photons. Finally, we do not include
any mechanism for accreting material onto the core, because the
dimensions of our finest numerical grid are comparable to the actual
radius of the core. The thermal energy carried by that material is
therefore also not accreted, remaining instead with the gas. In
combination, these factors ensure thermal energy is produced
efficiently, as flow features remain distinct enough to resolve the
strong shocks that develop during the passage of material through the
core's environment, and dissipated (or simply lost, as in the case of
accretion onto the core) inefficiently since no mechanisms are
included to do so. Instead, the thermal energy is converted again into
kinetic energy as the high pressure gas expands outwards in bubbles
and rejoins the background flow. We will explore the consequences of
energy loss mechanisms in section \ref{sec:thermo} below.

Variations in entropy are significant in our simulations because the
physical model includes only one mechanism by which its value can
change after a packet of gas enters the simulation volume.
Specifically, the hydrodynamic scheme itself generates entropy at
shocks as a consequence of non-linearities in the solution of the
fluid mechanics equations there. Therefore, any increase in entropy
is, by assumption, a very strong {\it a posteriori} indication that
shocks developed in the flow. Other entropy generation mechanisms that
might be present in the flow will exhibit different characteristics
than we observe in our simulations. For example, entropy generation in
turbulent flows would be characterized by a conservative cascade of
energy flow from large to small scale flow features, followed by
entropy and thermal energy generation at small scales. Such a cascade
is not observed, and we believe turbulence is not an important
characteristic of the flow activity present in our simulations.

The fluid entropy in the background flow shows essentially no visible
variation at different locations in the flow. Entropy of material near
the core increases as it undergoes irreversible heating in shocks, but
such heating is highly inhomogeneous and low entropy material is
frequently present at distances only a small fraction of the accretion
radius from the core itself. For example, comparing the bottom, high
resolution, panels of figures \ref{fig:cutout-mid-entr} and
\ref{fig:cutout-mid-vvect} (discussed below), we see that low entropy
`background' material (i.e. the dark blue shaded material in figure
\ref{fig:cutout-mid-entr}) in the upper right quadrant of the images
is falling rapidly towards the core. Despite its proximity, it remains
as essentially pristine background disk material until it falls to a
distance of only $\la 0.1$\ra\ (or equivalently, a few times Jupiter's
current radius, $R_J$) from the core. There, its motion is strongly
perturbed by the gravitational forces it experiences, which alter its
trajectory onto paths that intersect those of other nearby packets of
matter, generating shocks. As it re-enters the background flow, high
entropy material begins to mix with the un-shocked background
material, but the mixed material remains distinct from the background
until it leaves the simulation volume.

Figure \ref{fig:cutout-mid-temp} shows the material temperatures for
the same snapshots as seen in figure \ref{fig:cutout-mid-dens}.
Temperatures as high as 6000~K are common in the innermost envelope,
within distances of $\sim0.1$\ra (or as we note above, a few times
Jupiter's current radius, $R_J$), but quickly drop to a few hundred
degrees at larger separations. Most material at and outside the
accretion radius, \ra, is only slightly warmer than the background
200~K flow. Exceptions are materials that are clearly remnants of
dynamical activity nearer to the core which expand outwards, such as
the large mushroom shaped feature extending to the lower right in the
figure. Apart from the radial temperature variation assumed for the
background flow, visible temperature fluctuations of large magnitude
do not extend away from the core to nearly the distances that are seen
for the density variations. Nevertheless, material that has undergone
a close encounter with the core does remain warmer and can be
distinguished from the background flow even as it leaves the
simulation volume, some time after the interaction occurred.

It is of some interest to compare the distributions of temperature and
entropy near the core. Although large deviations are present, the
temperature distribution appears far more spherically symmetric than
does the entropy distribution. This is important because it indicates
that material may undergo adiabatic compression without being shocked
while falling inwards towards the core, and only suffer a shock deep
in the envelope where the activity is most dynamic. Implications of
such compression and shock behavior will be discussed in section
\ref{sec:chondrules}.

Figure \ref{fig:cutout-mid-vvect} shows velocity vectors of the flow
in the midplane at the same time as is shown in figure
\ref{fig:cutout-mid-dens} above. In conflict to the orbital
predictions from pure celestial mechanics that the flow of material
approaching the Hill volume will turn around on a `horseshoe' orbit,
matter approaching the outer portion of the Hill volume can be nearly
unaffected, often passing completely through its outer extent with
only a small deflection. At the time of this snapshot, the trajectory
of material radially outward of the planet passes directly through the
Hill volume nearly unaffected, while material radially inwards
experiences larger trajectory perturbations even at distances well
beyond the Hill radius. Although we do not include specific examples
as figures here, we note that as the flow evolves over time, this
pattern frequently reverses itself, exhibiting exactly the opposite
behavior--material inward of the planet is perturbed only slightly,
while material outwards is significantly affected to a distances of
2-3\rh. We will quantify these variations in more detail in section
\ref{sec:spin}, below.

In contrast with this intermittently active flow further away,
material is always very strongly perturbed on the scale of the
accretion radius, where large amplitude space and time varying
activity develops. This too conflicts with the orbital mechanics
picture, in which matter inside the Hill volume simply orbits the
core. Pristine disk material can enter the Hill volume from the
background flow and shocked, high entropy material can escape and
rejoin the background flow. Changes in the flow pattern occur on time
scales ranging from hours to years, with a typical encounter time
inside the accretion radius of less than a month. Additional
discussion of these variations is found in section \ref{sec:spin}
below.

\subsection{Quantitative metrics of the activity}\label{sec:metrics}

The figures and discussion in the last section demonstrate the fact
that the environment around the core is dynamically active in a
qualitative sense, but do not quantify any measure of that dynamical
activity apart from its overall morphology. Here we attempt to
characterize the activity using a more quantitative analysis. First,
we discuss the distributions of mass and temperature in the envelope
as a function of distance from the core, then, as integrated totals of
mass and angular momentum contained in the envelope at a given time.
We will use two metrics to describe the latter two quantities.
Specifically, we tabulate the total mass enclosed by a sphere of given
radius and the total angular momentum of that material, calculated
using the core as the origin of coordinates. 

\subsubsection{The distribution of the gas near the core, and its
temperature}\label{sec:1D-dist}

\begin{figure}
\includegraphics[width=8.5cm]{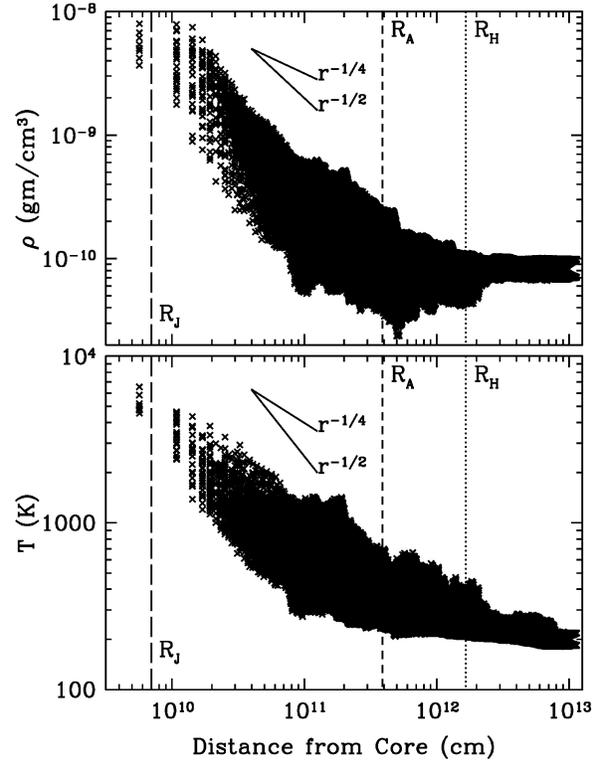}
\caption
{\label{fig:OneD-rhoT} The volume density (top) and temperature
(bottom) of the gas in our prototype model, plotted for each zone in
the grid as a function of distance from the core, at time $t=74$~yr
after the beginning of the simulation. The vertical dotted and short
dashed lines define the location of \rh\  and \ra, respectively, and
the long dashed line defines the present day radius of Jupiter, and is
included to provide scale. Solid lines define the slopes of $r^{-1/2}$
and $r^{-1/4}$ power laws.} 
\end{figure}

Because of the extremely non-homogeneous morphology and dynamically
active flow, little purpose is served by attempts to reduce the flow
variables to spherically averaged quantities. Nevertheless it remains
interesting to analyze the overall distribution and limits of various
quantities as they change as a function of radius. Such analyses are
useful both to illustrate differences from the flow in our simulations
from 1D models and, secondly, to quantify the extreme values that may
be expected at each radius and their significance in the context of
the overall evolution of the system.

Figure \ref{fig:OneD-rhoT} shows the density and temperature of each
grid zone in our prototype model, each as a function of distance from
the core. In both cases, the distributions are nearly flat at large
distances (i.e. $>>$\rh), varying only slightly at a given distance as
a consequence of the overall radial gradients of density and
temperature defining the background flow in the global disk model. A
quantity of material at large distances from the core exists that is
clearly distinct from the background, as a population of points with
somewhat higher temperatures, but is not visible in the density plot.
This signature is due to material which has been processed through the
core's envelope and returned to the background flow, but which has not
yet fully cooled or mixed with it. The material retains its higher
temperatures until it is carried out of the grid volume. An example of
such flow is illustrated in figures
\ref{fig:cutout-mid-dens}-\ref{fig:cutout-mid-temp}, where a hot
outflow is shown emerging to the lower right in the lower panels of
the figures. 

At distances smaller than the Hill radius, both density and
temperature begin to deviate more significantly from their background
values. At a given distance from the core, the maxima found in either
quantity approach factors of $\sim5-8$ larger than the minima at the
same distance, with the largest differences occurring inside the
accretion radius. Densities both increase above their background
values and decrease to well below the background, reflecting both the
compression of infalling material and the expansion of hot material
after it is processed deep in the core's gravitational potential well.
Both quantities increase all the way to the inner radial limit imposed
by our grid resolution, at a distance from the core of
$\sim5.6\times10^9$~cm, which is somewhat smaller than the present day
radius of Jupiter itself. For the innermost zones, densities reach
values of $\sim10^{-8}$~g/cm$^3$, nearly two orders of magnitude above
that in the background flow, while temperatures rise to 5-7000~K.
Examination of lower resolution variants of this simulation (i.e.
simulations {\it sg10} and {\it lo10}, not shown here), show lower
maximum densities and temperatures, indicating that the values reached
in figure \ref{fig:OneD-rhoT} are not converged, and suggesting that
still higher values might be reached near the core/envelope interface
if resolution were to be increased. 

Both distributions vary widely at any given radial distance from the
core, and neither can reasonably be fit to a 1D power law. In
particular, the density distribution cannot be fit to a $r^{-2}$ power
law, as would be expected from a spherically symmetric infall.
Instead, and to the extent that it follows a single profile at all, it
falls nearer to a profile between $r^{-1/4}$ and $r^{-1/2}$, as
indicated by the lines accompanying the figure. The much shallower
radial dependence presumably reflects the fact that no mass accretion
onto the core is permitted, so that material entering the envelope is
opposed by material already there, and by material returning to the
disk. 

The lower limits of the temperature distribution at each radius
provide an interesting flow metric because they define the population
of material at each radius which has undergone the least heating,
either due to compression or to shocks. Of great interest is that
material at temperatures below $\sim300$~K is not uncommon even at
distance as small as 10$^{11}$~cm, about 1/4 of the accretion radius,
while at other locations at the same distance, temperatures as high as
1-2000~K are present. The presence of low temperature material close
to the core demonstrates that some fraction of the background disk
material can penetrate deeply into the envelope while undergoing
essentially no processing by it, either due to shocks or to adiabatic
compression. This material is interesting because as it continues its
trajectory through the envelope, it may subsequently undergo some
compressional or shock event which heats it rapidly to temperatures
more typical of other material at these distances. A quantification of
what fraction of material undergoes such rapid heating, as opposed to
that undergoing some slower heating process or to that simply being
advected out of the envelope again, is beyond the scope of this study.
Its existence is important to note however, because if an appreciable
quantity of material can penetrate deeply into the envelope while
remaining cold, then be rapidly heated and ejected again into the
background flow, some signature of the processing may remain in
material available for study today. We discuss a preliminary study of
such a possibility in section \ref{sec:chondrules}.  

\subsubsection{The mass contained in the envelope as a function of
time}\label{sec:massdist}

\begin{figure}
\includegraphics[width=8.5cm]{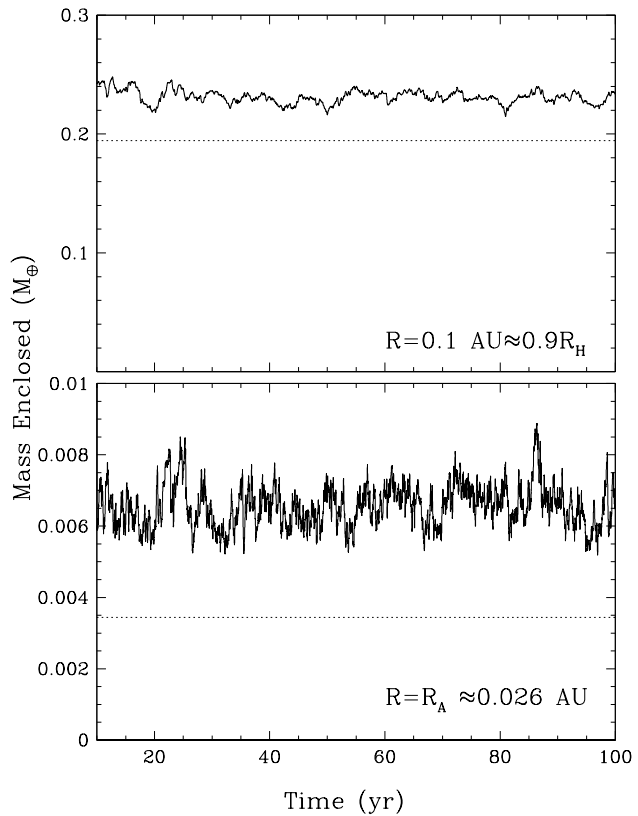}
\caption
{\label{fig:mass-encl} The mass enclosed by spheres of radius
$r=0.10$~AU$\approx0.9$\rh\ (top) and $r=$\ra$(=0.026$~AU) (bottom),
each as functions of time during the simulation. The first 10 years of
the evolution define the period when we first `grow' the core from
1\me\ to its final mass of 10\me, thus the envelope masses at this time
are not meaningful and are not included in the figure. The dotted
lines correspond to the enclosed mass at $t=0$, as characterized by 
the unperturbed disk background condition.}
\end{figure}

Figure \ref{fig:mass-encl} shows the mass enclosed by spheres of
radii, \ra\  and $r=0.10$~AU$\sim0.9$\rh, each as functions of time.
Except for the first ten year period of the simulation (omitted in the
plots), in which we artificially grow the core to its final mass, the
enclosed masses exhibit no secular mass accretion over the duration of
the simulation, consistent with our neglect of cooling. The amount of
mass enclosed within the accretion radius is quite small, averaging
$\sim.007$\me, while on the scale of the Hill radius, a quantity of
$\sim 0.24$\me\ collects near the planet. In comparison to the
`background flow' initial condition, the inner envelope mass has
increased by a factor $\sim1.8$, while the envelope mass as a whole
has increased by a factor $\sim1.2$. 

Variations in the enclosed masses with time are visible in both
curves, with magnitude $\sim10-20$\% for the smaller sphere and
$\sim5-10$\% for the larger. In neither case does the variation grow
or shrink as a function of time, again consistent with our neglect of
cooling. The temporal character of the variability, such that any sort
of periodicity can be defined at all, clearly favors higher frequency
oscillations close to the core and slower oscillations farther from
it. Variations on time scales well under a year are common for the
accretion sphere volume, and are often superimposed on longer time
scale features, as the overall morphology of the flow reacts to
changes in the larger scale flow. The existence of the variations over
time is a consequence of the dynamical activity discussed above, as
morphological features flow into or out of the respective spheres
defining our metrics. In turn, their time scales are consequences of
the shorter dynamical times deeper in the core's gravitational
potential well, as compared to those farther from the core. 

\subsubsection{The angular momentum of the gas around the
core}\label{sec:spin}

In addition to the mass and temperature distributions discussed in the
last sections, the distribution of angular momentum of the material in
the environment of the core can also provide insights into the
character of activity. Here, we describe the magnitude and the
variability in angular momentum distributions of material around the
core. In the figures below and the discussions accompanying them, we
will show both instantaneous values as functions of time, and time
averages and RMS deviations over the duration of each simulation. The
time dependent plots will provide insights into any periodicities that
are present in the results. Time averaged quantities will be useful
for comparisons between simulations with different physical
parameters, and in comparison to the present day values for Jupiter.
To enable such comparisons, we will present values normalized by
envelope mass, as specific angular momenta.

In the discussion below, we refer to the envelope's angular momentum
as its `spin'. We caution the reader however, that this term is not
strictly correct, since while some material may be in a bound orbit
around the core, other parts of the `envelope' for which spin is
tabulated, may instead be more accurately classified as part of the
background flow. Also, as an aid to reader's intuitive grasp of the
meaning of each of the three spin components, we note that the $z$
component corresponds to the `top'-like motion of the planet, and
corresponds to the largest component of the present day spin of the
planets in our solar system (excluding Uranus). The $x$ component
defines a spin axis outwardly directed along the line connecting the
star and the core and the $y$ component defines a spin axis pointing
forwards along the direction of motion of the core in its orbit.

\begin{figure*}
\rotatebox{-90}{
\includegraphics[width=13.5cm]{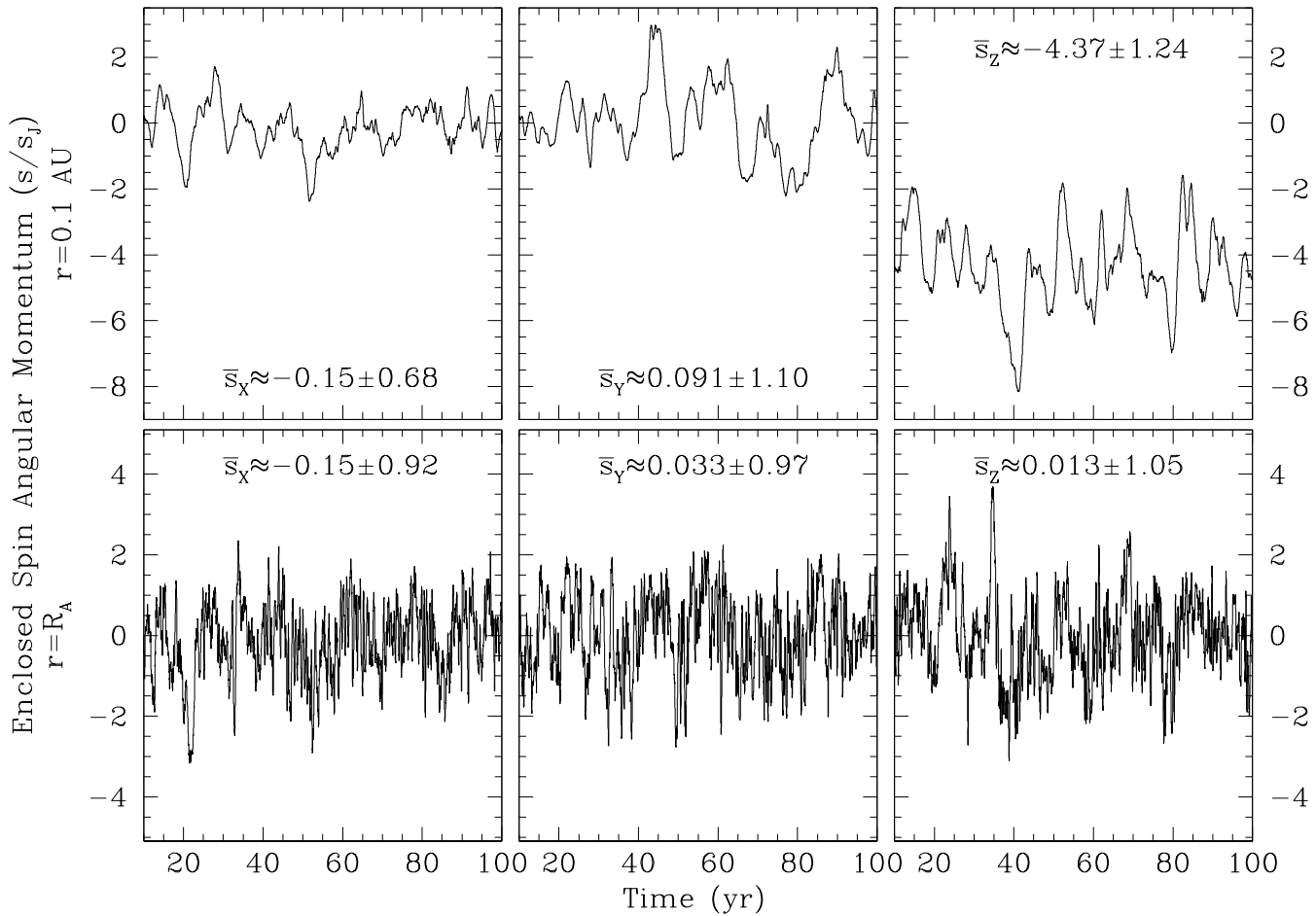} }
\caption
{\label{fig:spin-enc} 
The spin angular momentum per unit mass calculated around the core for
gas enclosed by a sphere of radius $r=0.1$~AU$(\approx0.9$\rh)
or by a sphere one accretion radius ($r=$\ra$(\approx0.026$~AU) in size
(top and bottom respectively). For the larger sphere, the net $z$ spin
has a negative bias due to the shear of the background flow. Both 
the $x$ and $y$ spins are centered near zero. All three exhibit large
amplitude swings in magnitude, including changes in sign: the sense of
rotation around the core reverses itself. In contrast, the spin
components enclosed by the accretion radius exhibit almost no
systematic trends except a continuing large amplitude and short term
variation. } 
\end{figure*}

In figure \ref{fig:spin-enc} we show the enclosed spin angular
momentum of gas relative to the core as a function of time, in each of
the three coordinate directions, and at the same two radii as the
enclosed mass above. For the larger sphere, the spin magnitude
oscillates, seemingly aperiodically, both above and below zero in the
the $x$ and $y$ directions on time scales of 10-20 yr. The peaks and
valleys of the oscillations extend to values as large as a few times
that of Jupiter's spin, normalized by envelope mass, and typically
falls in a range within $\pm1$~\sj\  of its mean. In the $z$
coordinate, a similar feature, with oscillations of similar or
slightly larger magnitude, is also present but does not cause the sign
to change because the mean lies significantly below zero. The fact
that the sign of the spin is uniformly negative is important because
it indicates that the `envelope' contains material which is merely
passing through the Hill volume as part of the background shear flow.
It is not a part of either a static or forwardly-rotating envelope, as
expected in a purely dynamical flow in which pressure plays no role.

On the smaller scale of the accretion radius, where we would expect
the matter flow to be influenced more by the core, the enclosed spin
per unit mass again lies near parity with present day Jupiter. Its
magnitude varies on time scales much smaller than a year, both above
and below zero (i.e. both one direction and the other), in all three
coordinate directions. The time variations for this smaller enclosing
sphere do not appear to be well correlated with any of the longer
variations visible in the larger sphere. In large part, this is a
consequence of the much larger total mass enclosed by the latter, so
that the variations on small scales are simply washed out.  

Combining the results of figure \ref{fig:mass-encl}, which shows that
the envelope contains only a tiny fraction of the present day mass of
Jupiter, with those of figure \ref{fig:spin-enc}, which shows spin
normalized to the envelope mass, we find that actual spin magnitudes
are only a small fraction of present day Jupiter's total spin. At such
an early stage in the evolution, such a value is not particularly
surprising. Close examination of the spin characteristics remains
interesting nonetheless, for several reasons. First, perhaps the most
important feature of figure \ref{fig:spin-enc} is not a exact values
that any spin component takes, but rather that the amplitudes of each
component (apart from the background shear of the disk included in the
$z$ spin calculation), are quite similar. This indicates that the spin
direction of the planet itself is not yet fixed. The era of giant
impacts between nascent cores may still continue during this phase of
planet formation, with little evidence remaining in the spin
characteristics of the planet in its final configuration.  

Secondly, the similarity between the spin magnitudes seen here and
those of Jupiter indicates that no angular momentum loss mechanism
must be invoked in the circumplanetary environment in order for
accretion to proceed. The most common such mechanism would, of course,
be a circumplanetary accretion disk through which circumstellar disk
material would pass as it accretes onto the core. No such disk appears
neccessary at this early stage of the planet's evolution.

Third, the envelope's spin direction does not appear to be
established: it changes sign in all three coordinate directions on
spatial scales comparable to \ra\ and it changes sign in $x$ and $y$
even on much larger scales. Also, although the $z$ coordinate spin in
the $r=0.1$~AU enclosing sphere varies widely, it varies around a
negative mean value. At this scale, a purely dynamical flow (i.e. that
of a set of `test particles', affected by gravitational and fictitious
forces only), would be expected to produce forward envelope rotation
(positive $s_z$) as mass continues to accrete. Instead, some portions
of the background material flow through the Hill volume, because the
purely dynamical picture is incomplete. The flow is not dominated
solely by the gravitational forces of the core and star combined with
the fictitious forces present in a rotating medium, but also by the
pressure forces of the gas itself. We will discuss the relative
importance of the hydrodynamic effects in more detail in section
\ref{sec:backt-signif}, below.

\section{Sensitivity of the activity to numerical and physical
parameters}\label{sec:sensitivity}

The dynamical activity shown in the results above persists as long as
we are able to run our simulations--typically about 100 years of
simulation time for each run. Unfortunately, that time is quite short
compared to the time scale for the envelope to grow in mass or for the
disk as a whole to evolve. We are therefore unable to answer many
important questions regarding the activity. In particular, in order to
make broadly applicable conclusions about the physical systems we are
modeling, we must show that they are common to a wide variety of
initial conditions that may be encountered in the circumstellar nebula
at various times in its evolution, that they are common to a variety
of physical models, and that they are not unduly contaminated by
consequences of our initial conditions. The results must not simply be
`start up transients' or other artifacts of initial conditions that do
not represent the true system with sufficient fidelity.

To fully address these concerns requires models far more sophisticated
than are presently possible and we defer discussion of any final
determinations of their answers to the future. Instead, in this
section we consider a number of simple parameters that can be varied
within the context of our models, and to which we may reasonably
suspect that our models may be sensitive. If our models vary their
behavior in these, admittedly crude, sorts of parameter studies, we
may justifiably conclude that they will be even more sensitive to the
much more varied and variable physical properties present over the
lifetime of a circumstellar disk. In addition, there may be
unaccounted for numerical issues besetting our simulations which cast
doubt on the results. If the simulations are comparatively insensitive
to the parameters varied here, we will make a first step towards the
physically meaningful conclusion that the dynamical activity discussed
above can be present at interesting levels at many different times
during circumstellar disk evolution. 

As a useful means to quantify similar and dissimilar behavior, we
adopt the spin of the envelopes, introduced in the last section, as a
metric suitable for our purposes here. We summarize the time averaged
quantities for each of the simulations discussed above and in the 
following sections, in table \ref{tab:spins}. As in table \ref{tab:sims},
the first column defines the simulation name. The next two pairs of
columns (columns 2/3 and 4/5) each specify the time averaged spins
and their variances of the envelope material enclosed by spheres of
$r=0.1$~AU and $r=$\ra, respectively, for each simulation. Both quantities
normalized to the angular momentum per unit mass of present day Jupiter.
For the series of simulations with varying background temperature
({\it tm05-tm40}), the sphere of radius \ra, is replaced by one of radius 
$r=0.025$~AU, for consistency, as discussed in section \ref{sec:backt-signif}.
For the {\it iso2} and {\it iso3} simulations, no time averages
are reported, due to their short duration. The last column specifies
the radial offset of the corotation radius from the planet that results
from each physical model. We explore the importance of variations
in this quantity below.

\begin{table}
\caption{Time Averaged $z$ components of Spins and Spin Variances}
\label{tab:spins}
\begin{tabular}{cccccc}
\hline
Label & $s_z/s_J$ & $\sigma_s$ & $s_z/s_J$ & $\sigma_s$ & CR offset        \\
      & $r=0.1$AU & $r=0.1$AU  & $r=$\ra   & $r=$\ra    & $\delta a$ (\rh) \\
\hline
lo10 &  -3.54       &  0.99    &   0.28    &  0.73      &  -0.219          \\ 
ft10 &  -4.63       &  1.31    &  -0.05    &  1.20      &  -0.039          \\
fl10 &  -4.40       &  0.91    &   0.005   &  0.97      &  -0.172          \\
br10 &  -4.75       &  0.93    &  -0.43    &  1.19      &  -0.203          \\
bp10 &  -4.83       &  1.31    &  -0.53    &  1.15      &  -0.344          \\
bs10 &  -4.48       &  1.27    &  -0.13    &  1.19      &  -0.203          \\
sg10 &  -4.68       &  0.98    &  -0.18    &  1.13      &  -0.203          \\
so10 &  -3.99       &  0.07    &  -0.17    &  0.03      &  -0.203          \\
sg20 &  -3.36       &  1.71    &  -0.03    &  1.93      &  -0.156          \\
b05h &  -2.75       &  1.70    &   0.60    &  1.67      &  -0.053          \\
tm05 &  -2.26       &  1.50    &   0.78    &  1.54      &  -0.053          \\
tm10 &  -4.26       &  1.43    &   0.11    &  1.41      &  -0.103          \\
tm20 &  -4.37       &  1.24    &   0.038   &  1.05      &  -0.203          \\
tm40 &  -5.53       &  0.76    &  -0.52    &  0.82      &  -0.39           \\
iso1 &  -2.59       &  0.31    &   1.44    &  0.42      &  -0.203          \\
iso2 &    -         &   -      &    -      &   -        &  -0.102          \\
iso3 &    -         &   -      &    -      &   -        &  -0.051          \\
adi1 &  -3.88       &  0.13    &  -0.22    &  0.15      &  -0.203          \\
\hline
\end{tabular}
\end{table}

\subsection{Sensitivity to the changes in the rotation
curve}\label{sec:rotcurv}

The dynamical activity seen above is clearly driven by interactions
between the background flow and the core, mediated by hydrodynamic
interactions deep in the core's gravitational potential well. We saw
that the background flow could be distinguished from the local
envelope to much smaller distances than are predicted by purely
dynamical models. We also saw that background material could penetrate
deeply into the core's potential well before undergoing substantial
thermodynamic evolution such as compression or shock heating. It is
therefore reasonable to ask whether features of the background flow
may influence the strength of other properties of the activity. Does
changing properties of the background flow change the behavior of
activity generated in our simulations?

Two important properties of the background flow are first, the overall
shear itself and, second, the radial location of `corotation' relative
to the orbit radius of the core itself. The first is largely a
consequence of the near Keplerian orbital flow in the disk and, while
other rotation curves may be of interest in some systems (e.g.
galaxies), they are of no practical importance in the present context.
The relative positions of the core and of the disk material orbiting
at the same frequency is of much greater interest. It may shift
signficantly due to the overall mass of the disk (and whether that
mass is included self consistently in a calculation of the rotation
curve), its temperature profile and its density profile. To what
extent will such shifts affect the flow around the planet itself?

Although only a model that includes all components of gravity on all
components of the system can be correct, including, excluding or
otherwise modifying one term or another will allow us to explore the
sensitivity of the results to that term. Here, we will include the
effect of gravity and pressure in several different ways in order to
explore the sensitivity of the results to the importance of the
rotation curve of the disk, and the offset it may have relative to the
orbit velocity of the planet. All experiments will include the
gravitational force of the star and the planet as a fixed background.
In separate simulations we will add additional sources of
gravitational force, component by component. Parameters for each of
these simulations, and the position of the co-rotation resonance
relative to the core for each of these simulations are tabulated in
Tables \ref{tab:sims} and \ref{tab:spins}. 

\begin{figure*}
\includegraphics[width=16cm]{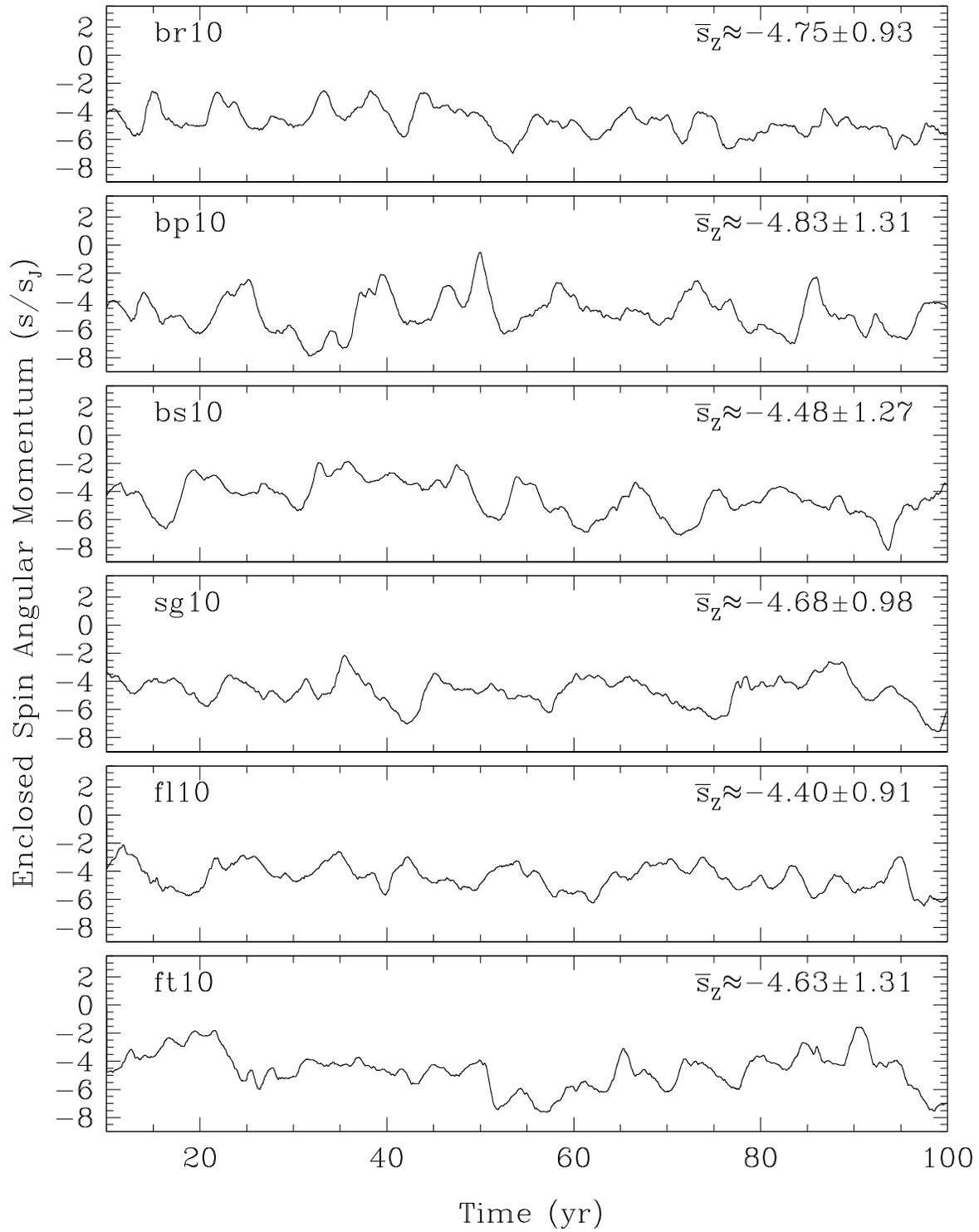}
\caption
{\label{fig:zspin-R0.1all} 
The $z$ spin angular momentum per unit mass calculated around the core
for gas enclosed by a sphere of radius, $r=0.1$~AU$(\approx0.9$\rh) in
size, for the model designated in the upper left corners of each
panel. The time averaged magnitude of the spin and its variance are
tabulated in the upper right corner of each panel.} 
\end{figure*}

\begin{figure*}
\includegraphics[width=16cm]{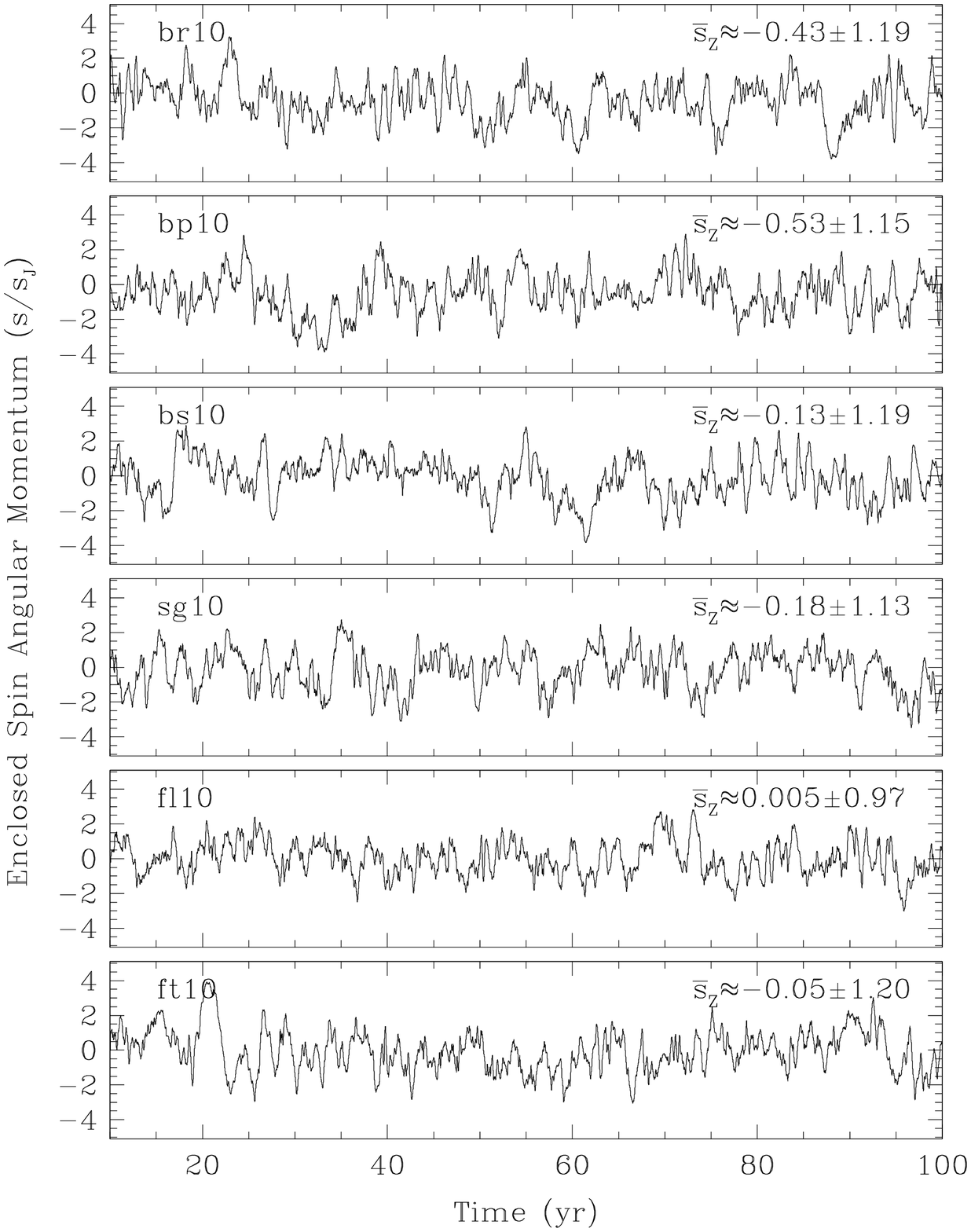}
\caption
{\label{fig:zspin-RAall} 
Same as for figure \ref{fig:zspin-RAall}, but for a sphere of radius 
$r=$\ra$(\approx0.026$~AU).}
\end{figure*}

In figures \ref{fig:zspin-R0.1all} and \ref{fig:zspin-RAall} we show
the $z$ component of spin per unit mass enclosed inside spheres of the
same two radii around the planet as are shown in figure
\ref{fig:spin-enc}, for a variety of different physical assumptions
about the underlying rotation curve. In every case, variations
qualitatively similar in both magnitude and time scale to those seen
for our prototype model are present. Quantified in terms of their
variances over time, the spins in each of the models vary by as much
or more than the present day normalized spin of Jupiter over time
scales much shorter than the 100 yr lifetime of our simulations. As
expected from their different initial conditions, differences appear
in the time averaged spin values, with a spread of about half of the
present day normalized spin of Jupiter between the maximum and minimum
seen in a given model. Of note is that the spread in the average
values is smaller than the variations seen over time in a single
model. No model stands out as particularly different from the others,
nor are the results of any model particularly different from that of
the prototype model discussed in section \ref{sec:prototype}. In fact,
the difference between the prototype's averages and those of the {\it
sg10} simulation, which uses an identical physical model but one less
grid, are comparable to the differences between it and any of the
other models with varying physical parameters.

We conclude that different physical conditions in the disk will
generate difference in the details of the envelope activity, but will
not affect its presence or overall magnitude.

\subsection{Sensitivity to the description of the core
itself}\label{sec:core-mass}

In addition to possible sensitivities to the background flow, we might
also expect dynamical activity to be sensitive to the mass of the core
and to the distribution of mass close to the core, each through the
gravitational forces they exert. Both can readily be probed by varying
the mass of the core and its gravitational softening length.
Increasing the core mass retains the shape of the potential well, but
changes its depth. Increasing the softening length changes the shape
of the potential well, while decreasing its depth (see
eq. \ref{eq:pm-grav}). The softening length serves as a probe of the
sensitivity to the mass distribution because, in addition to its use
to reduce numerical instabilities in a simulation, its presence in the
otherwise perfectly inverse square, Newtonian force law is effectively
equivalent to the statement that the core's mass distribution is
spatially extended on the scale of the softening length itself.

\begin{figure*}
\rotatebox{-90}{
\includegraphics[width=13.5cm]{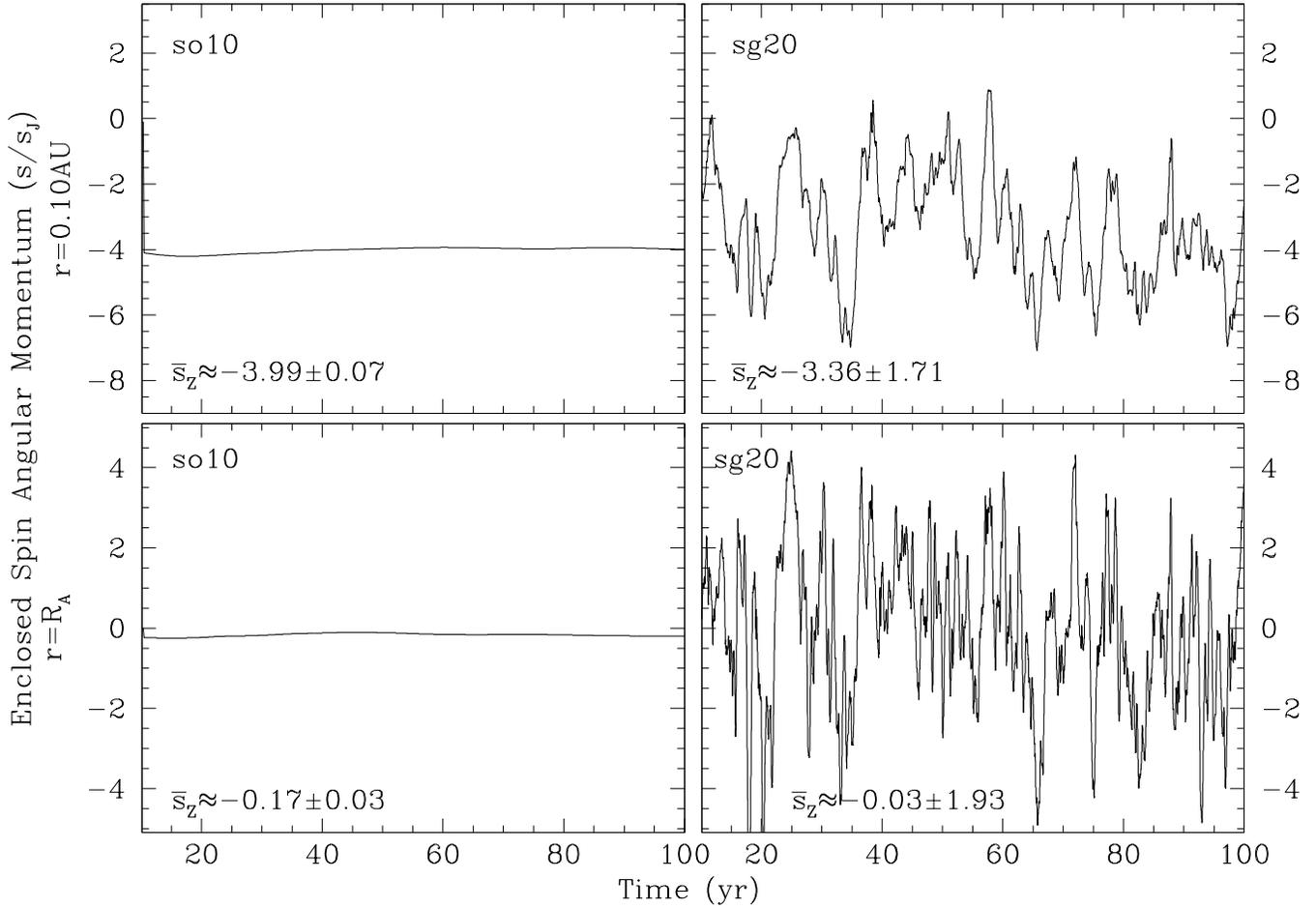}}
\caption
{\label{fig:zspin-grav} 
The $z$ spin angular momentum per unit mass of the envelope material
enclosed in spheres of radius $r=0.1$~AU($\approx0.9$\rh) (top) and of
$r=$\ra\ (bottom), each as functions of time. The left panels refer to
simulation {\it so10}, with increased gravitational softening, while
the right panels refer to simulation {\it sg20}, with increased core
mass.} 
\end{figure*}

Here, we test the sensitivity of our results with two models. In one
model we define a core mass of 20\me, which is twice our prototype
value, and in the second model, we define the gravitational softening
parameter, $\epsilon$ (eq. \ref{eq:pm-grav}), from $\epsilon=1\times
\delta x$ to $\epsilon=16\times \delta x$, where $\delta x$ is the
size of one grid cell on the finest (in this case, the fifth) mesh.
This increase corresponds to an increased spatial scale from $\sim
1.9$R$_J$, to $\sim 30$R$_J$, or $\sim$\rh/8. Figure
\ref{fig:zspin-grav} shows the $z$ spin of envelope material for these
two models. In both cases, the characteristics of the envelope
activity differ markedly from those of the prototype. With increased
softening, the variations are near completely absent. With increased
core mass, the magnitude of the variations increase by a factor of
available to disk material as it traverses through the deeper
gravitational potential well of the more massive core.

We conclude that characteristics of the core and envelope do affect
the magnitude of the dynamical activity, and that a deep, sharp
potential well is a requirement for the activity to be present. An
envelope will likely exhibit dynamical activity only when the core it
surrounds is sufficiently massive, and when the envelope itself is
sufficiently diffuse not to contribute substantially to the shape of
the potential well.

\subsection{Durability of the dynamical activity}\label{sec:durable}

Because the duration of our simulations is short and the initial
conditions are not especially tuned to begin close to any sort of
equilibrium or steady state flow, it is possible that the dynamical
activity is simply a consequence of inconsistencies in our initial
condition. If so, then we might expect the behavior of the flow
observed in a given simulation to change as a function of how long we
run that simulation. In order to investigate this possibility, we have
run a comparatively low resolution simulation, using the same physical
model as our prototype, for more than 1600 years of simulation time.

\begin{figure*}
\rotatebox{-90}{
\includegraphics[width=13.5cm]{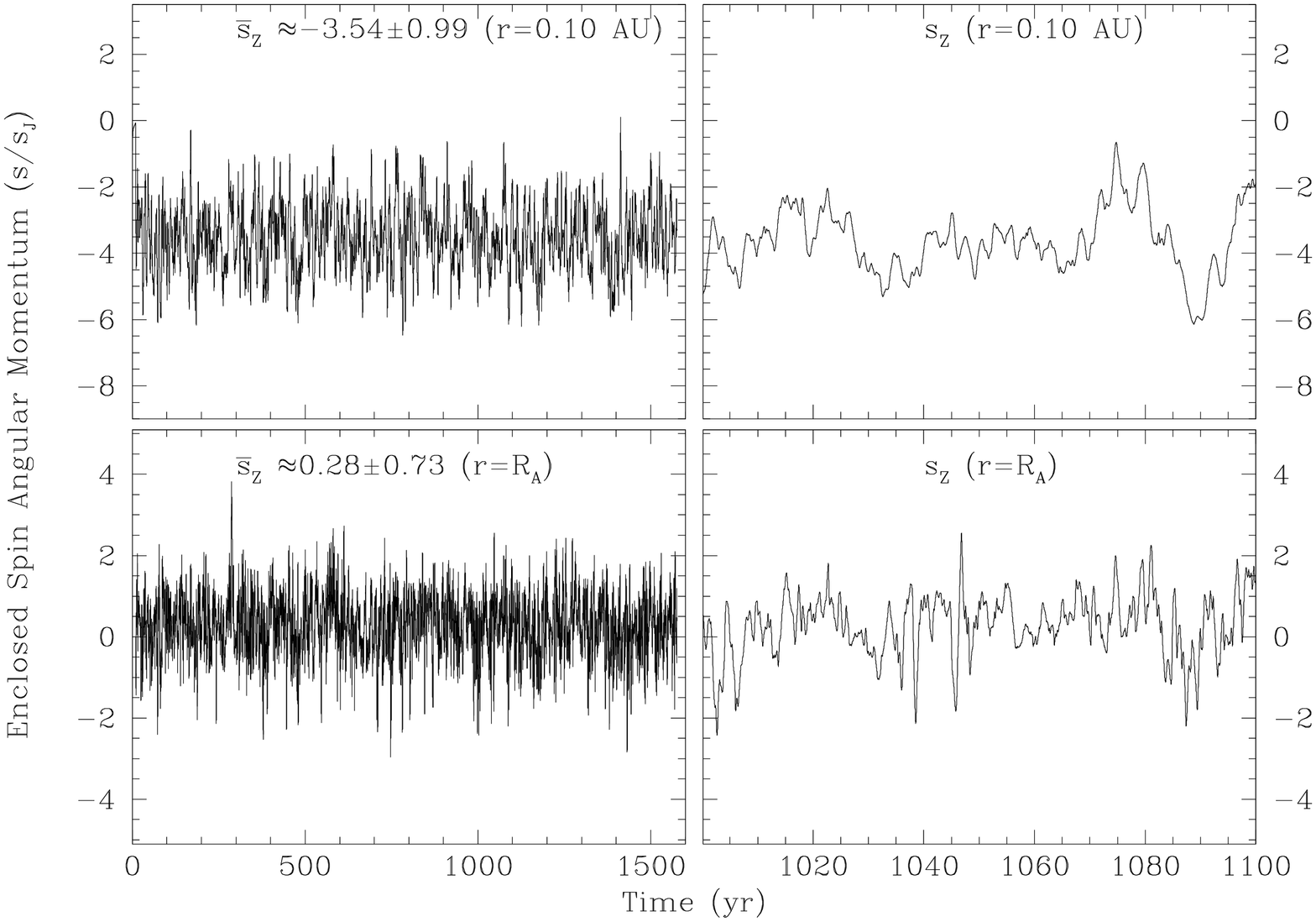}}
\caption
{\label{fig:zspin-lo10} 
The $z$ spin angular momentum of the envelope material in simulation
{\it lo10} as a function of time, enclosed in spheres of radius
$r=0.1$~AU($\approx0.9$\rh) (top) and of $r=$\ra\  (bottom). The left
panels show the variation over the entire span of the simulation,
while the right panels show the variation over a 100~yr segment. }
\end{figure*}

In figure \ref{fig:zspin-lo10} we show the $z$ spin of the envelope
material as a function of time, over the life of the simulation. As
for the high resolution models above, the magnitude of the spin
displays continuing activity, with no signs of decrease or modulation.
Inside the accretion radius, the spin direction changes on time scales
of $\la$~1~yr, as it does for our prototype. On the scale of the Hill
volume, the overall direction (backwards relative to the present day
spin of Jupiter) again retains a strong signature of the background
flow, but is again perturbed significantly by the dynamical activity.
In both cases, the variances around the means fall some $\sim20$\%
smaller than were seen in the higher resolution models, but remain
comparable to the normalized spin of present day Jupiter. An important
difference is that a positive mean spin exists for the material inside
\ra\  and, for the $r<0.1$~AU volume, a somewhat less negative overall
spin because of the opposing contributions of the inner envelope and
the background flowing through the outer envelope volume.

We attribute the differences in behavior to differences in resolution
because the smaller cells and time steps provided by the higher
resolution realization permit more rapid and larger amplitude
variations than in the lower resolution simulation. None of the
differences contribute to any overall growth or decay of the dynamical
activity. We therefore conclude that the dynamical activity is not due
to start up transients arising from inconsistencies in our initial
condition. 

\subsection{Sensitivity to thermodynamics}\label{sec:thermo}

We now turn to the question of how our results depend on the
thermodynamic treatment of the gas. There are two distinct physical
properties that may be important for the flow. They are first the
heating and cooling properties of the gas together with the equation
of state and second, the temperature of the background flow. In the
following two sections, we examine both of these possibilities in
turn. 

\subsubsection{Importance of heating and cooling, as approximated by
assumed equations of state}\label{sec:eos-signif}

Although we do not include any external heating or cooling directly in
our study, we can perform experiments comparing our ideal gas equation
of state to locally isothermal and locally isentropic equations of
state. Detailed discussions in \citet{DynII} and \citet{Pick03}
describe the physical interpretation of the effective heating and
cooling that can be placed on simulations that use such formulations.
For our purposes here, it is sufficient to note that both ordinarily
imply very efficient cooling. Simulations which employ a locally
isentropic equation of state effectively assume that all shock heating
is instantly radiated away. Simulations which employ a locally
isothermal equation of state effectively assume that all heating due
both to shocks and $PdV$ work is instantly radiated away. 

\begin{figure*} 
\rotatebox{0}{
\includegraphics[width=125mm]{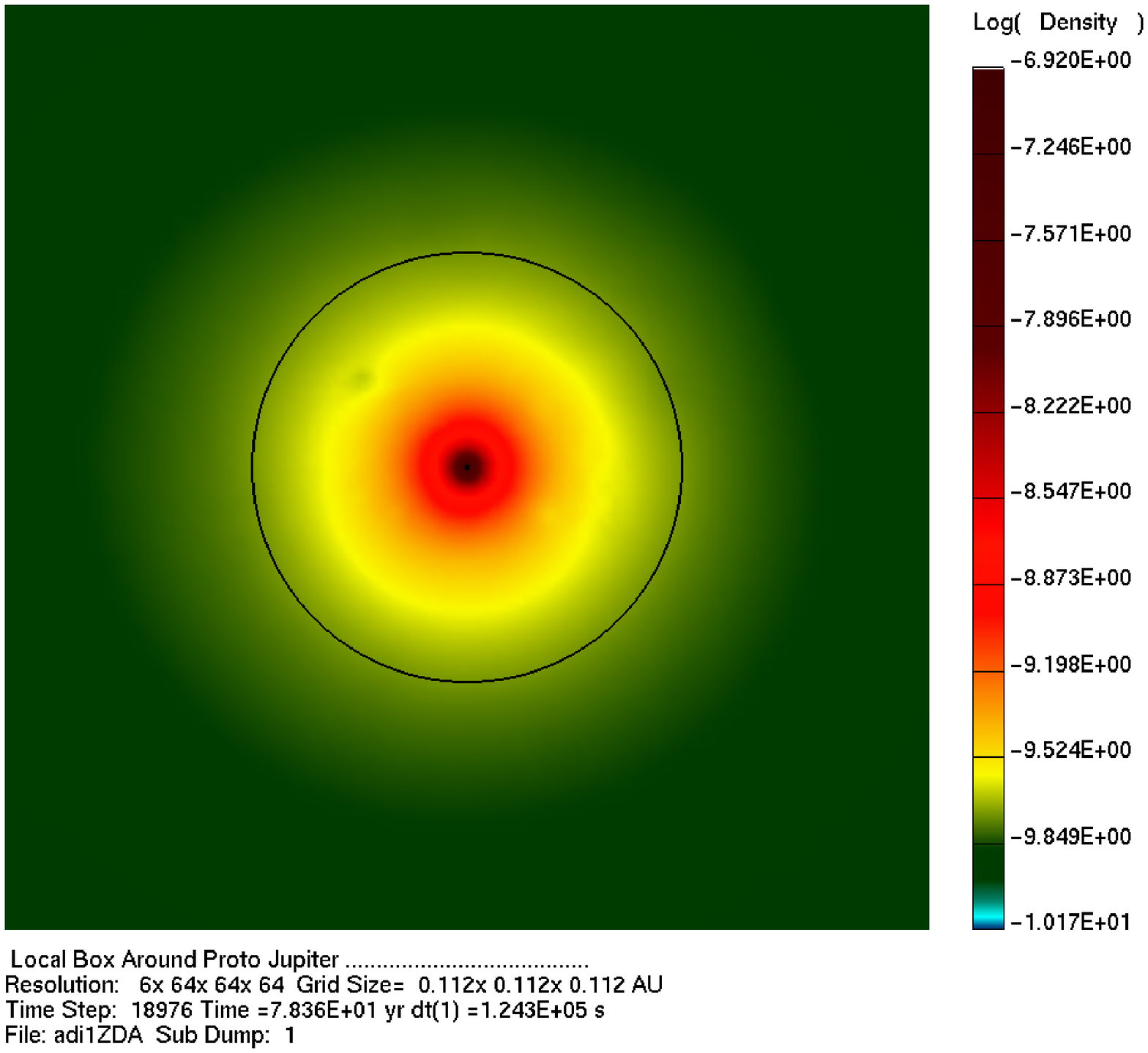}}
\rotatebox{0}{
\includegraphics[width=125mm]{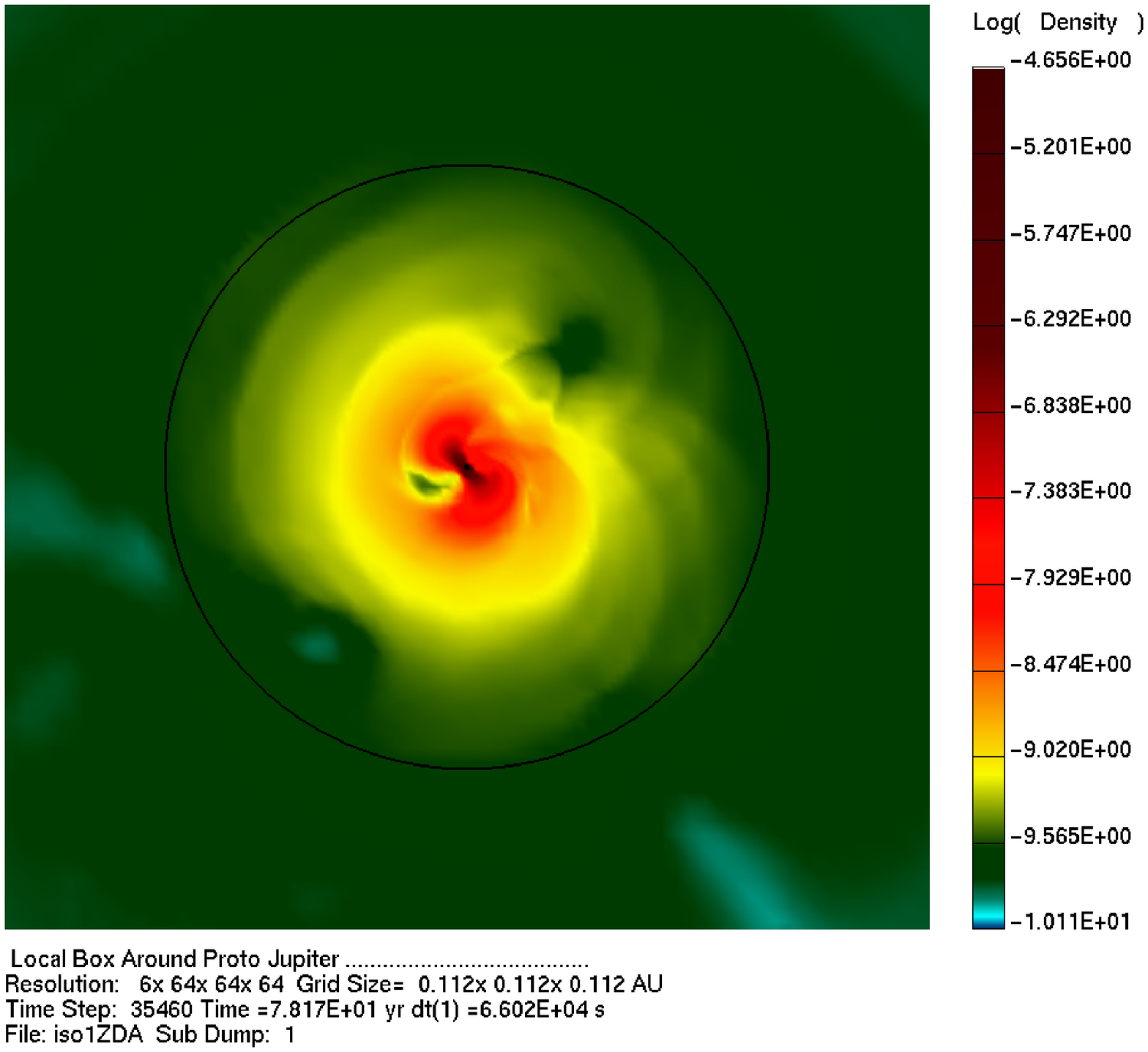}}
\caption{\label{fig:EOS-morph}
As in Figure \ref{fig:cutout-mid-dens}, but showing densities for the
locally isentropic (top) and isothermal (bottom) simulations.
The color scales are logarithmic and extend from $\sim 7\times
10^{-11}$ g/cm$^3$ in each panel to $\sim 10^{-7}$ and $\sim 2\times
10^{-5}$  g/cm$^3$ for the isentropic and isothermal simulations,
respectively. The black circles define the accretion radius, \ra,
for each simulation.}
\end{figure*}

Figure \ref{fig:EOS-morph} shows the density structure obtained from
locally isentropic and isothermal evolution. In each case, the
morphologies are far more uniform than are seen in the evolution with
an ideal gas treatment (as seen in figure \ref{fig:cutout-mid-dens}).
The density structure in the locally isentropic evolution appears
nearly spherically symmetric, with only small perturbations being
visible in the image, towards the upper left and lower right of the
core. Over time, perturbations of similar size and magnitude appear at
various positions in the envelope as the flow changes, but do not grow
substantially larger.

Locally isothermal evolution leads again to fewer dynamical features
in the flow than in our prototype simulation, but here some activity
remains visible, predominantly in the form of a barred spiral pattern
generated by material orbiting the core, along with some smaller
features which are remnants of prior spiral structures that have
merged or disintegrated as the evolution proceeds.   

The densities in the deepest parts of the envelope in both of the
simulations rise to values much higher than those seen in our
prototype. The highest central densities occur in the isothermal
evolution, at more than four orders of magnitude higher than in the
ideal gas evolution, and two orders of magnitude higher than the
isentropic evolution. The higher densities are the most visible
consequences of the implied cooling assumptions in each model, which
do not permit high temperatures to evolve deep in the envelope. In
consequence, pressure forces opposing the gravitational attraction of
the core are correspondingly reduced, which allows additional mass to
concentrate there.

We monitored the stability of the gas to numerically induced collapse
through violation of the Jeans criterion \citep{Truelove97} and,
although not strictly applicable to this system\footnote{The Jeans
analysis considers small, linearized perturbations of quantities away
from a smooth background, while the present configuration also
includes a large `external' potential gradient, due to the core.}, the
criterion was satisfied over the evolution of the simulation, using a
safety factor of $J\sim 4$. Nevertheless, at time $t \approx 85$~yr,
the gas near the core began to collapse in upon itself, and shortly
thereafter we terminated the simulation. Although a strict application
of the Jeans criterion is satisfied, we believe that this collapse
behavior is not due to any physical cause, but rather a numerically
induced consequence of the isothermal evolution and the high spatial
resolution.

\begin{figure*}
\rotatebox{-90}{
\includegraphics[width=13cm]{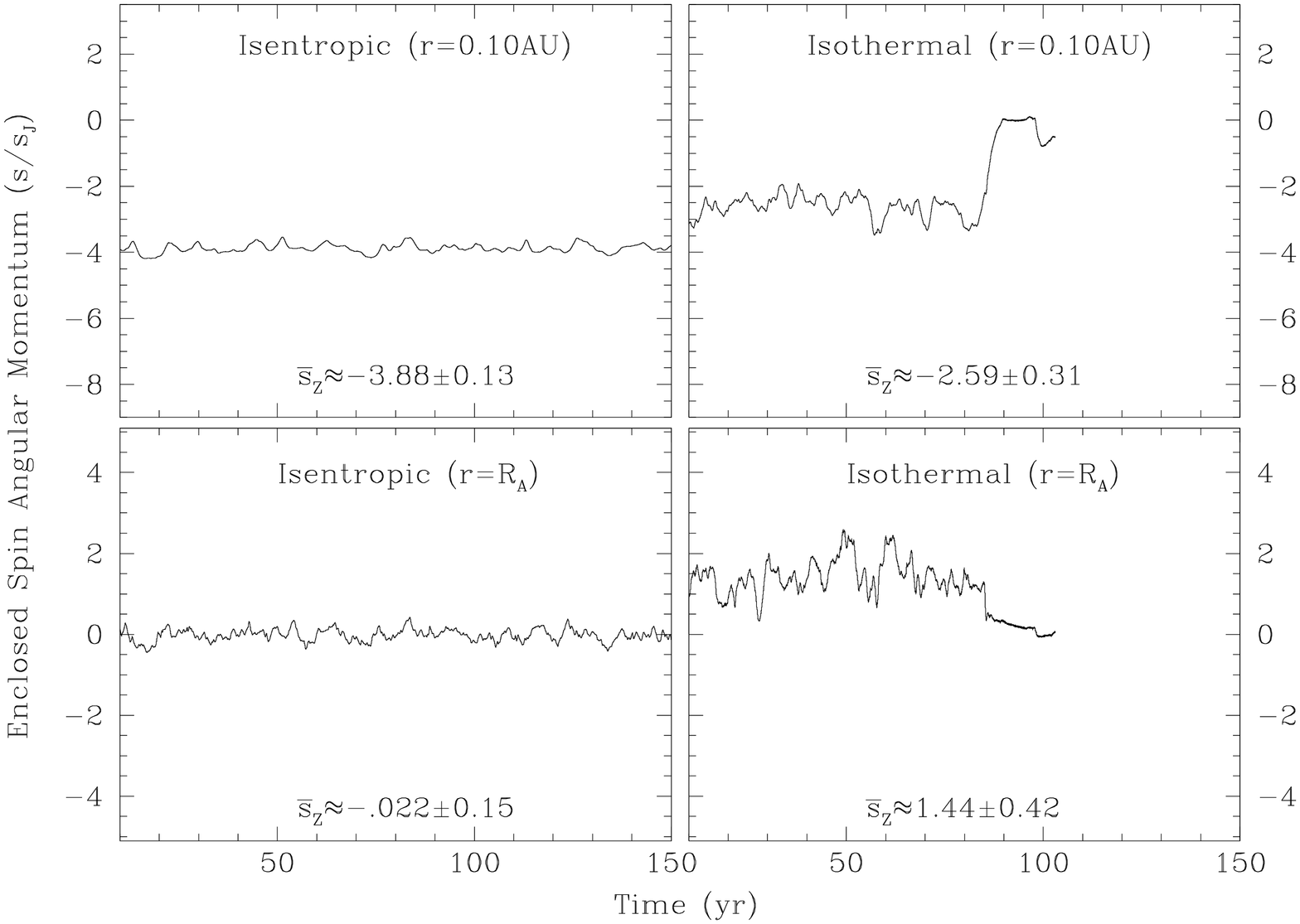}}
\caption
{\label{fig:zspin-EOS}
The $z$ component of spin angular momentum of the envelope material,
for the same two spherical volumes shown in figure \ref{fig:spin-enc}
and for the isentropic (left panels) and isothermal (right panels)
equation of state assumptions. Evolution of the isothermally evolved
simulation suffered catastrophic numerical errors after $\sim 80$~yr
of evolution and was terminated (see text). Average values appearing
in each panel are derived from the period spanned by [10,80]~yr for
the isothermally evolved simulation and [10,150]~yr for the
adiabatically evolved model.} 
\end{figure*}

Figure \ref{fig:zspin-EOS} shows the evolution of the $z$ spin of the
envelope material over time, for both simulations. In both, some
variability remains, particularly in the isothermally evolved model.
In comparison to the variations seen in figure \ref{fig:spin-enc},
however, the magnitudes are far smaller. Variations of only
$\la 0.2$~\sj\  are found in the isentropic evolution compared to
$\sim1$~\sj\  or more seen in figure \ref{fig:spin-enc}. The time scales
of the variations also appear somewhat longer, even for the smaller
volume accretion sphere, defined by $r<R_A$, though we have made no
attempt to quantify such time dependence.

Variations seen in the isothermally evolved simulation are larger than
the isentropic simulation, near $\sim0.3-0.4$\sj, but again remain
smaller than those seen in the prototype. For the larger volume
sphere, the spin remains consistently negative, as it does for all of
our other simulations, but its magnitude is much smaller. This
difference appears to be due to the fact that the spin of the inner
envelope is consistently positive, acting in the opposite sense to the
flow further from the core. Quantified by its behavior for $r\le$\ra,
the sign of the envelope spin is consistently positive: though its
magnitude still varies, the spin of the envelope is always oriented in
the same direction as the present day Jovian planets. Positive spin
orientation is an expected result for purely dynamical systems of
course (i.e. those without hydrodynamic effects), because of the
influence of the fictitious forces on the motion. The fact that we
recover a positive spin orientation {\it only} in this simulation,
where hydrodynamic feedback is largely suppressed through the
isothermal evolution assumption, provides strong support for our
conclusion that the presence and strength of the dynamical activity is
tied directly to hydrodynamic feedback effects on the flow. Further,
it serves as an important check on the physical model underlying all
of our simulations because under physical conditions where feedback is
removed, the expected dynamical result is recovered. 

Due to the changes in the behavior of the flow in these models, we
conclude that assumptions made regarding the energy balance will be of
critical importance for models of dynamical activity in the core's
environment. Of these assumptions, a complete model for the radiative
transport through the gas will be of primary importance deep in the
envelope. A necessary corollary is that a well determined dscription
of the material composition and size distribution over the range of
temperatures and densities will also be critical, because of their
influence on the opacities throughout the relevant envelope volume.

\subsubsection{Importance of the background
temperature}\label{sec:backt-signif}

In section \ref{sec:spin}, we concluded that an important physical
parameter in defining properties of the activity was the existence of
hydrodynamic feedback in the system, arising from pressure forces. As
the pressure and pressure gradients change in importance compared to
other forces acting on the flow, will the character of the activity
also change? Here we explore this question.

The most physically relevant parameter to vary in such a study is of
course the pressure itself, or its gradient, but neither are directly
accessible for use as numerical parameters in our simulations.
Instead, we vary the overall background temperature assumed for the
disk in different simulations. This quantity will serve as a proxy for
the overall magnitude of pressure effects through the relationship
defined by the equation of state. To address our question, we have run
in four variants of our prototype model (simulations {\it tm05, tm10,
tm20} and {\it tm40}), with background temperatures at the core's
orbit radius of $T=50$, $T=100$, $T=200$ and $T=400$~K. For the
$T=50$~K model, we also include a variant omitting local self gravity
of the disk ({\it b05h}).

We include two variants of the model with $T=50$~K background
temperature in order to investigate the consequences of a numerical
issue affecting simulation {\it tm05}. Due to its low background
temperature and the inclusion of local self gravity, mass accumulates
to an unphysical extent inside the simulation volume. We attribute
this behavior to the fact that the Jeans wavelength derived for the
material within the simulation volume appoximates the size of the
simulation volume itself, rather than to any physical cause. Though an
important consideration for other metrics, such as enclosed mass, in
terms of the effects this numerical defect may have on the metric
chosen in the present discussion (spin), we will find that the
differences between the two simulations are minimal. 

\begin{figure*}
\rotatebox{-90}{
\includegraphics[width=13cm]{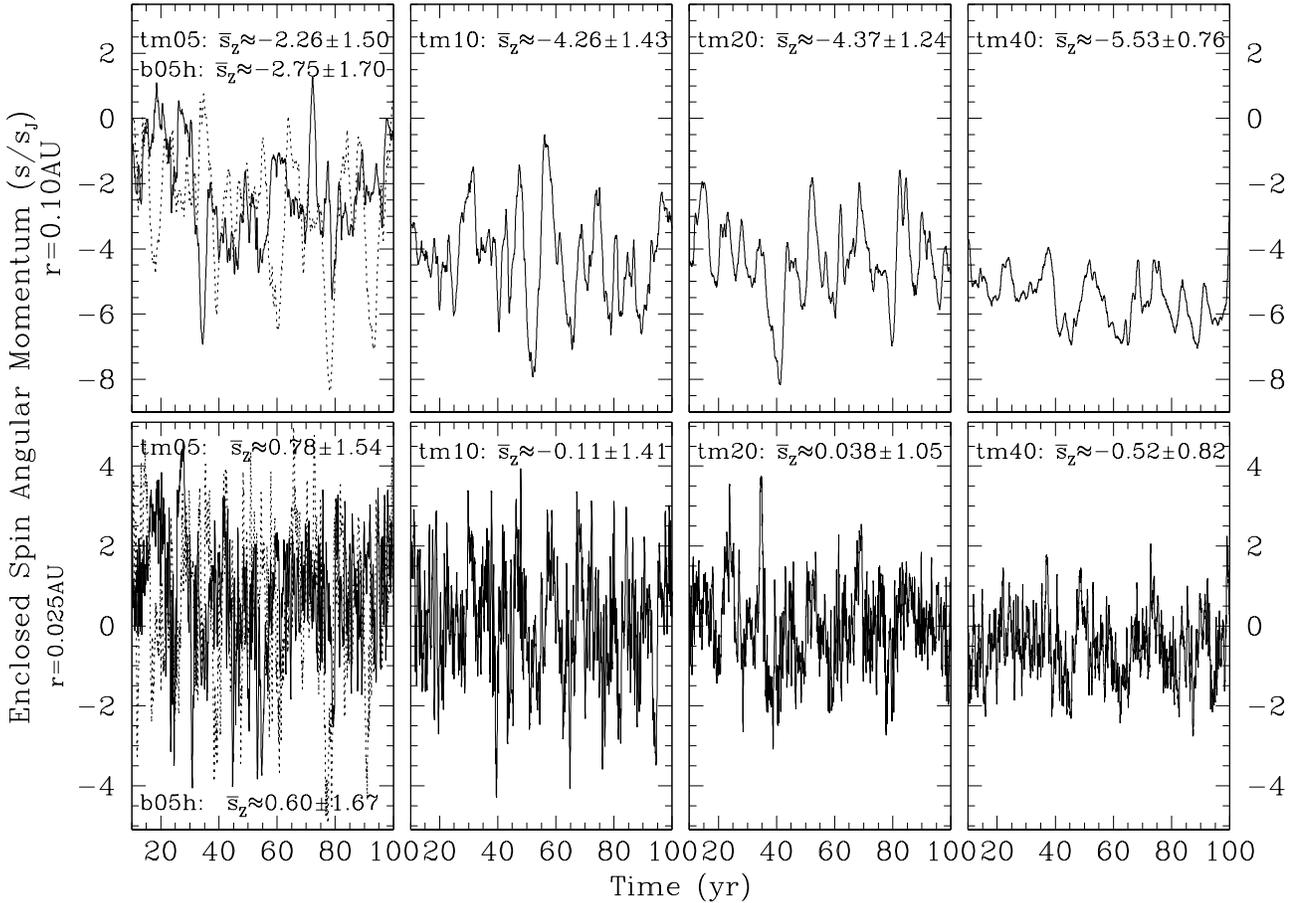} }
\caption
{\label{fig:zspin-backT} 
The $z$ component of spin angular momentum per unit mass as a function
of time, enclosed in spheres of radius $r=0.1$~AU($\approx0.9$\rh)
(top) and of $r=0.025$~AU (bottom). From left to right, the panels show
the simulations with background temperature $T=50$~K, $T=100$~K,
$T=200$~K and $T=400$~K, respectively. In the left-most panels, simulation
{\it tm05} is shown with solid curves, and {\it b05h} is shown with
dotted curves.} 
\end{figure*}

Figure \ref{fig:zspin-backT} shows the evolution of spin of the
envelope material in the $z$ coordinate over time, for each of these
simulations. From left to right, the figure shows the progression from
low to high background temperatures. For consistency with the size of
the sphere defined for the accretion radius in previous sections, we
plot the spin enclosed by a sphere of radius $r=0.025$~AU, rather than
$r=$\ra, since the latter quantity varies with temperature. The former
quantity is approximately the value of \ra\  at the $T=200$~K background
temperature of our prototype simulation.

Several notable features appear in the figure. First, the time
averaged spin values vary with background temperature. For the smaller
enclosing sphere in particular, the highest background temperature
models generate spins averaging well below zero (opposite the present
day spin directions of the major planets). For the simulation with the
lowest background temperature, the spin increases to an average nearly
80\% that of Jupiter's present day spin normalized by mass and in the
same direction. At larger radii, a similar effect occurs--the envelope
spin increases, becoming much less negative as temperature decreases.
The values of the spin averages fall well outside the range derived
from the study of different physical models in figures
\ref{fig:zspin-R0.1all} and \ref{fig:zspin-RAall}, above, and we
conclude that they signify a quantitative difference in the flow
behavior in comparison to the other models in our study.

The trend towards more negative spins at higher temperatures and more
positive values at lower temperatures correlates directly with the
changes in position of co-rotation relative to the core (see Table
\ref{tab:spins}). For simulation {\it tm40}, for example, the inwards
offset is largest of any in our study. By its nature, such an offset
exposes the core to material with correspondingly larger relative
velocities, and larger net angular momenta relative to the core. In
this case, the large negative angular momentum of the material
overwhelms the tendency of the system's fictitious forces to produce
prograde rotation around the core, even in the regions closest to the
core itself.

At the other extreme, the behavior seen in the lowest temperature
simulations is consistent with the prograde rotation we would expect
in the limit of a zero temperature, purely dynamical flow, expected
from an ensemble of massless test particles. Even so, for the $T=50$~K
simulation, where \ra$\sim$\rh, the spin enclosed within the Hill
volume remains negative, indicating that a significant component of
material at these distances is not `envelope' material at all, but
rather merely background disk material which is passing through that
volume of space. We conclude that for any temperature that we might
reasonably expect to be present in the disks of forming stars, some
disk material will enter the classical Hill volume and return again to
the circumstellar disk.

\begin{figure}
\includegraphics[width=8.5cm]{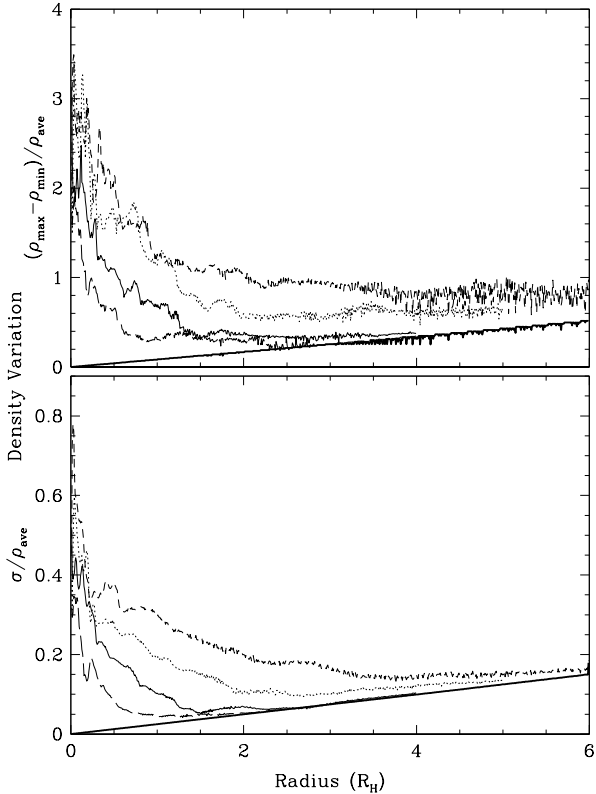}
\caption
{\label{fig:densvar} 
Variations in density as a function of distance from the core at
$t\approx 74$~yr (i.e. at the same time presented in figures
\ref{fig:cutout-mid-dens}-\ref{fig:cutout-mid-vvect}), for
simulations with different background temperatures, using the density
extrema (top) and the density variance (bottom) as metrics, both
normalized to the average density at each distance. The long dashed,
solid, dotted, and short dashed lines define the variations for
simulations {\it tm40, tm20, tm10} and {\it tm05}, respectively. The
heavy solid line defines the variations intrinsic to the background
flow, which are derived from the radial gradient of density as a
function of distance from the star.} 
\end{figure}

Second, at progressively lower background temperatures, the amplitudes
of the spin variations increase, particularly for the smaller
enclosing sphere (bottom panels) where the variation increases by
nearly a factor of two as measured by the rms deviations from the
averages. The amplitude and spatial extent of density variations in
the envelope shows a similar trend, as we show in figure
\ref{fig:densvar}. Within a distance of $\sim1/4$\rh from the core,
amplitudes of a factor of up 2-3 in normalized ``peak to peak''
variation and 30-70\% in variance are present in all four simulations.
At larger distances, significant differences appear between the
behaviors seen at different temperatures. While the variations in both
amplitude and spatial extent decrease with increasing temperature in
all four simulations, for the highest temperature realization they
become essentially indistinguishable from the background flow within a
distance of about 2\rh. At the other extreme, for the lowest
temperature realization, variations remain significant up to
$\sim4$\rh\ from the core, and are non-zero all the way to the edge of
the simulation volume, at 6\rh. The behavior of the middle temperature
realizations are bracketed by those at the high and low extremes.

We attribute the increases in variability, both in spin and in
density, to feedback generated by perturbed material as it returns to
the envelope for one or more additional encounters. Expelled material
will inevitably exhibit differences from the smooth background flow as
it enters the core's environment. When temperatures are low, flow
perturbations extend to larger distances and reach greater amplitudes
than at higher temperatures, due to the decreased importance of
pressure derived forces far from the core. These forces act to
restore the flow to its background state and, when they are small,
larger excursions may be sustained before the restoring forces reach
magnitudes large enough to return the flow to its unperturbed state.
Significantly, variability is not suppressed for any background
temperature studied, and we conclude that the activity is robust.

\subsubsection{Generalizing the results to other core
masses}\label{sec:gen_coremass}

The simulations in the last section consider only 10\me\ cores. The
results, however, can be generalized by framing them in terms of the
the accretion radius, to parameterize the spatial extent of the
activity, and the dimensionless ratio of the accretion radius to the
Hill radius:
\begin{equation}\label{eq:ratio}
{{R_A}\over{R_H}} = {{ G(3M_*)^{1/3}}\over{a_{\rm pl}}} 
                          {{M_{\rm pl}^{2/3}}\over{c_s^2}},
\end{equation}
to parameterize characteristics of the flow activity. The ratio is
inversely proportional to the background temperature, by way of the
squared sound speed term, $c_s^2$, and its relationship to temperature
through the equation of state. It is also directly related to the
core's mass present in the definitions of both \ra\ and \rh, resulting
in the proportionality of \mplan$^{2/3}$. Therefore, for a given
semi-major axis, we can specify characteristics of the dynamical
activity for any pair of $(T,M_{\rm pl})$ values. 

We can interpret the meaning of a given ratio by recalling the
physical meanings of the two quantities that define it. Specifically,
the accretion radius, as noted in section \ref{sec:radii}, defines the
location at which gas's thermal energy and gravitational energy of the
core become comparable in magnitude. In turn, the Hill radius (or more
generally, the full three dimensional Roche surface, for which the
Hill radius defines the largest separation from the core) defines the
locations at which the centrifugal potential energy arising from the
rotating frame and the combined gravitational potential energies of
the star and core become comparable. Their ratio, therefore, becomes a
measure of the relative importance of the rotating frame in comparison
to thermodynamic properties on the flow.  

In this context, we designate the region of parameter space where the
ratio \ra/\rh\ is below unity, as the ``thermodynamic flow regime'',
because the thermal energy of the background flow exceeds that of both
gravitational and centrifugal potentials at the Hill radius.
Therefore, the flow will be dominated by the pressure effects that
define the thermodynamic state of the gas, with only small influences
from the rotating frame. At smaller separations, effects of the
rotating reference frame will have still smaller influence, while
effects due to the gas's thermal properties increase. As gas falls
deeper into the core's gravitational potential well, hydrodynamic
feedback on the flow becomes important, as gravitational potential
energy is converted into thermal energy. We therefore expect flow
features similar to the ``non-stationary'' behaviors reported by,
e.g., \citet{R97,R99}, who report flow patterns around a point mass in
which the direction of rotation oscillates in time. 

We designate the region of parameter space where the ratio is larger
than unity, as the ``celestial dynamic flow regime'', because thermal
energy plays a relatively smaller role at the Hill radius, so that gas
flow more closely resembles that of a purely dynamical system of
massless test particles. Systems in this regime will be characterized
by prograde rotational flows near the core, a ``horseshoe'' orbit
region at distances outside the separatrix defined by the Hill volume,
and circulating orbits both inside and outside of the orbit position
of the core. Finally, for ratios near unity, thermal energies and
gravitational energies, along with the forces associated with each,
are of comparable magnitudes. We therefore may expect the flow to
exhibit features of both types of flow.

\begin{figure}
\rotatebox{-90}{
\includegraphics[width=6.5cm]{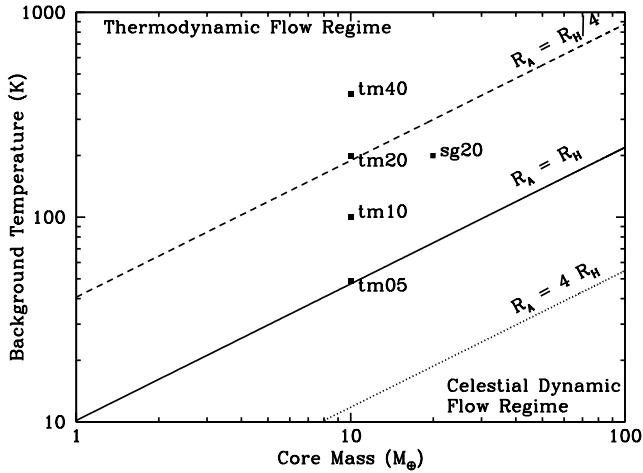}}
\caption{
\label{fig:ratios}
Lines of constant \ra/\rh, plotted in $(T,M_{\rm plan})$ space,
together with the loci of several of our simulations, as marked. Three
lines are plotted, showing ratios of unity (solid) and ratios where
\ra\ is four times larger (dotted) or smaller (dashed) than \rh. The
regimes designated as celestial dynamic and thermodynamic flow regimes, 
as defined in the text, are also shown.}
\end{figure}

Figure \ref{fig:ratios} displays three lines of constant radius ratio
in $(T,M_{\rm pl})$ space, each assuming semi-major axis of 5.2~AU, as
in our simulations. Following the discussion above, we expect flow
patterns of similar character at each point along these lines, and
different flow patterns as the ratio changes. Specifically, that
oscillating flows are ubiquitous for all of the simulations, as
expected from their placement in the thermodynamic flow regime. In
addition, for lower temperatures, where the ratio lies closer to
unity, the effects of the rotating frame play a larger and larger role
as the time averaged spin tends toward positive, or at least less
negative, values. Higher amplitude oscillations are also generated,
consistent with the increased importance of the celestial dynamic
effects due to the rotating frame. 

Of note in the figure is that for the low core masses typical of the
early stages of growth, and for temperatures of $\la 150-200$~K, as
expected in the region of core formation (i.e. the ``ice line'', where
solid densities increase due to the formation of ices), the flow
characteristics fall well within the thermodynamic flow regime.
Therefore, given a background temperature near that of ice formation,
we expect that a 1\me\ core will exhibit similar flow characteristics
to our simulation {\it tm40}, with an oscillating flow pattern present
in a limited volume close to the core, with low amplitude. At the same
temperature, a 3-5\me\ core would generate activity over a larger
fraction of its Hill volume, with similar behavior to that of our
prototype simulation {\it tm20}. Thermodynamic driving of the flow
will become unimportant only at very high core masses or very low
temperatures.

We conclude that thermodynamically active flows will be an important
characteristic of core environments during their early stages of
growth. The activity will only decrease in importance for objects
above a few tens of earth masses in size, as values of the radius
ratio increase to well above unity. Because we have not attempted
simulations with such high core masses however, we cannot determine
with certainty the masses for which the activity will become
negligible.

\section{Relevance of our models for chondrule
formation}\label{sec:chondrules}

An important implication of the temperature and density data derived
from our simulations, discussed in section \ref{sec:1D-dist}, lies
well outside of the study of dynamical phenomena that may develop in
the Jovian planet formation environment. Specifically, the densities,
temperatures and time scales that occur in the envelope may be
important for the theory of meteoritics and, more specifically, to
models of chondrule formation. 

Chondrules, and the meteorites that contain them, represent a large
fraction of the volume of meteorites that have fallen to the earth and
been studied. They are solidified, spherical melt droplets of
refractory materials typically up to a few millimeters in size.
Briefly summarized from \citet{PP4_Jones}, we note for our purposes
that chondrules underwent both a very rapid heating event to
$\sim$2000~K, followed quickly by a rapid cooling event of magnitude
50-1000~K/hr. Among a wide variety of models purporting to produce
such conditions, passage of solid material through nebular shocks is
currently favored as among the most likely \citep[see e.g.][]{DC02}.
Among this model's drawbacks are, first, that shocks that have the
right density, temperature and velocity characteristics are hard to
form and, second, that it is difficult to arrange for these shocks to
exist for a long enough time to produce enough bodies to match the
current observations. 

The parameter space for which chondrule production may occur in shocks
is bounded by pre-shock densities within a factor of a few of
$10^{-9}$~g~cm$^{-3}$, temperatures near 300~K and shock speeds near
6--7~km~s$^{-1}$ \citep[][Table 5]{DC02}. For these conditions, the
authors also note that enhancements in the particle density were
important for formation models. Concurrent work of \citet{CH02,iida01}
come to similar conclusions. Figures \ref{fig:cutout-mid-dens} and
\ref{fig:cutout-mid-temp} show that appropriate background conditions
exist in our simulations, and we propose that dynamical activity in
the Jovian envelope could provide a source for both shocks and
reversible compressive heating that remedy the shortcomings noted
above. 

Although the basic temperature and density conditions that occur in
our simulations can be seen, e.g., in figures
\ref{fig:cutout-mid-temp} and \ref{fig:OneD-rhoT}, ascertaining
whether short duration heating and cooling events are also present
requires additional analysis. We considered and discarded the option
of including test particles that could be passively advected with the
local flow, because of the significant added computational and storage
cost they would require. Instead, we rely on the similar--but not
identical--solution of `advecting' test particles through a fixed-time
snapshot of the conditions at a specific time.  

\begin{figure*}
\psfig{file=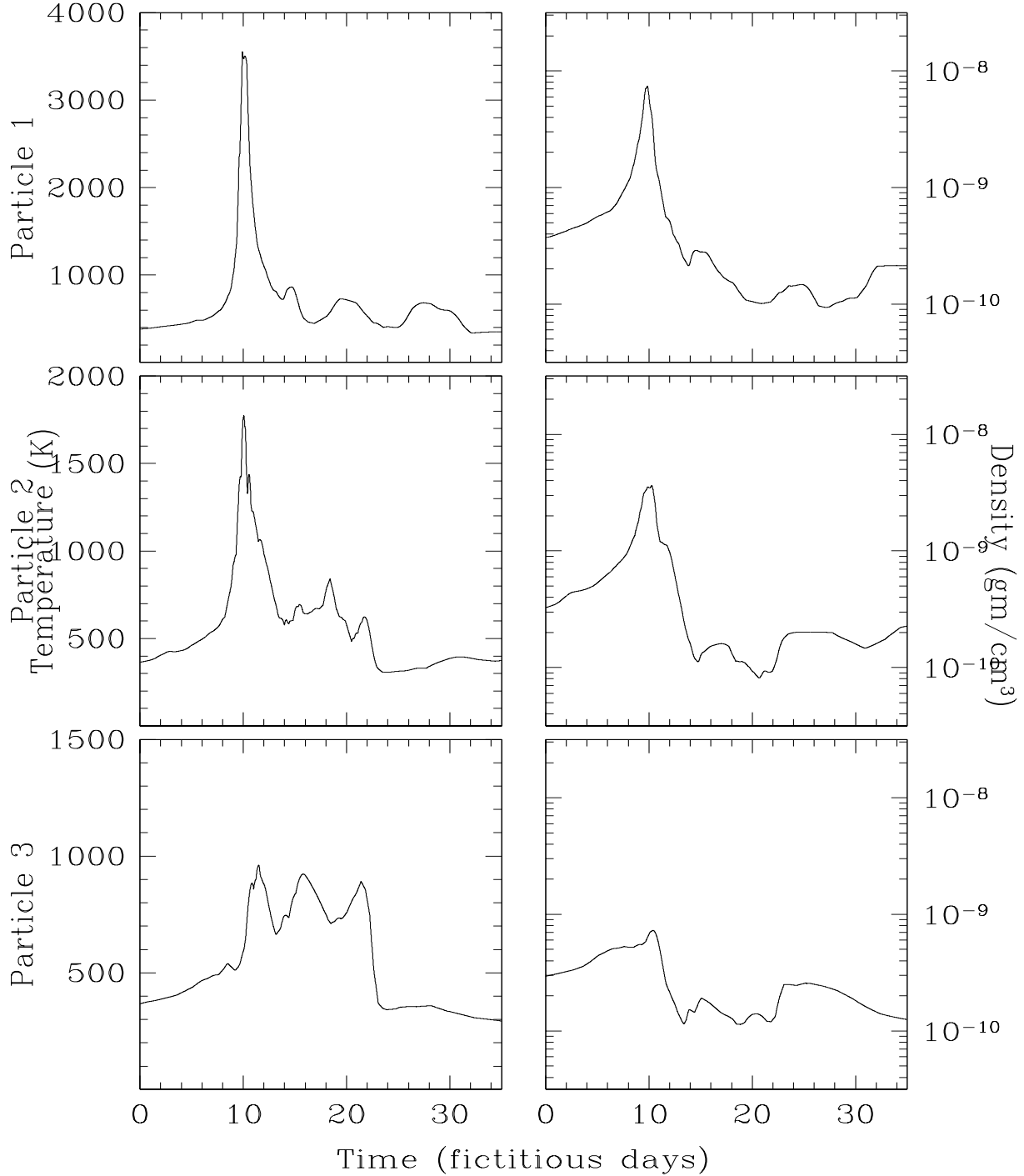,height=8.0in,width=7.0in,rheight=7.7in}
\caption{\label{fig:thermo-traj}
The temperature (left panels) and density (right panels) encountered
by three test particles as they are advected through a snapshot of the
simulation volume, each as functions of a fictitious time coordinate,
described in the text. The zero point for time has been arbitrarily
shifted in each case so that the peak temperature occurs at
$\sim10$~days.} 
\end{figure*}

To that end, we have performed a streamline analysis of the
trajectories of an ensemble of test particles injected into the
simulation restart dump of our prototype model, for the same time as
that shown for figures \ref{fig:cutout-mid-dens} and
\ref{fig:cutout-mid-temp}. Particles are placed at a set of locations
at the edge of the grid and allowed to advect through the simulation
volume using the local fluid velocity, linearly interpolated between
the grid zones adjacent to the particle at each time. Each particle is
given a time step based on that used to advance the gas at that
location, so that it advances forwards through a fictitious time
coordinate, passing through the volume at the rate of matter in the
local flow. Similar linear interpolations of density and internal
energy are used to derive the temperature from the equation of state. 

In Figure \ref{fig:thermo-traj}, we show temperatures and densities
for three test particles for which a passage through the environment
of the core has occurred. The particles shown were chosen to
illustrate the range of peak temperatures and densities that may be
encountered by slightly different trajectories through the envelope.
Each encountered a short duration heating and cooling event as they
passed through the dynamically active region close to the core, but
the magnitudes and duration varied in each case. In the top example
(test particle 1), the peak temperature rose to well over 3000~K and
the density to nearly $10^{-8}$~g~cm$^{-3}$ as the particle's
trajectory passed through the innermost regions of the envelope. The
conditions encountered by the particle 2 were much more moderate, with
peak temperature and density values of $\sim1800$~K and
$4\times10^{-9}$~g~cm$^{-3}$ respectively. Particle 3's trajectory
took it only through the outer portion of the envelope, so that it
encountered only much lower temperatures and densities, although in
this case for a much longer period of time than either of the other
two cases shown. All three particles encountered several days of
cooler processing near 500-800~K, and a visual scan of many other
similar events shows that such additional annealing is common.

The temperature peaks for test particles 1 and 2 offer widths of $\la
1$ day, with both a very rapid rise and fall. Close examination of
their trajectories reveal that the widths reflect essentially the
crossing time for a single grid zone in the simulation. Therefore,
although already quite narrow, they are actually likely to be
overestimates of the true widths that would be obtained from the same
models realized at still higher resolution. The dynamical activity in
the envelope, coupled with the temporally very narrow
temperature/density peaks, especially in cases similar to that of
particle 2, offer evidence that chondrule production could occur in
the environment of Jovian planet formation.

\section{Discussion}\label{sec:discuss}

\subsection{Summary of our main results}\label{sec:summary}

We have performed a suite of 3D numerical simulations of a cubic
`cutout' region around a 10\me\ point mass embedded in a circumstellar
disk, in a modified shearing box framework. We find that the flow is
not characterized by a simple one dimensional circumplanetary envelope
embedded in a near static background flow. Instead it is highly
dynamic, varying on time scales ranging from months to years far from
the core, down to hours or days, very close to it. 

We used the RMS amplitude of time variations observed in components of
specific angular momentum of material flowing around the core as a
quantitative proxy for this activity and studied its sensitivity to
changes in physical and numerical parameters used to model the
evolution. Specifically, we varied the character of the core, via its
mass and gravitational softening coefficient, and we studied the
sensitivity of the activity to artificial changes in the background
flow, imposed by changes in the overall temperature or density
gradients in the disk, imposed by changes in the background
temperature of the disk, or imposed by changes in the assumed equation
of state. We found that the amplitude is sensitive to the character of
the core, increasing with core mass and decreasing with larger
gravitational softening. Increases in the background temperature of
the disk cause decreases in both the amplitude and spatial extent of
the activity. Activity can also nearly be suppressed when we assume a
fixed equation of state--either isothermal or isentropic--which in
turn serves as a proxy for more complex heating and cooling processes
neglected in our current work. Each of these factors contribute to the
efficiency and character of hydrodynamic feedback on the large scale
flow by the material returning to the background after encountering
the core and we conclude that the origin of the activity is closely
linked to such feedback.

Finally, we demonstrate that an important consequence of the dynamical
activity is to generate very short duration heating and cooling events
in material that travels through the envelope. We compare conditions
in these events to those expected to be required to form chondrules,
and show that the ranges of temperatures and densities reached span
those for which chondrule formation is expected. They also extend to
longer duration and lower temperature events for which significant
annealing may occur, and we propose that the origin Jovian planets and
the origin of of chondrules and other annealed solids in the solar
nebula are linked, through the dynamical activity in the envelope. If
such a link can be established, then chondrules and annealed solids
represent a physically examinable record of the processes present
during the formation of Jovian planets.

\subsection{Potential implications of our
results}\label{sec:implications}

Our models represent a significant step towards a complete picture of
the formation of Jovian planets in the context of the core accretion
model. All of our simulations generate some dynamical activity and
only simulations evolved using fixed EOS evolution or large
gravitational softening produce activity of significantly lower
magnitude than in our prototype simulation and its many variations.
Given the severe and unphysical restrictions on the thermodynamics
imposed by fixed EOS evolution, we do not believe that the results of
those simulations closely resemble the actual evolution of any real
system. A large gravitational softening parameter is similarly
restrictive, because it implies a spatially extended and relatively
massive envelope, at very early phases of the core's growth. We
therefore believe that more sophisticated models will show that
dynamical activity will be present at levels important enough to play
a role in overall formation process of Jovian planets. If so, our
results mean that the standard core accretion model for Jovian planet
formation will require revisions, to include the effects of such
dynamical activity.

Given this statement, we now speculate on what those effects may be,
and their contribution to the planet formation process. Most
importantly, we expect the mass accretion rate to differ from that
derived in the context of the standard core accretion model, because
in that model the accretion is throttled by the rate of energy loss
from the outer surface of the envelope. When the rate is low, pressure
forces build up and choke off further accretion, when it is high,
accretion proceeds more quickly. However, if the envelope has no
stable outer surface, high pressures at any point on the `surface'
(such that one can be defined at all) will not hinder material flow
through an adjacent region where they are low. Indeed, such high and
low pressure regions may simply be consequences of material flowing
into or out of the core's environment. They may also be consequences
of thermal energy generation from shocks created by interacting
material flows. In this case, an accurate model of the evolution will
require an accurate accounting of the contributions from such energy
generation and advective terms, to both the mass accretion and energy
loss rates of the material in the core's environment.

Depending on its strength, dynamical activity may either delay
accretion, by repeatedly disrupting the formation of the envelope, or
enhance it, by driving instabilities in the flow. In the context of
our current models, we cannot determine which outcome is more likely,
or if both occur at different times. If the envelope is disrupted on
timescales short compared to its cooling timescale, then our present
models, which neglect cooling entirely, will provide an accurate
picture of the long term flow pattern. Any material which does cool
will be swept away, rather than becoming a permanent part of the
core's envelope. If instead instabilities occur which do {\it not}
entirely disrupt the envelope, a likely consequence will be rapid mass
and angular momentum transport through the envelope onto the core.
Instabilities are known to develop in rotating polytropes
\citep{PDD,TIPD}, morphologically similar to the planet formation
environments discussed here, but whether those or other, similar
instabilities can be triggered in the present context awaits further
study. Also, the consequences of such rapid accretion are unclear. How
quickly would a massive envelope develop in which further activity did
not play a large role?

The scenario we present offers a number of attractive advantages over
other models of Jovian planet formation. Unlike both the class of
formation models using global gravitational instabilities in the
circumstellar disk and of many previous models of core accretion, a
massive disk is not required at any time during the evolution. In
particular, no massive disk is required to exist for the long period
of time required for the classic picture of core accretion, for which
most gas accretion occurs only at the conclusion of a long hydrostatic
growth period. Assuming that the activity results in enhanced mass
accretion rates, it also provides a formation mechanism which avoids
one of the most critical unsolved issues in standard models of core
accretion, that the core does not grow quickly enough to avoid
migrating through the disk and onto the star. Finally, our scenario
naturally provides a mechanism for producing very short duration
heating and cooling events with thermodynamic characteristics similar
to those expected to produce chondrules and annealed solids.
Production in this scenario will endure for a significant fraction of
the formation time scale of Jovian planets (itself a significant
fraction of the disk lifetime), resulting both in a large yield of
objects and allowing both processing and reprocessing events to occur,
both factors consistent with the meteorite record.

\subsection{Comparisons to other work}\label{sec:comparisons}

The work described in this paper can be profitably compared to three
main categories of research on Jovian planets. Together they make up a
spectrum of models ranging on one extreme, to multidimensional models
which explore the interaction of a circumstellar disk with a massive
object embedded within it, using relatively simple physical models
\citep{lsa99,DHK02,dhk03,dkh03,BLOM}, and on the other extreme, to one
dimensional models which explore the core accretion paradigm using
highly detailed physical models
\citep{Pol96,AMB04,HBL05,HPK08,TerHein11,liss09}. In the middle lies a
class of work most similar to ours
\citet{AB09,Raf06,PapNel05,MKIM08,TOM12}, in which special attention
is paid to the interactions between the envelope and the surrounding
disk.

The main focus of most previous multi-dimensional work has been on
large scale interactions between the core and the disk as a whole, in
order to study the migration of the core through the disk and
determine parameters related to its survival. In this context, many
details of the structure of the envelope become less important, and
approximations are made to sidestep the more difficult numerical and
physical issues. In comparison, our focus has been primarily on the
interaction of the envelope with the core and the immediately
surrounding disk material. Accordingly, we provide only a rudimentary
description of the background disk, through a boundary condition and
through its gravitational effect on the material inside our simulation
volume. We concentrate our spatial resolution on the envelope region
only, thereby permitting much finer resolution of the flow features.
The work presented here represents the highest resolution simulations
studying the formation of Jovian planets so far published. Our work
also includes a physical model which, while necessarily more simple
than required for a complete picture of the evolution, at least avoids
some of the most restrictive assumptions of previous models.

As noted in section \ref{sec:prototype}, the morphologies found in our
work differ from those of the previous studies, exhibiting both more
dynamical activity and less disk-like or spiral structures in the
circum-core environment. The reasons for the differences between the
morphology most likely lie in differences in the physical assumptions
made in each. Specifically, the most important assumptions include
those regarding the thermodynamic treatment for the gas, those
regarding its distribution around the core and those regarding the
continued mass accretion onto the core itself. We now discuss the
physical implications of each of these assumptions.

We employ an ideal gas equation of state to close the system of
hydrodynamic equations. Additionally, no sources of heating or cooling
are included apart from shock heating generated by the dynamical
interactions of the flow on itself. In contrast, many previous workers
have chosen an isothermal equation of state, which artificially (and
very efficiently) removes thermal energy generated by those same
interactions. High temperatures and pressures are therefore
suppressed, and the feedback loop that would otherwise exist as the
hot gas perturbs incoming material cannot form. As we showed in
section \ref{sec:eos-signif}, our simulations do not exhibit dynamical
activity when an isothermal equation of state is used, consistent with
the lack of such activity in other work. A complete model will of
course include some cooling effects due to radiation, which will
undoubtedly act to suppress the activity we see, however the magnitude
of this effect is difficult to estimate from the currently available
literature. 

Although some calculations have been performed (AB09) which include
radiative cooling, they do not supersede our results due to
limitations implied by the resolution afforded by the simulations
themselves. Their spatial resolution, although high in relative terms,
is still much lower than is required resolve the flow sufficiently at
small distances from the core to capture the extremely small scale
features that characterize the shocks and dynamical activity close to
the core. For example, tests of the shock capturing characteristics of
the PPM algorithm in use in our simulatons, show that shocks can be
resolved with a maximum of 2-3 grid zones, corresponding to distances
of $\la 2 R_J$ at the grid zoning used in our simulations. In
contrast, the SPH (`Smoothed Particle Hydrodynamics') method used by
AB09 requires at least 5-10 particle smoothing lengths to resolve a
shock \citep[see, e.g., section 6 of][]{price12}. While AB09 quote a
minimum smoothing length over their simulations comparable to our zone
size, their overall particle count is quite low, implying that very
few particles have such small sizes, while most are much larger.
Coupled with the comparatively large artificial viscosity required by
the method, we believe that the features we observe in our simulations
would not have been observable in theirs, independent of any
considerations regarding cooling. 

Our work assumes that no gas has accumulated around the core. This
assumption manifests explicitly in our initial conditions, for which
we define only a `bare' core embedded in an unperturbed background
flow, and implicitly through our specification of the gravitational
softening of the core. In addition to its normal purpose as an
aid to sidestep various numerical issues in finitely resolved
simulations, using non-zero gravitational softening is equivalent
to the assumption that the core's mass is distributed over a finite
spatial volume, corresponding to the spatial scale of the softening
(see, e.g., \citet{dkh03} for a particularly detailed discussion of
such softened potentials). The gravitational softening used our
simulations is smaller than employed in many other works, where
simulations with values of $\sim$\rh/10--\rh/5 are typical
\citep[e.g.][]{dkh03}. In contrast, the largest softening we use in
any simulation is $\sim$\rh/8 (for simulation {\it so10}), while in
all others at the same resolution it is a factor 16 smaller. Given the
results of section \ref{sec:core-mass}, where we showed that softening
values this large are sufficient to suppress dynamical activity, we
conclude that other published work would not have seen such activity,
due to the differences between their assumptions and ours.

Further, the effect of permitting the core to accrete essentially all
material moving into a small volume around it is to suppress any
buildup of gas in the envelope. Such buildup would otherwise provide a
back pressure and either stall further accretion or, as happens here,
provides a supply of material with which additional infalling material
can interact. For simulations of cores with sufficient mass, well
above that expected to cause the onset of `core instability'
\citep{PP4_WGL}, unimpeded accretion may be a physically relevant and
important process to consider. In the present case however, where we
study the evolution around only 10\me\ cores with very low mass
envelopes, such an assumption is not valid. Unimpeded accretion is
also physically inconsistent with the existence of the massive
envelope implied by the large gravitational softening parameters,
since any such envelope would quickly be accreted into the core
itself.

Our spatial resolution is nearly fine enough to resolve the core (see
section \ref{sec:code}). In this context, implementing a `black hole'
accretion model, in which all material moving into a predefined volume
around the core is accreted, would be inconsistent with the known
character of the core (i.e. that it is solid). Such material becomes
unavailable to interact with other material falling onto the core at
later times. Therefore, models in which accretion is permitted may
unintentially suppress the feedback loop that generates the vigorous
dynamical activity we observe in our simulations. 

Previous 1D work focuses primarily on the secular growth of the
envelope and core over its long term (10$^5$-10$^6$ year) evolution.
It therefore includes detailed treatments of physical processes
relevant over long time scales, while assuming that short term
dynamical fluctuations average out over time. In comparison, our
models include a simpler physical model but fully 3D spatial
resolution. Due to the computational cost of 3D models however, our
simulations extend over an very short period of the total evolutionary
history of the planet's growth. 

Each of these 1D models still requires that a boundary condition be
applied at the outer edge of the envelope. To the meager extent that
defining an envelope can be meaningful in the simulations we present,
our dynamical results are broadly consistent with the condition
assumed in the \citet{liss09} simulations, in which the boundary was
defined at a radius of $\sim$\rh/4 from the core. The \citet{liss09}
assumption is based on the result of dynamical calculations somewhat
similar to our own \citep{dkh03}, in which only tracer particles
closer than \rh/4 from the core remained near it. Our simulations do
not form any envelope structure, and material in the volume that they
define as most strongly bound to the core is most dynamically active
in our simulations. We do however, concur with the result that
material at larger separations (still within the Hill volume) is
largely composed of background disk material whose trajectory has
merely caused it to pass near the core, rather than become bound to
it. This same result is inconsistent with that of \citet{TerHein11},
who find that the volume of gas that is bound to the core will grow to
fill its entire Roche lobe even if it initially fills only some
fraction of it. 

An important concern regarding all 1D models is the issue of whether
or not the assumption that hydrodynamic properties of the system
actually do average out over time, so that a hydrostatic evolutionary
model is accurate. Do hydrodynamic effects force the evolution to
proceed in a direction that it otherwise would not have taken,
considering only the hydrostatic properties of the system? Our 3D
models illustrate that the formation environment for Jovian planets
cannot be adequately described by any hydrostatic treatment. Such a
treatment would imply that material and energy transfer between the
envelope and disk occurs through a stable boundary, with material far
from that boundary being only indirectly affected. On the other hand,
our simulations show that the interactions between disk and envelope
frequently disrupt the envelope entirely, so that its growth must
begin again from scratch. Also, with such vigorous activity, cooling
by any process becomes far less relevant, since the cooled material
does not remain near the core for long enough to affect the long term
evolution of the envelope. Since our simulations extend over only a
tiny fraction of the total formation history of Jovian planets and
also do not incorporate a sufficiently complete physical model of the
evolution, we cannot yet speculate as to the fraction of a planet's
formation history such a statement will be true. In the following two
sections, we discuss the implications of such activity on current core
accretion models, and the components of physical models that we
believe will be required to make such conclusions.

\section{Unanswered questions}\label{sec:questions}

The physical model included in our work is a necessarily simplistic
view of the actual environment of forming Jovian planets and, because
of that, our work generates more questions than it answers about the
growth of Jovian planets and solar system formation. In this section,
we point out a number of such simplifications in our models, discuss
what we believe is the relative importance of each in the context of
Jovian planet formation, and pose a number of questions that can be
addressed as they are removed.

\subsection{Simplifications that affect the thermodynamic state of
the envelope}\label{sec:thermostate}

We made two simplifications in this work which appear to be critical
to making definitive connections between our results and the physical
systems they are intended to model. Each of these factors contribute
directly to models of the internal structure of a hydrostatic (or
nearly so) envelope, determining its energy balance and its density
and temperature structure as a function of radius. They are critical
components of both hydrodynamic and hydrostatic models in this
context. Before any detailed analysis of the conditions will be of
lasting value for theories of planet formation, each must be
accurately accounted for, particularly in the deeper regions of the
envelope where dynamical activity is most vigorous.

First, no account has been taken of the radiative cooling and heating
processes that undoubtedly occur in the envelope, beyond our models
employing fixed equations of state. As discussed in the context of the
possibilities for gravitational instabilities to form planets
\citep{DBMNQR_PP5}, use of fixed equations of state as proxies for
cooling models greatly restricts the behavior of the energy balance in
the gas. Whether or not these restrictions correctly represent the
actual heating and cooling behavior of the Jovian envelope and its
environment, or instead artificially suppress the activity which
should actually be present, must be addressed in follow-up work.

In our simulations, the envelope volume is characterized both by high
temperatures and densities and by rapid changes and large gradients of
the same quantities. It seems likely that including the effects of
radiative transfer will tend to smooth out temperature gradients
throughout the system. How much so, and to what extent are the
temperatures and densities artificially altered from their actual
values by our physical model? To what extent will the dynamical
properties of the system will change when they are included? Will the
dynamical activity continue? Will shocks still develop in the flow?
Preliminary indications with the comparatively extreme restrictions of
locally isothermal and locally isentropic equations of state suggest
that at least some activity remains, so we remain hopeful that our
current treatment is approximately correct.

Unfortunately, the material opacities needed to model radiative
transport accurately are poorly constrained, since they arise largely
from the properties of ice and dust, both in terms of grain sizes
distribution and in terms of their composition. In many regions
throughout the envelope, solids may be vaporized either temporarily or
permanently, causing opacities to change by orders of magnitude over
small spatial and temporal scales. When solids are removed as opacity
sources, the envelope may become optically thin so that radiative
cooling becomes more efficient, particularly closer to the core where
temperatures are highest. Will radiative cooling in these regions
redistribute or remove enough energy from the gas to change the
activity that occurs when radiative processes are neglected? Moreover,
when they reform, grains with similar chemical compositions may have
quite different opacities than originally. Radiative transfer models
which incorporate this grain history will likely require the advection
of tracers in the simulations, which model different chemical
compositions of different solid species. This is challenging, both
in the context of developing a physical model and numerically. To what
extent will such models affect the dynamics? 

Second, no account has been taken of the temperature and density
dependence of the equation of state. Including a complete equation of
state will also cause important changes to the flow because, at
various temperatures, the adiabatic exponent, $\gamma$, of the gas
will differ from the constant value of $\gamma=1.42$ that we have
assumed. While this value will be approximately correct for
temperatures relevant for the background circumstellar disk material
(i.e. $T\sim 100-1000$~K), where at various temperatures rotational
and vibrational modes of hydrogen molecules play a greater or lesser
role, it becomes much less so for higher temperatures and densities
characteristic of the gas deep in the envelope near the core. Above
$T\sim 1500-2000$~K, dissociation and ionization become increasingly
important, and the effective adiabatic exponent falls to values as low
as $\gamma\sim1.1$, before increasing again to $\gamma=5/3$ when the
gas becomes fully dissociated. As shown in figure
\ref{fig:cutout-mid-temp}, the envelope temperatures span this range.
For a given compression, a lower adiabatic exponent means that
pressure increases by a smaller amount. Therefore, our fixed exponent
treatment may artificially increase pressure gradients beyond the
physically relevant levels they should reach, thereby artificially
stimulating dynamical activity to a greater than realistic extent. On
the other hand, lower exponents mean that a given compression event
can increase the mass density to a higher values, perhaps to levels at
which self gravitating instabilities can develop. Which overall effect
will dominate?

Once these thermodynamic effects are included, it will be of interest
to run simulations for far longer in time, in order to explore
questions relating to whether activity continues indefinitely or
instead decays with time. Our current models simulate only a 100~yr
segment of the Jovian planet formation history, during a time when the
envelope did not contain much mass. Because there were no energy loss
mechanisms in our current models, no mass accretion into the envelope
could be expected. Will dynamical activity continue to overwhelm net
mass accretion onto the core, will activity continue while superposed
upon an overall net accretion rate (and at what rate?), or will the
activity simply decay over time to levels unimportant for the
evolution?

Over such long timescales, A third simplification of some significance
to the thermodynamics will also play a role. Namely, the effect of
heating due to planetesimals infalling into the envelope. Over the
short periods simulated in our models, such heating is not likely to
play a large role in the dynamical state, because only a few events
occur during such a short time and because the energy input from any
single event is not a large fraction of the internal energy of the
envelope. Such conditions will not be true over longer periods, during
which many more events occur. How will dynamical activity change when
this additional heating source is accounted for? In our current
simulations, disk material enters the envelope then leaves it again,
carrying with it thermal energy generated in shocks or other
interactions with material near the core. Will a similar outcome hold
for the heating generated by an infalling planetesimal, or will the
additional thermal energy be trapped in the envelope, escaping only
slowly to the background disk?

Longer simulations will permit studies not only of the accretion rate,
but also of the envelope's behavior as it becomes more massive. Does
activity depend on having only a low mass envelope? Does activity
enhance the net accretion rate or reduce it? An answer to this
question is important in the context of the overall survival of a
proto-planet because the observed lifetimes of the circumstellar disks
out of which planet form may only be 2--4 million years \citep{HLL01},
while core accretion models \citep{Pol96} require as many as $\sim$6
million years even in a relatively massive massive disks (2--3 times
the minimum solar nebula).

\subsection{Simplifications in the initial conditions and the
physical realization of the disk and core}\label{sec:simpleIC}

In addition to the simplifications in the thermodynamic treatment, our
models also study only a limited range of the parameter space of
interest in terms of the initial conditions explored. They employ
simplified treatments of several physically relevant components of the
overall system, such as the treatment of the core itself, and the
vertical structure of the disk.

In this work, we have studied the flow around only 10\me\ cores, but
such objects must first, of course, grow from smaller masses. Will
dynamical activity occur in the flow around lower mass cores? The
discussion in section \ref{sec:gen_coremass}, would suggests that
activity will occur, but would be less vigorous for the smallest cores
for conditions typical in the parts of the disk where cores are
expected to form. Is there a lower limit below which activity does not
play an important role in the evolution? If the character of mass
accretion also changes markedly with the strength of the dynamical
activity, then an important observable consequence of such a limit
would be the present day masses of the Jovian planets in our solar
system. Can such a correlation be made?  

Due largely to limits imposed by computational cost, we have
approximated the core as a softened point mass with no actual size or
surface. With only a factor of $\sim2$ increase in spatial resolution
from the highest employed here, simulations would begin to resolve the
core and some, more physical, accounting must then be made of its
structure. We forsee that a simple solid surface realized by a
reflecting boundary condition will be the next step, and will provide
a physical picture adequate to model the influence of the core's
spatial extent on the flow. Of interest will be the question of
whether or not interactions with the surface enhance dynamical
activity, perhaps by being reflected off of it to interact with other
infalling material. Also, the gravitational potential well reaches its
maximum depth at the core's surface. Resolving the flow there means
that the temperatures and densities seen close to the core will
represent the most extreme of those expected in any dynamical flow.
Will some fraction of such material also be ejected into the
background disk flow? Will it carry signatures of its passage through
the core environment?

A still more complex treatment of the surface would require models of
the interaction between the material in the core and the flow. To what
extent does mass exchange between the core and envelope occur? Does
high-$Z$ core material mix into the envelope, (and perhaps also back
out into the disk itself?), thereby enriching it, or does it remain
largely unaffected by the envelope activity? Does such mixing change
with envelope mass?

Indirectly, the accretion rate will also influence the migration rate
of the core through the disk. As its mass increases, the mutual
gravitational torques between the core and disk will increase in turn
and the core interacts more strongly. Whether the stronger interaction
leads to increased migration will depend on details of the mass
distribution around the core. Accounting only for torques from
Lindblad resonances \citep[see e.g.,][]{Ward97a}, migration will
accelerate. Accounting for corotation torques as well, migration may
decelerate or even reverse direction \citep{PAM2,PAM3}.

While omitting gravitational forces in the $z$ coordinate precludes a
description of the disk's vertical structure, we expect that the
influence of that structure on the envelope activity will not be
large. Here again, the activity is limited only to the immediate
environment of the core, a factor of several in spatial extent smaller
than the disk's scale height. A possibly more important consequence of
such a treatment will be the possibility of mass being ejected
entirely out of the disk, to fall later on some region distant to the
core.  In this case, the question of how large the accretor must be
before it first begins to deplete mass from an entire vertical column
of the disk. For an accretor of that size, accretion may be reduced
due to the loss of material accreting onto the poles of the
proto-planet and a gap may form due to the accretion of most of the
material near the planet, also slowing additional migration through
the disk.

Of somewhat lesser importance for the near term are the geometric
approximations imposed by our choice of coordinate system and our
omission of background gravitational forces in the $z$ (vertical)
coordinate of our disks. Mitigating the importance of both cases, is
the fact that activity is limited to a comparatively small volume
close to the core. Therefore, the coordinate system itself plays only
a small role. 

\subsection{Is there a connection between models of Jovian planet
formation models and models of chondrule
formation?}\label{sec:consequences}

As noted above, a consequence of dynamical activity in the envelope is
that material is not irrevocably bound to the core after first
entering the envelope environment. Instead, some fraction of the
material returns to the circumstellar disk from which it originated.
Will this material retain a signature of its passage through the hot,
circumplanetary region? The possibility is intriguing because if so,
some such material could still exist today, potentially providing a
record of the conditions present during the formation of Jovian
planets that could be directly observed. In fact, a common
class of meteorites--chondrites--exist \citep[see e.g.][]{PP3_PalBoy}
and are found throughout the solar system, which exhibit signatures of
high temperature processing on time scales of only a few hours or
days. Can the formation of Jupiter actually be linked to the formation
of chondrules observed in the meteorite record though a model such as
ours? Can it be linked to the formation of other materials, such as
the annealed silicates found in comets \citep{HD02}, for which the
required temperatures and densities are lower?

Although the results from our initial streamline analysis are
promising, they are no substitute for an investigation of the
trajectories of specific packets of material through the system. In
order to establish any real connection, we must perform a much more
detailed analysis of the conditions encountered by such packets. Is a
particle's thermodynamic trajectory the same when it is advected
through an actual time dependent flow, as opposed to the fictitious
advection through a fixed flow that we have performed? Will more
detailed analysis show that the shocks that are generated fit into the
required density/temperature/velocity parameter space? One concern
already apparent is that the flow velocities of material flowing
through the shocks (1-2 km~s$^{-1}$, as estimated from the directly
available fluid flow velocities themselves) are uncomfortably low
compared to those quoted by \cite{DC02} and \cite{iida01}. It seems
unlikely that the velocities will be increased as dramatically as that
by any of the improvements to the models we might make. Finally, what
fraction of material encounters conditions appropriate for chondrule
formation during its passage, in comparison to material that instead
encounters regions that are inappropriate? And what fraction of the
total budget of solid material in the solar nebula undergoes such
processing? Equivalently, do conditions appropriate for chondrule
formation exist for a large or small fraction of the disk lifetime, so
that a larger or smaller fraction of solids encounter those conditions
during their evolution? Finally, how do processed materials get from
where they form (near 5~AU) to their final locations, in meteorites
throughout the inner solar system?

\subsection{Future Directions}\label{sec:future}

We believe the most profitable course following this work will be to
investigate the character of activity that is generated for cores of
different masses, in order to confirm directly the scaling
relationship discussed in section \ref{sec:gen_coremass}. In parallel
with this effort, models employing an accurate equation of state for
the circumstellar material will be of great interest. Such an
equation of state must include the various molecular and atomic states
of hydrogen, which will likely cause changes to the strength of the
hydrodynamic feedback through the changes they imply for the ratio of
specific heats, $\gamma$. Ultimately however, we expect that models
including radiative transfer will be required to accurately model the
full process of mass flow and accretion in the neighborhood of forming
planetary cores. In future work, we hope to explore all of these  
questions.

\appendix

\section{Differences from the standard shearing sheet
formalism}\label{sec:equation-diffs}

The differences between the standard shearing sheet model and our
modifications to it are to be found in the equations of motion in the
$x$ and $y$ directions, equations \ref{eq:force-x} and
\ref{eq:force-y}, corresponding to an underlying radial and azimuth
coordinate frame that is rotating with the angular speed of the core. 
Additional differences are found in the specification of the boundary
treatments, as discussed in section \ref{sec:bound} above. To
illustrate the differences in mathematical formalisms we show here the
approximations underlying both our formalism and the shearing sheet
and their differences, starting from the equations of motion in a
non-rotating cylindrical coordinate frame.

In an inertial, cylindrical coordinate frame, the equation of motion
in the radial and azimuthal directions are
\begin{eqnarray}\label{eq:r-cyl-iner}
      {{\partial (\rho V_r       )}\over{  \partial t   }} + 
      {{\partial (\rho V_r V_r   )}\over{  \partial r   }} +  
      {{\partial (\rho V_r V_\phi)}\over{r \partial \phi}} +  
      {{\partial (\rho V_r V_z   )}\over{  \partial z   }} -
      {{ \rho V_\phi^2 }\over{ r}}
      & & \nonumber \\
    = -       {{\partial p   }\over{\partial r}}
      - \rho {{\partial \Phi }\over{\partial r}},
      & &
\end{eqnarray}
and
\begin{eqnarray}\label{eq:phi-cyl-iner}
      {{\partial (\rho V_\phi       )}\over{  \partial t   }} + 
      {{\partial (\rho V_\phi V_r   )}\over{  \partial r   }} +  
      {{\partial (\rho V_\phi V_\phi)}\over{r \partial \phi}} +  
      {{\partial (\rho V_\phi V_z   )}\over{  \partial z   }} +
      {{ \rho V_r V_\phi }\over{ r}}
      & & \nonumber \\
    = -       {{\partial p   }\over{r \partial \phi}}
      - \rho {{\partial \Phi }\over{r \partial \phi}},
      & &
\end{eqnarray}
where the capital letters on the velocities are used to denote the
inertial frame velocities.  In equation \ref{eq:r-cyl-iner}, the $\rho
V_\phi^2/r$ term defines an analogue of the centrifugal force which,
mathematically, comes from the fact that the full time derivative of
the momentum includes components that account for the fact that
direction vectors $\hat r$ and $\hat \phi$ are dependent on position
and through them, also time. In equation \ref{eq:phi-cyl-iner}, the
$\rho V_r V_\phi/r$ term plays a similar role, as a near analogue of
the Coriolis force.

When translating to a frame rotating at angular velocity $\Omega_{\rm
fr}$, we must include centrifugal and Coriolis terms of the form
$\mathbf \Omega_{\rm fr}\times(\mathbf\Omega_{\rm fr}\times \mathbf r)
- 2\mathbf\Omega_{\rm fr}\times\mathbf v$. Additionally, the meanings
of each of the variables as defined above changes, to reflect the
moving coordinate system. For purposes of clarity, we note the
identifications $(V_r \rightarrow v_r, V_\phi \rightarrow r\Omega_{\rm
fr} + v_\phi, V_z \rightarrow v_z)$ for each of the coordinate
direction. In other words, lower case variable names denote quantities
in the moving frame. Then, the form of equations \ref{eq:r-cyl-iner}
and \ref{eq:phi-cyl-iner} change to include the additional terms as
follows:
\begin{eqnarray}\label{eq:r-cyl-rot}
      {{\partial (\rho v_r       )}\over{  \partial t   }} + 
      {{\partial (\rho v_r v_r   )}\over{  \partial r   }} +  
      {{\partial (\rho v_r v_\phi)}\over{r \partial \phi}} +  
      {{\partial (\rho v_r v_z   )}\over{  \partial z   }} -
      {{ \rho v_\phi^2 }\over{ r}}
      & & \nonumber \\
      -{2\rho v_\phi \Omega_{\rm fr} } -\rho r \Omega_{\rm fr}^2 
    = -       {{\partial p   }\over{\partial r}}
      - \rho {{\partial \Phi }\over{\partial r}},
      & &
\end{eqnarray}
and
\begin{eqnarray}\label{eq:phi-cyl-rot}
      {{\partial (\rho v_\phi       )}\over{  \partial t   }} + 
      {{\partial (\rho v_\phi v_r   )}\over{  \partial r   }} +  
      {{\partial (\rho v_\phi v_\phi)}\over{r \partial \phi}} +  
      {{\partial (\rho v_\phi v_z   )}\over{  \partial z   }} +
      {{ \rho v_r v_\phi }\over{ r}}
      & & \nonumber \\
      +{{2\rho v_r \Omega_{\rm fr}}}
    = -       {{\partial p   }\over{r \partial \phi}}
      - \rho {{\partial \Phi }\over{r \partial \phi}}.
      & &
\end{eqnarray}

The cross term in equation \ref{eq:r-cyl-rot} arises from the Coriolis
term, $2\mathbf\Omega_{fr}\times \mathbf v$, while the first and third
terms come from the centrifugal forces due to the curvi-linear
coordinate system and the rotating frame, respectively. Together, the
three terms in eqution \ref{eq:r-cyl-rot} form a perfect square
identical to the corresponding single term in equation
\ref{eq:r-cyl-iner}. We complete the specification of the equations of
motion defined in equations \ref{eq:force-x} and \ref{eq:force-y}, by
replacing the cylindrical coordinate variables $(r,r\phi)$ with
$(x,y)$.

In both the treatment above and the shearing sheet, the identification
of the coordinates pairs $(x, r)$ and $(y, r\phi)$ is made in each of
the hydrodynamic equations. The shearing sheet approximation goes
further than this by omitting the Coriolis-like term due to the
curvature of the coordinate system (i.e. the term $\rho v_r v_\phi/r$,
above) and expands the sum of the centrifugal force and the
gravitational force of the central star in a Taylor series, for which
the zeroth order term is assumed to be zero, and only the linear term
is retained. \citet{HGB95} state the form of the equations to be
solved in the shearing sheet approximation, cast as partial derivatives
of velocities. For comparison to the form used here, we recast these
equations in momentum form, omitting all terms relating to magnetic
fields, which we neglect entirely:
\begin{eqnarray}\label{eq:force-x-shearing}
      {{\partial (\rho v_x    )}\over{\partial t}} + 
      {{\partial (\rho v_x v_x)}\over{\partial x}} +  
      {{\partial (\rho v_x v_y)}\over{\partial y}} +  
      {{\partial (\rho v_x v_z)}\over{\partial z}} 
      & & \nonumber \\
      - 2 {\rho v_y} \Omega_{\rm fr} 
    = -  {{\partial p   }\over{\partial x}}
      +    2 \rho q \Omega^2 x
      & &
\end{eqnarray}
\begin{eqnarray}\label{eq:force-y-shearing}
      {{\partial (\rho v_y    )}\over{\partial t}} + 
      {{\partial (\rho v_y v_x)}\over{\partial x}} +
      {{\partial (\rho v_y v_y)}\over{\partial y}} +
      {{\partial (\rho v_y v_z)}\over{\partial z}}  
      & & \nonumber \\
     +{{2 \rho v_x \Omega_{\rm fr} }}
    = -      {{\partial p    }\over{\partial y}}
      & &
\end{eqnarray}
where $q= -{\partial \ln \Omega_{\rm fr}}/{\partial \ln r}\approx
3/2$, for a nearly Keplerian disk. The last term on the right hand
side of equation \ref{eq:force-x-shearing} defines the first order
approximation to the effective potential in which the gas flows noted
above. Rather than use this Taylor expansion approach, we retain the
full form of the background potential of the star and disk, assuming
only that it is constant in time. This permits a direct specification
of the gravitational potential to be made, rather than an analytic
approximation. It also permits transient velocity fluctuations that
temporarily upset the assumption that centrifugal and gravitational
forces are in equilibrium to be accounted for more fully.

\section{Numerical issues regarding treatment of the
boundaries}\label{sec:bound-probs}

The simplest method of producing a flow-out boundary in either the
inner or outer $x$ (`radial') direction would be to reproduce the
flow variables of the last interior grid zone in the boundary cells
adjacent to that zone and just outside the grid. We have found this 
method unsatisfactory for our simulations because the background state 
is effectively a dynamical equilibrium condition rather than a static
equilibrium condition. Implementing the simplest flow-out boundary
changes the gradients there (e.g. the pressure gradient partially
responsible for the underlying rotation curve), and is equivalent to
violating the dynamical equilibrium condition. The effect of an
initially small outflow at the negative $x$ boundary, corresponding to
the inner radial boundary of the underlying disk, is to flatten the
gradient and further strengthen the outflow. At the positive $x$
boundary a flattened gradient has the opposite effect: the effect of
an initially small outflow is to temporarily create a very strong,
local {\it inflow} condition.

Since the major gradients of the background flow are in the radial
direction ($x$ direction in the simulation cube), the same concern
does not apply to the $y$ boundaries, however they can still affected
by another problem. If the disparity between the smooth background
state and the flow near the boundary is large, the simulated flow can
still be adversely affected. In particular, the fixed condition at the
boundary means that if some volume of gas flows towards that boundary
but does not approximately match the conditions there, it may either
be accelerated as it leaves the grid, causing mass further inside the
simulation volume to be similarly accelerated into the prematurely
evacuated volume, or it may be decelerated, effectively `piling up' at
the boundary and perhaps reentering the flow.   

We have closely monitored the flow in our simulations for consequences
of this effect and have observed it in only one (labeled {\it tm05} in
table \ref{tab:sims} above). We believe that the failure of this
simulation is due primarily to the fact that the assumed background
temperature was very low. As a consequence the Jeans wavelength of the
mass in the grid became roughly comparable to the linear dimension of
the simulation cube itself. As the planet drew additional mass into
its influence, the perturbation on the local potential grew as well
and further amplified the effect. The conditions just inside the
boundary were then no longer similar to those just outside it and
violated our smooth background assumption, resulting in matter turning
back into the simulation volume and perturbing the evolution still
further. A separate simulation, neglecting the local disk self
gravity, did not suffer from this defect, and we discuss it in section
\ref{sec:backt-signif} below as a more physically realistic
alternative to simulation {\it tm05}, with a nearly identical physical
model to its higher temperature cousins.

\section*{Acknowledgments}
We thank the anonymous referee for generous suggestions to improvement
the manuscript. We also thank G. D'Angelo for his comments, which we
used to clarify a number of the arguments made herein. We acknowledge
the support of the University of Edinburgh Development Trust. The
computations reported here were performed using the UK Astrophysical
Fluids Facility (UKAFF) using time allocations during 2003-2004, which
also provided financial support to AFN. Part of this work was carried
out under the auspices of the National Nuclear Security Administration
of the U.S. Department of Energy at Los Alamos National Laboratory
under Contract No. DE-AC52-06NA25396, for which this is LA-UR-12-26434.


\begin{thebibliography}{}

\bibitem[Alibert \etal(2004)]{AMB04} Alibert, Y., Mordasini, C.,
Benz, W. 2004 A\&A 417, L25

\bibitem[Ayliffe \& Bate(2009)]{AB09} Ayliffe, B. A., Bate, M. R.,
2009, MNRAS, 393, 49 (AB09)

\bibitem[Balbus, Hawley \& Stone(1996)]{BHS-MRI96} Balbus, S. A.,
Hawley, J. F., Stone, J. M., 1996, ApJ, 467, 76

\bibitem[Bate \etal(2003)]{BLOM} Bate, M. R., Lubow, S. H., Ogilvie,
G. I., Miller, K. A., 2003, MNRAS, 341, 213

\bibitem[Beckwith \etal(1990)]{BSCG} Beckwith, S. V. W., Sargent, A.
I., Chini, R. S. \& G\"usten, R., 1990, AJ, 99, 924

\bibitem[Boss(1995)]{Boss95} Boss, A.P., 1995, Science, 267, 360

\bibitem[Binney \& Tremaine(1987)]{GalDyn} Binney, J.,
Tremaine, S., 1987, Galactic Dynamics, Princeton University
Press: Princeton

\bibitem[Ciesla \& Hood(2002)]{CH02} Ciesla, F. J. \& Hood, L. L 2002,
Icarus, 158, 281

\bibitem[Colella \& Woodward(1984)]{ColWood84} Colella P.,
Woodward, P. R., 1984, J. Comp. Phys., 54, 174

\bibitem[D'Angelo \etal(2002)]{DHK02} D'Angelo, G.,
Henning, T., Kley, W., 2002, A\&A, 385, 647

\bibitem[D'Angelo \etal(2003a)]{dkh03} D'Angelo, G., Henning, T.,
Kley, W., 2003, ApJ, 586, 540

\bibitem[D'Angelo \etal(2003b)]{dhk03} D'Angelo, G., Kley, W.,
Henning, T., 2003, ApJ, 599, 548

\bibitem[Desch \& Connolly(2002)]{DC02} Desch, S. J., Connolly, H. C.
Jr., Meteoritics \& Planetary Science, 37, 187

\bibitem[Durisen \etal(2007)]{DBMNQR_PP5} Durisen R. H., Boss, A. P.,
Mayer, L., Nelson, A. F., Quinn, T. \& Rice, W. K. M. 2007, In
Protostars and Planets 5, p607-622, ed. Reipurth, B., Jewitt, D. \&
Keil, K. University of Arizona Press:Tucson 

\bibitem[Foglizzo \& Ruffert(1999)]{FR99} Foglizzo, T., Ruffert, M.,
1999, A\&A 347, 901

\bibitem[Goldreich \& Lynden-Bell(1965)]{GLB65} Goldreich, P.,
Lynden-Bell, D., 1965 MNRAS, 130, 7

\bibitem[Haisch \etal(2001)]{HLL01} Haisch, K. E., Lada, E. A., 
Lada, C. J., 2001, ApJL, 553, 153

\bibitem[Harker \& Desch(2002)]{HD02} Harker, D. E. \& Desch, S. J.
2002, ApJ 565, 109

\bibitem[Hawley, Gammie \& Balbus(1995)]{HGB95} Hawley, J. F., Gammie.
C. F., Balbus, S. A., 1995, ApJ, 440, 742

\bibitem[Helled, Podolak \& Kovetz(2008)]{HPK08} Helled, R., Podolak
M., Kovetz, A., 2008 Icarus, 195, 863

\bibitem[Hubickyj \etal(2005)]{HBL05} Hubickyj, O., Bodenheimer, P.,
Lissauer, J. J., 2005, Icarus, 179, 415

\bibitem[Iida \etal(2001)]{iida01} Iida, A, Nakamoto, T. \&
Susa, H. 2001, Icarus, 153, 430

\bibitem[Jones \etal(2000)]{PP4_Jones} Jones, R. H., Lee,
T., Connolly, H. C. Jr., Love, S. G. \& Shang, H. 2000, In Protostars
and Planets IV, pp. 927-962, ed. Mannings, V., Boss, A. P. \&
Russell, S. S., University of Arizona Press: Tucson

\bibitem[Lissauer \etal(2009)]{liss09} Lissauer, J. J., Hubickyj, O.,
D'Angelo, G., Bodenheimer, P., 2009, Icarus, 199, 338

\bibitem[Lubow \etal(1999)]{lsa99} Lubow, S. H. Seibert, M.,
Artymowicz, P., 1999, ApJ, 526, 1001

\bibitem[Machida \etal(2008)]{MKIM08} Machida, M. N., Kokubo, E.,
Inutsuka, S., Matsumoto, T. 2008, ApJ, 685, 1220 

\bibitem[Masset(1997)]{masset_thesis} Masset, F., PhD Thesis, University
of Paris

\bibitem[Miyoshi \etal(1999)]{MTTI} Miyoshi, K., Takeuchi, T., Tanaka, H.,
Ida, S., 1999, ApJ, 516, 451

\bibitem[Nelson \etal(1998)]{DynI} Nelson, A. F., Benz, W.,
Adams, F. C., Arnett, W. D., 1998, ApJ, 502, 342

\bibitem[Nelson \etal(2000)]{DynII} Nelson, A. F., Benz, W.,
Ruzmaikina, T. V., 2000, ApJ, 529, 357

\bibitem[Nelson \& Benz(2003)]{JovI} Nelson, A. F., Benz, W., 2003, 
ApJ, 589, 556

\bibitem[Palme \& Boynton(1993)]{PP3_PalBoy} Palme, H., Boynton, W. V.,
in Protostars and Planets III, ed. Lunine, J. I., Levy, E. University
of Arizona Press: Tucson

\bibitem[Papaloizou \& Nelson(2005)]{PapNel05} Papaloizou, J. C. B.,
Nelson, R. P., 2006, A\&A, 433, 247

\bibitem[Peplinski, Artymowicz \& Mellema(2008a)]{PAM2} Peplinski, A.,
Artymowicz, P., Mellema, G., 2008a, MNRAS, 386, 179

\bibitem[Peplinski, Artymowicz \& Mellema(2008b)]{PAM3} Peplinski, A.,
Artymowicz, P., Mellema, G., 2008b, MNRAS, 387, 1063

\bibitem[Pickett \etal(1996)]{PDD} Pickett, B. K., Durisen, R. H.,
Davis, G. A., 1996, ApJ, 458, 714

\bibitem[Pickett \etal(1998)]{Pick98} Pickett, B. K., Cassen, P.,
Durisen, R. H., Link, R., 1998, ApJ, 504, 468

\bibitem[Pickett \etal(2000)]{Pick00} Pickett, B. K., Cassen, P.,
Durisen, R. H., Link, R., 2000, ApJ, 529, 1034

\bibitem[Pickett \etal(2003)]{Pick03} Pickett, B. K., Mejia, A. C.,
Durisen, R. H., Cassen, P. M., Berry, D. K., Link, R., 2003, ApJ, 
590, 1060

\bibitem[Pollack \etal(1996)]{Pol96} Pollack, J. B., Hubickyj, O.,
Bodenheimer, P., Lissauer, J. J., Podolak, M., Greenzweig, Y.,
1996, Icarus, 124, 62

\bibitem[Price(2012)]{price12} Price, D. J.. 2012, J. Comp. Phys, 231, 759

\bibitem[Rafikov(2006)]{Raf06} Rafikov, R., 2006, ApJ, 648, 666

\bibitem[Ruffert(1992)]{R92} Ruffert, M., 1992, A\&A, 265, 82

\bibitem[Ruffert(1994)]{R94} Ruffert, M., 1994, ApJ, 427, 342

\bibitem[Ruffert(1997)]{R97} Ruffert, M., 1997, A\&A, 317, 793

\bibitem[Ruffert(1999)]{R99} Ruffert, M., 1999, A\&A, 346, 861

\bibitem[Terquem \& Heinemann(2011)]{TerHein11} Terquem, C.,
Heinemann, T., 2011, MNRAS, 418, 1928

\bibitem[Tanigawa, Ohtsuki \& Machida(2012)]{TOM12} Tanigawa, T.,
Ohtsuki, K., Machida, M. N., 2012, ApJ, 747, 47 

\bibitem[Toman \etal(1998)]{TIPD} Toman, J., Imamura, J. N.,
Pickett, B. K., Durisen, R. H., 1998, ApJ, 497, 370

\bibitem[Truelove \etal(1997)]{Truelove97} Truelove, J. K.,
Klein, R. I., McKee, C. F., Holliman, J., H., Howell, L. H.,
Greenough, J. A., 1997, ApJL, 489, 179L

\bibitem[Ward(1997a)]{Ward97a} Ward, W. R. 1997, Icarus, 126, 261

\bibitem[Ward \& Hahn(2000)]{PP4_WH} Ward, W. R., Hahn, J., M.,
2000, In Protostars and Planets 4, ed. Mannings, V., Boss, A. P.
\& Russell, S. S., University of Arizona Press: Tucson

\bibitem[Wuchterl \etal(2000)]{PP4_WGL} Wuchterl, G., Guillot, T.,
Lissauer, J. J., 2000, In Protostars and Planets 4, ed. Mannings, V.,
Boss, A. P. \& Russell, S. S., University of Arizona Press: Tucson

\end{thebibliography}
\end{document}